\pdfoutput=1
\documentclass[11pt]{arxiv-article}

\usepackage{simeontex-susy}
\usepackage{macros-susy}
\usepackage{definitions}

\usepackage{cite}
\allowdisplaybreaks[2]
\newcommand*{\univ}{^{\text{[univ]}}}

\RequirePackage{pifont}

\RequirePackage{hyperref}

\allowdisplaybreaks[2]
\preprint{\texttt{IPMU18-0059\\CALT-TH-2018-014}}

\newcommand{\OfficialTitle}{
  Universal correlation functions in rank 1 SCFTs
}

\title{
  \Huge\color{Thoughtless}
  \textbf{
      \dosserif
      \OfficialTitle
    }  
}

\hypersetup{pdfauthor={Simeon Hellerman and Shunsuke Maeda and Domenico Orlando and Susanne Reffert and Masataka Watanabe},pdftitle={\OfficialTitle}}

\author{%
  \begin{minipage}{.8\linewidth}
    \begin{center} \dosserif
      {\small
        \textbf{Simeon Hellerman}\textsuperscript{\ding{72}}, \textbf{Shunsuke Maeda}\textsuperscript{\ding{72}}, \textbf{Domenico Orlando}\textsuperscript{\ding{74}\ding{73}}, \textbf{Susanne Reffert}\textsuperscript{\ding{73}} and \textbf{Masataka Watanabe}\textsuperscript{\ding{72}}
         }
    \end{center}
    \authorBlock{\ding{72}}{Kavli Institute for the Physics and Mathematics of the Universe (WPI)\\
      The University of Tokyo\\
      Kashiwa, Chiba 277-8582, Japan}
    \authorBlock{\ding{74}}{INFN sezione di Torino and Arnold--Regge Center\\
    via Pietro Giuria 1, 10125, Turin, Italy}
    \authorBlock{\ding{73}}{Albert Einstein Center for Fundamental Physics\\
      Institute for Theoretical Physics\\
      University of Bern,\\
      Sidlerstrasse 5, \textsc{ch}-3012 Bern, Switzerland}
  \end{minipage}
}

\date{}

\newcommand*{\JJM}{{\color{Blue}\mathcal{J}}}
\def\ampname{{\cal Y}}
\def\D{{\Delta}}
\def\ll{_}
\def\muchlessthan{<\hskip-.07in <}
\def\L{\Lambda}
\def\cc{\,}
\def\t{\tau}
\def\th{\theta}
\def\lrmns#1{_{\rm {#1}}}
\def\dilaton{{\color{Sorbus} \tau}}
\def\b{\beta}
\def\Phb{\Phi^\dagger}
\def\kk{\kappa}
\def\upsns#1{^{[{#1}]}}
\def\tb{{\bar{\tau}}}
\def\bfuc{{\color{Sorbus} \beta}}

\newenvironment{PurpleEnv}%
{\color{Purple}}%
{\color{Black}}
{\color{Blue}}%
{\color{Black}}
{\color{Red}}%
{\color{Black}}

\def\o{\omega}

\def\purple#1{\begin{PurpleEnv} {#1} \end{PurpleEnv}}

\def\r{\rho}
\def\d{\delta}

\def\opt#1#2{\redd{[{#1}/{#2}]}}

\def\l{\lambda}

\def\Da{\Delta a}
\def\gg{\nabla}

\def\reddit#1{#1}
\def\ok{\blue{\checkmark}}
\begin{document}

\numberwithin{equation}{section}

\begin{titlepage}

  \newgeometry{top=23.1mm,bottom=46.1mm,left=34.6mm,right=34.6mm}

  \maketitle

  \thispagestyle{empty}

  \abstract{\normalfont \noindent
Carrying to higher precision the large-$\mathcal{J}$ expansion of
~\cite{Hellerman:2017sur}, we calculate to all orders in $1/\mathcal{J}$ the power-law corrections
to the two-point functions $\mathcal{Y}_n \equiv |x - y|^{2n\Delta_{\mathcal{O}}}
\langle  {\mathcal{O}}^n(x) \bar{\mathcal{O}}^n(y) \rangle$ for generators $\mathcal{O}$ of
Coulomb branch chiral rings
in four-dimensional $\mathcal{N} =2$ superconformal field theories.  We show
these correlators have the universal large-$n$ expansion
\[
\log(\mathcal{Y}_n) \simeq \mathcal{J} \mathbf{A} + \mathbf{B} + \log(\Gamma( \mathcal{J} + \alpha + 1))  ,
\]
where $\mathcal{J} \equiv  n \Delta_{\mathcal{O}}$ is the total $R$-charge of $\mathcal{O}^n$, the $\mathbf{A}$ and $\mathbf{B}$ are theory-dependent coefficients, $\alpha$ is the coefficient of the Wess--Zumino
term for the Weyl $a$-anomaly,  and the $\simeq$ denotes equality
up to terms exponentially small in $\mathcal{J}$. 
Our methods combine the structure of the Coulomb-branch
\ac{eft} with the 
supersymmetric recursion relations. However, our results
constrain the power-law corrections to all orders,
even for non-Lagrangian theories to which the recursion
relations do not apply.  For the case of
$\mathcal{N} = 2$ \acs{sqcd}, we also comment on
the nature of the exponentially small corrections, which
can be calculated to high precision in the double-scaling limit
 recently discussed by Bourget \emph{et al.} in~\cite{Bourget:2018obm}.  We show
the exponentially small correction is consistent
with the interpretation of the \ac{eft} breaking down
due to the propagation of massive \acs{bps} particles over
distances of order of the infrared scale $\abs{x - y}$.    %
    }

\vfill

\end{titlepage}

\restoregeometry
\setstretch{1.15}

\tableofcontents

\newpage
\acresetall

\section{Introduction}
\label{sec:introduction}

Dealing analytically with strongly coupled theories is difficult. However, working in a sector of fixed large quantum number $\JJM$ associated to a global symmetry leads to important simplifications. Starting from a strongly coupled \ac{cft}, it allows us to write down an approximately scale-invariant \ac{eft} in which most terms are suppressed by inverse powers of $\JJM$~\cite{Hellerman:2015nra}. The low-energy physics is governed by one or several Goldstone fields which encode the quantum fluctuations around the fixed-charge ground state. The resulting large-quantum-number expansion is very sensitive to the vacuum structure of the theory. In the non-supersymmetric cases of the critical $O(n)$ vector model~\cite{Hellerman:2015nra, Alvarez-Gaume:2016vff} and the $SU(n)$ matrix models~\cite{Loukas:2017lof, Loukas:2017hic}, there is a unique fixed-charge ground state which is homogeneous in space and the anomalous dimension of the lowest operator of charge $\JJM$ comes with a leading $\JJM^{3/2}$ scaling. The large-$R$-charge expansion of the $\mathcal{N}=2$ superconformal $W = \Phi^3$ theory which has no moduli space displays the same properties and the same leading $\JJM^{3/2}$ scaling, which signals the spontaneous breaking of \ac{susy}~\cite{Hellerman:2015nra}. Things are very different for \acp{scft} with a non-trivial moduli space. The moduli space of vacua implies a degenerate spectrum when the curvature of the manifold on which the \ac{scft} lives vanishes, and consequently the curvature is always relevant in the large-quantum-number expansion. The leading  $\JJM$ behavior of the conformal dimension is $\Delta_{\JJM} \sim +1 \cdot \abs{\JJM}^1$~\cite{Hellerman:2017veg, Hellerman:2017sur}.

\medskip
In theories without a small loop-suppressing parameter, in all cases where it can be checked against other methods,
such as Monte Carlo simulations on the lattice~\cite{Banerjee:2017fcx}, the conformal bootstrap~\cite{Rattazzi:2008pe,El-Showk:2014dwa,ElShowk:2012ht,Rychkov:2016iqz,Simmons-Duffin:2016gjk,Alday:2016njk} (and references therein), and exact supersymmetric
methods~\cite{Hellerman:2017veg,Hellerman:2017sur,Gerchkovitz2017,Baggio:2014ioa,Baggio:2014sna,Baggio:2015vxa}, the large-quantum-number expansion converges very well to the correct answer.

In this paper we focus on the last of these, carrying to higher order the previous results of~\cite{Hellerman:2017sur} in an ${\cal N} = 2$ superconformal gauge theory
in four spacetime dimensions. In~\cite{Hellerman:2017sur},  the \(1/\JJM\) expansion of correlation functions of chiral primary operators of dimension $\D$ and
$R$-charge $\JJM = \D\lrm{total} = n \D\ll\co$ were calculated, where  $\co$ is a generator of the holomorphic
coordinate ring of a one-complex dimensional Coulomb branch.%
\footnote{In cases where the conformal dimension $\Delta_{\cal O}$ is fractional,
there may be interesting oscillatory corrections to our asymptotic formula with a period set by the denominator of $\Delta_{\cal O}$.  It would be interesting to explore this question but for purposes of the present paper we simply choose $n$ such that ${\cal J} = n \Delta_{\cal O}$ is an integer. See ~\cite{Argyres:2018urp,Caorsi:2018zsq} for related work.}
We compute all the terms in the asymptotic expansion, and get an explicit universal result for any theory with a one-dimensional Coulomb branch, even for models without marginal couplings. Where we can compare our results with results from localization techniques, we find beautiful agreement and can even see the leading non-universal corrections from numerical data.

\medskip
The \ac{eft} of the Coulomb branch was used in~\cite{Hellerman:2017sur} to estimate the two-point
function
\bbb
 |x - y|\uu{ 2 n\D\ll\co}\cc\left< (\co (x) )\uu n
\cc  (\bar{\co}(y) )\uu n \right> \equiv  \ampname\ll n .%
\eee
The $\ampname\ll n$ are defined as correlation functions, %
\begin{equation}
  \ampname\ll n = |x - y|\uu{ 2 n\D\ll\co}\cc \frac{Z\ll n}{Z_0}\ ,
\end{equation}
where $Z\ll n$ is the path integral with insertions of the operators and 
$Z\ll 0$ is the path integral function without insertions.
Up to the Weyl transformation of the insertions and
the Weyl anomaly of 
the partition function itself, the path integral is the
same on any conformally flat space. In particular
the expectation values $\ampname$ transform
covariantly under Weyl transformations, with the Weyl
anomaly canceling when one divides $Z\ll n$ by $Z\ll 0$.
It is convenient to perform
some calculations in the conformal frame of the sphere
$S\uu 4$, in which we can compare our results with
those of supersymmetric localization, \it e.g. \rm~\cite{Gerchkovitz2017}.
 
Translating to the conventions of~\cite{Gerchkovitz2017}, the absolute normalization of the path integral with
and without insertions is defined as
\bbb
Z\ll 0 \equiv \exp{q\ll 0}\ ,
\xxn
Z\ll n \equiv \exp{q\ll n}\ ,
\xxn
\ampname\ll n \equiv \abs{x - y}\uu{ 2 n\D\ll\co} 16^{-n} G\ll{2n} = \abs{x - y}\uu{ 2 n\D\ll\co}\cc \exp{q\ll n - q\ll 0}\ .
\eee
Indeed, the quantity $Z\ll n \equiv \exp{q\ll n} \equiv Z\ll 0 \cc \ampname\ll n $ is a particularly
natural object in our way of computing the two-point function, as it can be thought of as a path integral with sources,
which can be computed with relative ease in the \ac{eft} of the Coulomb branch.

In~\cite{Hellerman:2017sur}, the expansion of $q\ll n$ was carried to order $\log(\JJM)$, yielding the result
\begin{equation}
  \begin{aligned}
    \log(\ampname\ll n) %
    & = q\ll n - q\ll 0 = \JJM\cc \log(\JJM)+   n \cc \log({\bf N}\ll\co)  + (\a + \hh)\cc \log(\JJM) + \order{\JJM^0}\\
    &= \log(\JJM!) +   n \cc \log({\bf N}\ll\co) + \a\cc \log(\JJM) + \order{\JJM^0}\ ,
  \end{aligned}
\end{equation}
where $\a$ is the coefficient of the Wess--Zumino term for the Weyl $a$-anomaly.
The constant ${\bf N}\ll\co$ is
a normalization constant that can be absorbed into the normalization of $\co$ itself, though for theories with a marginal
coupling the operator $\co$ has dimension $\D = 2$ and there is a natural normalization of $\co$ traceable to its relation
with the actual marginal operator which is its descendant.  

In this note we shall use the \ac{eft}
of the Coulomb branch directly to calculate the
large-$\JJM$
expansion of $q\ll n$ up to
and including order \(1/\JJM\).  
From there, we make a number of observations about
the subleading corrections to the large-$\JJM$ behavior
of $q\ll n$.

Our main result is that the correlators for any theory with a one-dimensional Coulomb branch (whether it has a marginal
coupling constant or not) have a universal large-\(\JJM\) behavior controlled by the formula
\begin{equation}
  \begin{aligned}
    q\ll n &\simeq
    \JJM\log(\JJM) + \pqty{\a + \hh} \cc\log(\JJM) + \pqty{{\bf A} - 1} \JJM + {\bf B} + \log \sqrt{2\pi} + \sum\ll{m \geq 1} \frac{\hat{K}\ll m(\a)}{\JJM^m} \\
 &\simeq \JJM\cc {\bf A} + {\bf B} + \log[ \cc \G(\JJM + \a + 1) \cc]\ ,
\end{aligned}
\end{equation}
where ${\bf A}$ and ${\bf B}$ are theory-dependent coefficients.
The \( \simeq\) indicates the presence of non-universal corrections that are exponentially small in \(\JJM\).
The simplicity and universality of the large-$\JJM$ behavior is
special to the case of a Coulomb branch chiral ring of dimension $1$.
For a Coulomb branch of dimension $\geq 2$ the operator of lowest $U(1)$  is not generally unique, and the behavior of the two-point function of powers of a given
generator may be much more complicated.
In the
one-dimensional case, the correlators can be computed to all orders in $\JJM\uu{-1}$
in the Coulomb branch effective action, which has the simple form:

\bbb
S\lrm{eff} = \int \cc \sqrt{|g|} \cc d\uu 4 x \cc {\cal L}\lrm{eff}\ ,
\xxn
{\cal L}\lrm{eff} = {\cal L}\lrm{kin} + {\cal L}\lrm{super-WZ}
+ {\cal L}\lrm{sources} + {\cal L}\lrm{D-terms}\ .
\een{UniversalRankOneActionPRECAP}
where ${\cal L}\lrm{kin} $ is a free kinetic term for a vector multiplet, given in \rr{FreeKineticTerm}; the source term is 
the negative logarithm of the operator insertions, 
\bbb
 {\cal L}\lrm{sources} \equiv - n\cc {\tt log}[{\cal O}]\cc \d(x - x\ll{\co})
- n \cc {\tt log}[\overline{{\cal O}}]\cc \d(x - x\ll{\overline{\co}})\ ; 
\eee
and ${\cal L}\lrm{super-WZ}$ is the
supersymmetrized Wess-Zumino term for the Weyl a-anomaly and the $U(1)\lrm R$-symmetry, of
which the relevant details are collected in \eqref{RestrictedBEOActionInPieces}, \eqref{SuperWZB}-\eqref{SuperWZH}.

In the case of \(\mathcal{N} = 2\) \ac{sqcd} with four flavors we can estimate the leading corrections using numerical results from localization and we find that
\begin{equation}
  \eval{\log(\ampname\ll n)}^{\acs{sqcd}}_{\text{localization}}  - \eval{\log(\ampname\ll n)}^{\acs{sqcd}}_{\ac{eft}} \approx 1.6\, e^{- \pi \sqrt{\JJM/(2\Im \tau)}} = 1.6\, e^{- \sqrt{\pi \lambda }/2}
\end{equation}
with high accuracy in the double scaling limit \(\JJM \to \infty\), \(\Im \tau \to \infty\) with \(\JJM/\Im \tau = \lambda/ (2 \pi)\) fixed as suggested in~\cite{Bourget:2018obm}.

\bigskip

The outline of our reasoning is as follows:
\bi
\item ${\cal N} = 2$
superconformal invariance is sufficiently restrictive to 
forbid any possible higher-derivative $F$-terms
in the \ac{eft} of the Coulomb branch,
for theories of rank one.
As a result, the entire Wilsonian action on the Coulomb
branch is given by the tree-level effective kinetic term,
the supersymmetrized Wess--Zumino term for the 
spontaneously broken Weyl invariance, and 
unknown $D$-terms which do not affect
correlation functions of chiral primaries. 
\item{The F-term content of the \ac{eft} only depends on the $a$-anomaly coefficient
of the underlying \ac{cft}, which we parametrize
as in~\cite{Hellerman:2017sur}, following the normalization in \ac{aefj}
\bbb
\a \equiv 2 (\Delta a)\ups{\acs{aefj}}\ ,
\xxn
\Delta a \equiv a\lrm{CFT} - a\lrm{EFT}\ ,
\xxn
a\ups{\acs{aefj}} \equiv {5\over{24}}\cc {a\over{a\ll{U(1){\rm ~vector~multiplet}}}}.
\eee
It follows that all the correlators, regardless of the details, depend only on \(\alpha\) in some theory-independent formula, modulo terms that are either affine in \(\JJM\), coming from the normalization of the external operator and the sphere partition function or nonperturbatively small in \(\JJM\), coming from the breakdown of the \ac{eft}.
}
\item{In terms of $\a$, the order $n\uu{-m}$
term in $q\ll n$ can be expressed as
\begin{equation}
\eval{q\ll n}_{\order{n^{-m}}} = \frac{K\ll m (\a)}{n^{m}} \ ,
\end{equation}
where $K\ll m(\a)$ is a polynomial
defined by quantizing the
effective theory of the Coulomb branch with
Wess--Zumino coefficient $\a$, regularized
and renormalized to preserve the spontaneously
broken superconformal symmetry, see Sec.~\ref{sec:diagr}.}
\item{
For values of \(\alpha\) that can be realized as superconformal gauge theories with a marginal coupling, the theory has to obey recursion relations, which lead to ``effective recursion relations'' for the power-law terms, which are algebraic rather than differential, because these terms must be independent of \(\tau\) and \(\bar{\tau}\), because of point the absence of higher-derivative $F$-terms.
Using the recursion relations we can compute all the polynomials $K\ll m(\a)$. This leads to a unique formula for all the power-law corrections, for any value of \(\alpha\), \emph{for theories that obey the effective recursion relations}
\begin{equation}
  K_m(\a) \equiv \D\ll\co\uu{-m} \cc \hat{K}\univ\ll m(\a) \ ,
\end{equation}
where the \(\hat K\univ_m\) satisfy
\begin{equation}
  \dv{\hat K\univ_{m+1}(\alpha)}{\alpha} = - m \hat K\univ_{m}(\alpha)
\end{equation}
and are the coefficient of \(\JJM^{-m}\) in the asymptotic expansion of \(\log \Gamma(\JJM + \alpha + 1)\):
\begin{equation}
\hat{K}\univ\ll m(\a) \equiv \hat{P}\univ\ll {m+1}(\a) \equiv \eval{ \log \Gamma(\JJM + \alpha + 1)}_{\order{\JJM^{-m}}} = \frac{(-1)^{m+1}}{m \pqty{m+1}}B_{m+1}(\alpha + 1),
\end{equation}
where \(B_{m+1}\) is the Bernoulli polynomial of degree \(m + 1\), see Sec.~\ref{sec:lagtheo}.
}
\item{Since unitarity does not appear to be a logical
necessity for the validity of the recursion relations,
we argue that the existence of an infinite series
of ${\cal N} = 2$ superconformal $SU(2)$ gauge theories 
with nonunitary matter, marginal gauge
coupling, and distinct values of $\a$, proves
that $\hat{K}\ll m(\a)$ is universally completely
determined by the existence of this series.
In other words, for any value of \(\alpha\) the power-law corrections can be shown to obey the \emph{effective} recursion relations, to all orders in \(\JJM\) (but not nonperturbatively) and
\begin{equation}
  \hat{K}\ll m(\a) = \hat{K}\univ\ll m(\a) = \frac{(-1)^{m+1}}{m \pqty{m+1}}B_{m+1}(\alpha + 1) \ ,
\llsk\forall m \geq 1\ ,
\end{equation}
for any theory with a one-dimensional Coulomb branch,
whether it has a marginal coupling constant or not. This is discussed in Sec.~\ref{Universality}.
}
\item We can do a simple consistency check by calculating the classical terms, without reference to any completion, unitary or nonunitary, with or without marginal coupling.  We find that the \(\alpha^{m+1} / \JJM^m\) term agrees with our proven formula.  %
\item We can also check against direct calculations in the case of \(\mathcal{N}=2\) \ac{sqcd}, where the correlators can be computed directly by localization. We find fantastic agreement and the precision is limited only by the omission of instanton corrections. We can remedy that by considering the limit suggested in~\cite{Bourget:2018obm}, where we find sufficiently precise agreement that we can even see the leading nonuniversal correction, which strongly suggests an interpretation in terms of virtual \ac{bps} dyons propagating over macroscopic distances, see Section~\ref{sec:SQCD}.
\ei

\medskip
The plan of this paper is as follows. In Section~\ref{sec:diagr} we discuss the building blocks of the Feynman diagrams which appear for the expansion of the two-point functions at order $1/\JJM$, discussing the basic setup in Section~\ref{sec:setup}, the normalization of the observables in Section~\ref{sec:normalization} and the $\alpha$-dependence of the observables in Section~\ref{sec:alpha-dep}. In Section~\ref{sec:ex-diag}, we give concrete examples of diagrams appearing at order $n^0$ and $n^{-1}$.
In Section~\ref{NonuniversalCorrectionDiscussion}, we discuss the universality for power-law corrections and the nonuniversality for exponential corrections.
In Section~\ref{sec:lag-corr}, we derive the $\JJM\uu{-m}$ corrections in Lagrangian theories, using recursion relations to determine
all correlation functions of Coulomb branch chiral primaries for ${\cal N} = 2$ \acp{scft} with a marginal coupling $\t$. In Section~\ref{sec:examples}, we treat the concrete examples of Abelian gauge theory without matter, ${\cal N} = 4$ \ac{sym} with $G = su(2)$ and ${\cal N} = 2$ super-\acs{qcd} with $G = su(2)$ and $N\lrm F = 4$.
In Section~\ref{Universality}, we argue that our result for the power-law corrections should apply to any value of $\a$, for Coulomb-branch chiral-primary correlators
in any rank-one theory with any value of $\a$, whether or not it has a marginal coupling. In Section~\ref{sec:UVreg}, we discuss manifestly ${\cal N} = 2$ superconformal \ac{uv} regulators with marginal couplings which are constructed by adding ``ghost'' hypermultiplets with reversed spin-statistics which allow us to
regulate any one-dimensional Coulomb-branch \ac{eft} with $\a$-coefficient satisfying $\a \in {3\over 2} - {4\over 3} \cc \IZ$.  In Section~\ref{sec:univPoly}, we use this fact to write the power-law corrections for $\a \in {3\over 2} - {4\over 3} \cc \IZ$.
In Section~\ref{sec:SQCD} we compare the universal \ac{eft} behavior with the results from $S\uu 4$ localization, finding an excellent agreement.
In Section~\ref{sec:conclusions}, we close with a brief discussion of our findings.
A lot of the technical details of the above discussion is relegated to the Appendix. The solution to the recurrence relations is detailed in Appendix~\ref{sec:solve-recurrence}.  The ${\cal N} = 2$ supersymmetrization of the Weyl anomaly action is given in Appendix~\ref{sec:Weyl}. The nonexistence of higher-derivative $F$-terms on conformally flat space is argued in Appendix~\ref{sec:noF}. The ${\cal N} = 2$ superconformal gauge dynamics with ghost hypermultiplets which is used in Section~\ref{Universality} is explained in Appendix~\ref{GhostHyperAppendix}. 
The saddle point value of the classical action is shown in Appendix~\ref{sec:saddle-point}. Finally, in Appendix~\ref{sec:numerics}, the numerics for the comparison to the localization result are given.

\section{Diagramatics and quantization of the EFT}
\label{sec:diagr}

As in~\cite{Hellerman:2015nra, Hellerman:2017veg, Hellerman:2017sur},
we consider the effective theory of the Coulomb branch,
as an effective theory in which conformal invariance
is treated as an exact nonlinearly realized dynamical symmetry, and ${\cal N} = 2$ Weyl invariance as an exact 
symmetry of the dynamical fields and background fields 
together.

As explained in Appendix~\ref{sec:noF}, rank one theories
are special in this regard, as the effective theory
admits no superconformally invariant $F$-terms beyond
the kinetic term and super-\ac{wz} terms, 
respecting the ${\cal N} = 2$ superconformal symmetry.
All higher-derivative terms allowed by the symmetries are $D$-terms.
We can treat our theory as a Wilsonian effective action with a cutoff $\L$ satisfying
\bbb
E\lrm{IR} \muchlessthan \L \muchlessthan |\phi|\ .
\eee
We can then compute quantum effects by regularizing and renormalizing the theory with local counterterms that remove
all dependence on the cutoff scale $\L$.

The absence of superconformal $F$-terms means that any
counterterms consistent with the symmetries must be
$D$-terms.  Since two-point
functions can in principle be computed by supersymmetric
localization, however, they are necessarily independent of
$D$-terms.  It follows that all power-law $1/n^m$ corrections to the logarithm of the correlator are necessarily independent of the details of the microscopic theory, depending only on the $\a$-coefficient.

\subsection{Setup}\label{sec:setup}

Begin by representing the correlator as
\bbb
\ampname\ll n = |x - y|\uu{ 2 n\D\ll\co}\cc Z\uu{-1}\ll 0 \times Z\ll n\ ,
\eee
where
\bbb
Z\ll n \equiv \exp{q\ll n} 
\eee
is the path integral with sources $- \JJM\cc\log(\phi\lrm{hol})$
and $- \JJM\cc \log(\phb\lrm{hol})$ (with $\JJM = n\D\ll\co$) inserted at $x$ and
$y$, respectively.  The sources can also be taken to be
smeared over a scale $\e = O(\L\uu{-1})$ if desired, though this makes no difference to the result in the end, 
as the field $\phi$ is nonsingular with respect to
itself and the limit $\e\to 0$ is nonsingular.

Here we have taken $\phi\lrm{hol}$ to be normalized such
that $\phi\uu{\D\ll\co} = \co$ exactly. 
This is the superfield that we take to be holomorphic
in the background fields, if one is to turn on any background
fields such as a marginal coupling.  For instance, in a
Lagrangian theory, $\phi\lrm{hol}$ is the superfield whose 
effective kinetic term is
\begin{equation}
{\cal L}\lrm{kinetic} = -i\times \text{(const.)} \times \int \cc d\uu 4\th\ll{{\cal N} = 2} \cc \t \Phi\lrmns{hol}\sqd + ({\rm h.c.}).
\end{equation}
For purposes of quantizing the effective theory of the Coulomb branch, it is more convenient to work in terms
of the field $\phi\lrmns{unit} \equiv \sqrt{\Im(\t)}$, whose kinetic term is
\begin{equation}
  \begin{aligned}
{\cal L}\lrm{kinetic} &= \text{(const.)} \times \int \cc d\uu 4\th\ll{{\cal N} = 2} \cc \Phi\lrmns{unit}\sqd + \text{(h.c.)}\\
&= |\pp\phi\lrmns{unit}|\sqd + \text{fermion kinetic} + \text{gauge kinetic}
\end{aligned}
\end{equation}
in Euclidean signature.  In general, when we write
$\phi$ without a subscript indicating the normalization,
we shall always be referring to the field $\phi \lrmns{unit}$.
More generally, the relationship between $\phi\lrmns{hol}$
and $\phi\lrmns{unit}$ can be written as
\begin{equation}
  \phi_{\text{hol}} =  {\bf N}\ll\co \phi_{\text{unit}} .
\end{equation}
In terms of the unit-kinetic-term superfield, the correlation
functions of $\phi$ are
\bbb
\ampname\ll n = Z\ll 0\uu{-1} \times Z\ll n = 
 {\bf N}\ll\co\uu{\reddit{2} \JJM} \times 
Z\ll n\uprm{unit}\ ,
\eee
where $Z\ll n\uprm{unit}$ is just the path integral with
insertions of the unit normalized field,
\bbb
Z\ll n\uprm{unit} \equiv \bigg \langle \cc \phi\uu{n\D\ll\co}\lrm{unit}(x)
\cc \phi\uu{n\D\ll\co}\lrm{unit}(y) \cc \bigg \rangle .
\eee

\subsection{Normalization of the observables}\label{sec:normalization}

As discussed in~\cite{Hellerman:2017sur}, no superconformal effective term can contribute to orders higher than $\JJM^0$ in the expansion of the correlation function. The \ac{wz} terms arising both for the Weyl symmetry and $U(1)$ $R$-charge are needed to compensate the difference between the anomaly coefficients of the underlying \ac{cft} and the \ac{eft} of the Coulomb branch. They cannot be written as superconformal terms in superspace, because they explicitly break both the Weyl symmetry and $R$-symmetry of the action. 
We must therefore write the \ac{wz} terms while being very careful and explicit about their normalization. 

We start with the \ac{wz}-term of the action as given in \ac{ks}~\cite{Komargodski:2011vj} (abbreviated henceforth with the superscript $KS$):
\begin{multline}\label{eq:KS-WZ}
S\ups{\acs{ks}}_{WZ}  =  -\Da^{[KS]} \int d^4x\, \sqrt{-{g}}\cc
\bigg[\dilaton\,E_4^{[KS]}
 +\bigg(4\,\Big(R^{\mu\nu}-\frac{1}{2}R\,g^{\mu\nu}\Big)\nabla_\mu\dilaton\,\nabla_\nu\dilaton  \\
  - 2\,(\nabla\dilaton)^2\Big(2\,\Box\dilaton - (\nabla\dilaton)^2\Big) \bigg) \bigg].
\end{multline}
In a Coulomb branch \ac{eft}, there is also a mismatch
of the $U(1)_R$ gravitational anomaly coefficients
between the \ac{eft} and the full underlying \ac{cft}, proportional
to the Weyl $a$-anomaly mismatch~\cite{Hellerman:2017sur}.
This anomaly mismatch can
only be canceled by a gravitational Green--Schwarz mechanism
involving the Goldstone mode $\beta$ of the spontaneously broken
$U(1)_R$.
Under supersymmetry the dilaton and axion form a chiral multiplet
$\tau + i \beta$.  For a rank-one theory there is a unique candidate 
for the axiodilaton.
In terms of the $\phi\lrm{hol}$ coordinate,
the complex axiodilaton $\dilaton + i \b$ of~\cite{Schwimmer:2010za,Komargodski:2011vj,Bobev:2013vta} is
\bbb
\dilaton + i \b = - \log(\phi / \m)\ .
\eee
If we want to specify how our dilaton depends on
background fields, we need to specify whether we 
are talking about the holomorphic dilaton, unit
dilaton, or something else.  

The holomorphic axiodilaton is defined as~\cite{Bobev:2013vta}
\bbb
(\dilaton + i \b)\lrm{holo} = - \log(\phi\lrm{holo} / \m)\ ,
\eee
while the unit-normalized axiodilaton is
\bbb
(\dilaton + i \b)\lrm{unit} = - \log(\phi\lrm{unit} / \m)\ ,
\eee
where the two differ by
\bbb
(\dilaton + i \b)\lrm{holo} = (\dilaton + i \b)\lrm{unit} \reddit{ - \log {\bf N}_\co}\ .
\eee
The quantity $\reddit{{\bf N}_\co}$ depends on the background
fields (such as complex marginal couplings $\t$) in a nonholomorphic way, and therefore we must keep
track of it if we are trying to do things like
compute the un-normalized correlators $\exp{q\ll n} \equiv Z \times
\ampname\ll n$ as a function of $\t,\tb$.  For
constant $\t,\tb$, only the overall normalization of
$\exp{q\ll n}$ is affected by the difference between
$(\dilaton + i \b)\lrm{unit}$ and $(\dilaton + i \b)\lrm{holo}$,
because the undifferentiated axiodilaton enters
the action of~\cite{Bobev:2013vta} only through the Euler density.
However even apart from this, it is valuable to keep track
of the difference, because one may want to compute,
for instance, in backgrounds with position-dependent
marginal couplings, in which case ${\bf N}\ll\co$ becomes position-dependent and its gradients enter the effective
action.

In the present paper, we will only ever consider 
constant gauge coupling $\t$, and therefore we will
work in terms of the unit-normalized axiodilaton,
taking care to add the extra term to the action of~\cite{Bobev:2013vta}:
\begin{equation}
 {\cal L}\lrmns{BEO}[(\dilaton + i \b)\lrm{holo}]
 = \reddit{(\Delta a)^{[KS]}} \times \log({\bf N}\ll\co)
\cc E\ll 4 + {\cal L}\lrmns{BEO}[(\dilaton + i \b)\lrm{unit}],
\end{equation}
and \({\cal L}\lrmns{BEO}\) is given in Appendix~\ref{sec:Weyl}.
When we refer to the super-dilaton 
without specifying, we will always be referring to the
unit-normalized rather than holomorphic dilaton,
as the unit-normalized axiodilaton is the more natural
object from the point of view of the large-$\JJM$ expansion.

In terms of the unit-normalized chiral superfield 
\bbb
\phi \equiv \phi\lrm{unit} \equiv \m\cc \exp( - \dilaton - i \b)
\equiv \m\cc \exp( - (\dilaton + i \b)\lrm{unit})\ ,
\eee
the full effective action for the path integral with sources is
\bbb
S\lrm{eff} = \int \cc \sqrt{|g|} \cc d\uu 4 x \cc {\cal L}\lrm{eff}\ ,
\xxn
{\cal L}\lrm{eff} = {\cal L}\lrm{kin} + {\cal L}\lrm{super-WZ}
+ {\cal L}\lrm{sources} + {\cal L}\lrm{D-terms}\ .
\een{UniversalRankOneAction}, as mentioned earlier in \rr{UniversalRankOneActionPRECAP}.  The $D$-terms will not affect our considerations at all,
as we are computing observables invariant under
some subset of the supersymmetry, and therefore
unaffected by $D$-terms.%

We also note
that in a more general Coulomb branch, of rank
more than $1$, we would expect higher-derivative
$F$-terms, of the type studied in~\cite{Argyres:2003tg, Argyres:2004kg, Argyres:2004yp, Argyres:2008rna}.  In a theory
of rank $1$, as we show in Appendix~\ref{sec:noF}, 
there are no superconformally invariant higher-derivative
$F$-terms at all.  It is this simplification that permits
the extraordinarily detailed calculation of correlation
functions to all orders in $1 / n$ that we are able
to perform in the present article.

We wish to emphasize particularly that the absence of
higher-derivative $F$-terms, means that the action
\rr{UniversalRankOneAction} is in effect an almost
\ac{uv}-complete action: There are no \acl{uv}
divergences to any order in $1 / \JJM$ perturbation theory,
which affect the correlation function.  Or more precisely,
for a superconformally invariant regulator, there are
no \ac{uv}-divergences in $1 / \JJM$ perturbation theory affecting the protected correlation functions;
for a non-superconformally-invariant
regulator, any \ac{uv} divergences will be proportional
to powers of $|\L| / |\phi|$, and can be
subtracted in a canonical way according to the criterion
of restoring superconformal
invariance of the quantum effective action.
We will now use this almost-\ac{uv} completeness of the F-term sector of the \ac{eft}, to derive an all-orders \(1/\JJM \) expansion for the chiral primary two-point functions.

\subsection{$\alpha$-dependence of the observables}
\label{sec:alpha-dep}

We can read off the form of the $\a$-dependence
from the form of the action as written in terms of 
$\phi\lrm{unit}$.  Modulo $D$-terms, the only terms in 
the action
are of order $|\phi|\uu 2$, $\JJM\cc \log|\phi|$,
and $|\phi|\uu 0 \cc \a$.  If we define $\hat{\a} \equiv \a / \JJM$, then the whole action, written in terms
of ${\hat{\a}}$ and $\JJM$, is strictly of order $\JJM\uu 1$,
modulo logarithms of $\JJM$.
Thus $\JJM\uu{-1}$ becomes a uniform loop-counting parameter of the theory: The parameter $\JJM$ only
occurs together with ${1/{\hbar}}$, so long as we
write the action in terms of $\JJM$ and $\hat{\a}$, the power
of parameter $\JJM$ exactly counts the number
of loops in a diagram:
\bbb
(\JJM-{\rm scaling~of~a~diagram}) \propto  \JJM\uu{[1-( {\rm number~of~loops})]}\cc F(\hat{\a})\ .
\eee
Since we are computing the
partition function and counting the source terms as part of the action itself, the diagrams we are computing are vacuum diagrams, with no ``external'' lines.  

Concretely, the classical solution
for $\phi$ is of order $\JJM^{1/2}$ and if we split
the classical solution into $\phi = \phi\lrm{classical} 
+ \phi\lrm{fluc}$, then we can decompose 
the free+ source +  the super-\ac{wz} into \ac{vev} and fluctuations; each vertex with $f$ fluctuations 
scales as $\JJM\uu{1-f/2}$ at fixed $\hat{\a}$, and
by the usual counting, a connected vacuum diagram, after
contracting all fluctuation lines, must
have scaling $\JJM\uu{1-( {\rm number~of~loops})}$.  As usual we ignore $D$-terms, of which the correlators
are independent.

To find
the $\JJM$-scaling at fixed $\a$ of a diagram, simply
turn the $\hat{\a}$'s back into $\a / \JJM$'s, which 
gives an extra factor of \(1/\JJM\) for each $\a$-vertex.
This gives
\bbb
(\JJM-{\rm scaling~of~a~diagram}) \propto \JJM\uu{-m}\ ,
\xxn
m \equiv (\#~{\rm of~loops}) + (\#~{\rm of~}\a-{\rm vertices}) - 1\ .
\een{DiagramaticnScalingFormula}
From formula \rr{DiagramaticnScalingFormula} 
two important properties of
correlation functions in the Coulomb-branch
\ac{eft} as a function of $\a$
are immediately clear:
\bi
\item{The term $\hat{K}\ll m / \JJM\uu m\in q\ll n$ is a polynomial
in $\a$ of order $m+1$; and}
\item{The terms $\hat{\a}\uu{m+1} / \JJM\uu m$ are composed
of tree diagrams only, and can be read off from the (negative of the) saddle point of the classical Wilsonian action
including both sources and super-\ac{wz} term.}
\ei
Therefore we can write the $\order{n\uu{-m}}$ term
in $q\ll n$ as
\begin{equation}
  \eval{q\ll n}_{\order{n\uu{-m}}} = {{K\ll m}\over{n\uu m}} 
\end{equation}
where
\begin{equation}
  K\ll m = K\ll m(\a) = P\ll{m+1}(\a)\ ,  
\end{equation}
 is a polynomial of order $m+1$ in
$\a$, with the leading term determined by the 
action of the classical saddle point with super-\ac{wz}
term included.

It is convenient to eliminate the explicit dependence on \(\Delta_{\mathcal{O}}\) and rewrite \(q_n\) in terms of a series expansion in the charge \(\JJM\) with coefficients \(\hat K_m(\alpha)\) defined by
\begin{equation}\label{eq:Kcoeff}
  \eval{q\ll n}_{\order{n\uu{-m}}} = \frac{K_m(\alpha)}{n^m} = \frac{\Delta_{\mathcal{O}}^m K_m(\alpha)}{\JJM^m} = \frac{\hat K_m(\alpha)}{\JJM^m} = \frac{\hat P_{m + 1}(\alpha)}{\JJM^m}
\end{equation}
where the polynomials \(\hat P_{m + 1 }(\alpha)\) are simply
\begin{equation}
  \hat{P}\ll{m+1}(\a) \equiv \D\ll\co\uu m \cc P\ll{m+1}(\a)\ .
\end{equation}

\subsection{Examples of diagrams}
\label{sec:ex-diag}

The nature of the one-point vertices depends how the
field $\phi$ is broken up into ``background classical
solution'' and ``fluctuation''.  The simplest starting point
is to break up $\phi$ into classical solution
and fluctuation, where the classical solution is
the solution at $\a = 0$, with the \ac{wz} term ignored.
This solution was written down in~\cite{Hellerman:2017sur}
and on $\IR\uu 4$ takes the form
\begin{equation}
  \label{eq:classical-solution}
  \begin{gathered}
    \begin{aligned}
      \phi_{\text{cl}}(x) &= \frac{e^{i \beta_0} \abs{x_1 - x_2}}{2 \pi \pqty{x - x_2}^2} \JJM^{1/2} \, , &
      \bar \phi_{\text{cl}}(x) &= \frac{e^{-i \beta_0} \abs{x_1 - x_2}}{2 \pi \pqty{x - x_1}^2} \JJM^{1/2} \, ,
    \end{aligned} \\
    \abs{\phi_{\text{cl}}(x)} = \frac{\abs{x_1 - x_2}}{\abs{x - x_1} \abs{x - x_2}} \frac{\JJM^{1/2}}{2\pi}.
\end{gathered}
\end{equation}
Note that these expressions depend on the conformal
frame.  In the conformal frame of the cylinder they become
\begin{equation}
  \label{eq:classical-solution-cylinder}
  \begin{gathered}
    \begin{aligned}
      \phi_{\text{cl}}(x) = \frac{e^{t / r} }{2 \pi r} \JJM^{1/2} \, , &&
      \bar \phi_{\text{cl}}(x) = \frac{e^{-t / r}}{2 \pi r} \JJM^{1/2} \, ,
    \end{aligned} \\
    \abs{\phi_{\text{cl}}(x)} = \frac{1}{2 \pi r} \JJM^{1/2}.
\end{gathered}  
\end{equation}
The dilaton $\dilaton$ is just constant and the axion $\b$
is linear in time,%
\begin{align}
\dilaton\lrmns{cl}\ups{\text{cylinder}} &= \reddit{\log(2 \pi \mu R) - \frac{1}{2} \log(\JJM)}\\
\b\lrmns{cl}\ups{\text{cylinder}} &= \reddit{ i \frac{t}{R}}\ ,
\end{align}
which are always properties of a helical classical solution.

Since the $\phi\lrm{classical}$ solves the classical
\ac{eom} exactly at $\a = 0$, there are no ``external'' lines
without $\a$-vertices.  With this organization
of $\phi$ into background plus fluctuation, there 
do indeed exist one-point vertices, but
each is proportional to $\hat{\a}$ and therefore
comes with an extra power of $\JJM\uu{-1}$.  So nontrivial
tree diagrams do exist in this organization of diagrams,
with each one-point vertex carrying an $\a$-factor,
which suppresses the weight of the diagram by $\JJM\uu{-1}$.

Let us examine the diagramatics of the first few
terms in the ${1/n}$ expansion of 
$q\ll n$.

\paragraph{Order $n\uu 0$.}

At order $n\uu 0$ we have only the determinant, with
no $\a$-vertices at all; that is, just the 
fluctuation determinant of the free vector multiplet
action with logarithmic sources proportional to
$\JJM = n \cc \D\ll\co$.  The sources make this
determinant nontrivial, but its value is
already known directly by expanding the 
free-field partition function
\begin{equation}
  q\ll n\uprm{free}
  \equiv \log(Z\ll n\uprm{free}) = {\bf A}\cc n + {\bf B} +\log( \Gamma( \D\ll\co n + 1) ) = \log(\Gamma(\JJM + 1)) 
\end{equation}
to order $n\uu 0$.%
Indeed, all diagrams without $\a$-vertices,
are simply terms in the expansion of  $\log( \Gamma(\D\ll\co n + 1)) = \log(\Gamma (\JJM + 1) )$: the order $\a\uu 0 \cc n\uu{-m}$ term in $q\ll n$, is just the $n\uu{-m}$ term 
in the
Stirling series of $\log(\Gamma(\JJM + 1))$:
\begin{equation}
  \label{eq:alpha-0-coefficient}
  \eval{q_n}_{\order{ \alpha^0 n^{-m}}} = \eval{ \log (\Gamma(\JJM + 1))}_{\order{\JJM^{-m}}} = \frac{(-1)^m B_{m+1}}{m \pqty{m + 1}}
\end{equation}
where \(B_{m+1}\) is the Bernoulli number.

At order $n\uu 0$ there is also a ``diagram'' with 
one $\alpha$-vertex that has no external lines at all:
This is just the evaluation of the (negative of the Euclidean) \ac{wz} term
on the classical solution, whose $\log(n)$ piece
was computed in~\cite{Hellerman:2017sur}.  In that paper we
did not compute the non-logarithmic contribution to $q\ll n$
at order $n\uu 0$.  Indeed this term by itself
is ill-defined due to the conformal anomaly; only
the difference $q\ll n - q\ll 0$ is well-defined.  While the
$n\uu 0$ term in
$q\ll n - q\ll 0$ is well-defined and in principle computable,
the computation requires a somewhat careful matching
of conventions and renormalization schemes between
the sphere partition function and the large-$\JJM$
partition function, and we do not pursue it in the present paper.  
We focus instead on large-$n$ limits of
differences $q\ll{n+1} - q\ll n$ between adjacent correlators,
which are also well-defined and amenable to
direct analysis at large $n$.

From equation \rr{DiagramaticnScalingFormula} we
see that there are no further contributions at 
order $n\uu 0$.

\paragraph{Order $n\uu {-1}$.}

Consulting equation \rr{DiagramaticnScalingFormula},
we see that the order $n\uu {-1}$ contribution to $q\ll n$
contains three distinct types of diagram:
\bi
\item{Two-loop diagrams with no $\a$-vertices;}
\item{One-loop diagrams with one $\a$-vertex; and}
\item{Tree-level diagrams with two $\a$-vertices.}
  \ei
  \newcolumntype{C}{>{\centering\arraybackslash} m{3cm} }
  \begin{table}
    \centering
    \begin{tabular}{lcCC}
      \toprule
      description                                & term                            & \multicolumn{2}{c}{diagrams}                                                    \\
      \midrule
      Two-loop with no $\a$-vertices    & \({\hat K_{1,0}}\) & \includegraphics{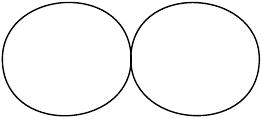} & \includegraphics{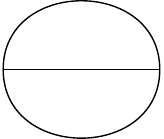} \\
      One-loop with one $\a$-vertex     & \({\hat K_{1,1} \alpha}\) & \includegraphics{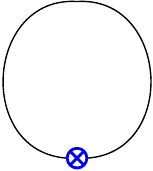} & \includegraphics{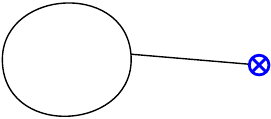} \\
      \\
      Tree-level with two $\a$-vertices & \({\hat K_{1,2} \alpha^2}\) & \multicolumn{2}{c}{\includegraphics{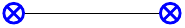}}                                \\
      \bottomrule
    \end{tabular}
    \caption{Diagrams appearing at order $1/\JJM$.}
    \label{tab:diagrams-n-1}
  \end{table}
We shall call these respective contributions 
\begin{equation}
  q\ll n\cc \bigg |\ll{O(n\uu{-1})} \equiv  \frac{\hat K_1}{\JJM}  \equiv
 \frac{\hat{P}\ll 2(\a)}{\JJM} 
\equiv \sum\ll{a = 0}^2 \frac{\hat{K}\ll{1,a} \a\uu a}{\JJM}
\end{equation}
\begin{itemize}
\item The two-loop diagram would be cumbersome to compute directly, but we do not have to: We know it is equal to the $\JJM\uu{-1}$ term in the expansion of $\log(\Gamma(\JJM + 1))$. So from Eq.~\eqref{eq:alpha-0-coefficient} we have:
  \begin{equation}
    \hat{K}\ll{1,0} = \frac{B_2}{2} =  + {1\over{12}}\ .
  \end{equation}
\item We calculate $\hat{K}\ll{1,2}$ in Section~\ref{sec:saddle-point}
of the Appendix, and the result is in Eq.~\eqref{eq:universal-saddle-value} (for \(m = 1\)):
\bbb
K\ll{1,2} \equiv \D\ll\co\uu{-1}\cc \hat{K}\ll{1,2}= +{1\over 4} ~{\rm for~the~case~}\D\ll\co = 2\ ,
\eee
which according to \rr{eq:Kcoeff}
means
\bbb
\hat{K}\ll{1,2} = 2\times K\ll{1,2} = + \hh\ .
\eee
\item The value $\hat{K}\ll {1,1}$ comes from of the
one-loop diagram with a single $\a$-vertex.  This diagram
is also somewhat tedious to compute directly and 
we will not need to do so.  
Instead, we will %
infer the value of $\hat{K}\ll{1,1}$ based on the known values
of $\hat{K}\ll{1,0}$ and $\hat{K}\ll{1,2}$ together with the
value of $\hat{K}\ll 2$ for $\a = 1$, which is realized by
${\cal N} = 4$ \ac{sym} and has a simple closed-form
expression for $q\ll n$.
\bbb
q\ll n\uu{{\cal N} = 4~{\rm SYM}} = {\bf A}\cc n + {\bf B}
+ \log( \Gamma(2n+2))\ .
\eee
This theory has $\D\ll\co = 2$ and $\a = 1$ (see~\cite{Hellerman:2017sur} for the normalization of the
$\a$-coefficient) and so using Stirling's formula we have
\begin{equation}
\eval{q\ll n}_{\order{1/n}} = K\ll 1\uprm{{\cal N} = 4 ~{\rm SYM}} = \frac{13}{24} .
\end{equation}
Then, using the definition of \(P_{2}\) we have
\begin{equation}
  \begin{aligned}
    \frac{13}{24} &= K\ll 1\uprm{{\cal N} = 4 ~{\rm SYM}} =
    K\ll 1 \cc \bigg |\ll{\D\ll\co = 2,~\a = 1} = P\ll 2(\a = 1) \cc \bigg |\ll{\D\ll\co = 2} 
     = {1\over 2} \cc
    \hat{P}\ll 2(\a = 1) = \\
    & = \frac{1}{2} \pqty{\hat{K}_{1,0} + \hat{K}_{1,1} + \hat{K}_{1,2}  } 
     = {7\over{24}} + \hh \cc \hat{K}\ll{1,1}.
  \end{aligned}
\end{equation}
This gives
\begin{equation}
  \hat{K}\ll{1,1} = + \hh\ .
\end{equation}
\end{itemize}
The final result is that the general form of the coefficient of the \(1/\JJM\) term is
\begin{equation}
  \label{eq:K1}
\hat{K}\ll 1 = \hat{P}\ll 2(\a) = {{\a\sqd}\over 2} + {\a\over 2}
+ {1\over{12}} = \hh \cc \bigg [ \cc \a\sqd + \a + {1\over 6}  \cc \bigg ]\ .
\end{equation}

Later we will be interested in the case of ${\cal N} = 2$
\ac{sqcd} with $N\ll f = 4$, which has $\a = {3\over 2}$
and $\D\ll\co = 1$.  This corresponds to a value of
\bbb
\hat{K}\ll 1(\a = {3\over 2}) = + {{47}\over{24}}\ ,
\xxn
\hat{K}\ll 1 (\D\ll\co = 2,~\a = {3\over 2}) = + {{47}\over{48}}\ .
\eee
We will be able to check this value against
correlation functions computed by
recursion relations starting from the sphere partition function
$Z = Z\ll 0 = \exp{q\ll 0}$.

\subsection{Universality for power law corrections versus nonuniversality for exponential corrections}
\label{NonuniversalCorrectionDiscussion}

Combining the results from~\cite{Hellerman:2017sur} and Sec.~\ref{sec:alpha-dep}, we find that the
correlation functions $Z\uu{-1} Z\ll n = \exp{q\ll n - q\ll 0}$
take the form 
\begin{equation}
  \label{AsymptoticExpansionOfqn}
    q\ll n \simeq
    \JJM\log(\JJM) + \pqty{\a + \hh} \cc\log(\JJM) + \pqty{{\bf A} - 1} \JJM + {\bf B} + \log \sqrt{2\pi} + \sum\ll{m \geq 1} \frac{\hat{K}\ll m(\a)}{\JJM^m} ,
\end{equation}
where $\JJM = n\D\ll\co$ and $\hat{K}\ll m = \hat{P}\ll{m+1}(\a)$ are some universal coefficients independent
of the microscopic details of the underlying
microscopic \ac{cft} including the dimension $\D\ll\co$
of the generator of the Coulomb branch and depending 
only on $\a$ as a theory-independent
polynomial of order $m+1$.  The
coefficients ${\bf A}$ and ${\bf B}$ can
depend on the theory overall, 
on the normalization of the operators
$\co$, on the marginal parameters $\t$ if any,
and on the renormalization scheme. The coefficients
$\hat{K}\ll m$, on the other hand, are independent
of the renormalization scheme and marginal
couplings, and can be computed in the effective theory
with no counterterm ambiguities because no
superconformal $F$-terms exist of order $\JJM\uu{-1}$
or higher, a property special to theories with
one-dimensional Coulomb branches.

The $\simeq$ in the formula indicates that
the formula above should be understood only
as an asymptotic expansion in $\JJM$: The 
\ac{eft} can be valid only
up to amplitudes associated with propagation of
massive particles over the infrared scale.
In a conformal theory, the mass $M$ of the lowest massive 
excitation must be given by $M = \kk\lrm M\cc |\phi|$
where $\kk_M$ is a dimensionless parameter depending on 
the theory overall and on the marginal couplings.  In familiar cases we know $\kk\lrm M$
depends on the gauge coupling, as $\propto
g\lrm{YM}\uu {+1}
\propto \Im(\t)\uu{-1/2}$ at weak coupling.  Thus
we expect \ac{eft} to break down
due to effects of size $\exp( - \kk\ll M L)$ where $L
\equiv \kk\ll L\times r$,
with $r$ being the radius of the sphere and $\kk\ll L$
being a dimensionless number.  Then the size
of the exponentially small effects signaling the breakdown
of the \ac{eft}, should go as $\exp(- \kk\ll M \kk\ll L \cc r \abs{\phi})
= \exp(- \kk \sqrt{\JJM})$, where
\bbb
\kk =  2 \pi \cc \kk\ll M \kk\ll L .
\eee
Here we have used the identity in Eq.~\eqref{eq:classical-solution-cylinder}:
\bbb
\JJM = 4\pi^2 \cc r\sqd\cc |\phi|\sqd
\eee
for the classical helical ground-state solution.

Since $\kk$ contains a factor of $\kk\lrm M$, and
since $\kk\lrm M$ is theory dependent (in particular,
depending on marginal parameters), we do
\rwa{not} expect the theory-independence of the $n\uu{-m}$ terms to extend to the exponentially small corrections.
We shall return to this point later on.

\section{Lagrangian theories}
\label{sec:lagtheo}

\subsection{Derivation of the $\JJM\uu{-m}$ corrections in Lagrangian
theories}
\label{sec:lag-corr}

In~\cite{Papadodimas:2009eu, Baggio:2014sna, Baggio:2014ioa, Baggio:2015vxa, Gerchkovitz2017},
Coulomb branch correlation functions were analyzed
for ${\cal N} = 2$ \acp{scft} with a marginal
coupling $\t$.  In these, the correlation functions
were shown to obey recursion relations with respect
to the coupling constant $\t$:
\begin{equation}
  \label{eq:recursion-relations}
  \del \bar \del q_n(\tau, \bar \tau) = \exp[ q_{n+1}(\tau, \bar \tau) - q_n(\tau, \bar \tau)] - \exp[ q_{n}(\tau, \bar \tau) - q_{n-1}(\tau, \bar \tau)] .
\end{equation}
In particular, for ${\cal N} = 2$ superconformal
gauge theories with $SU(2)$ or $SO(3)$ gauge
group, the recursion relations in Eq.~\eqref{eq:recursion-relations} are sufficient to determine
all correlation functions of Coulomb branch chiral
primaries
for any value of $\a$ where the \ac{eft} can be completed  by a superconformal gauge theory.%

In Appendix~\ref{sec:solve-recurrence}, we show that the recursion relations, when they apply,
fix the $\JJM\uu{-m}$ power-law corrections uniquely
for a given value of $\a$.
In the derivation, we use only
the fact that $q\ll n$ has the asymptotic
expansion~\eqref{AsymptoticExpansionOfqn} with $\hat{K}\ll m$ depending only on $\a$ and not on
$\t,\tb$ or $\D\ll\co$,
a property
which follows from the properties of the Coulomb-branch
\ac{eft} as discussed above.  We see in Eq.~\eqref{eq:recursion-result-gamma} that the recursion
relations uniquely fix
\begin{equation}
  \label{eq:Reflando}
  q_n \simeq \JJM \mathbf{A} + \mathbf{B} + \log( \Gamma( \JJM + \alpha + 1))
\end{equation}
and the coefficients in the perturbative expansion are those in Eq.~\eqref{eq:recursion-result-Bernoulli}:
\begin{equation}
  \label{eq:coefficients-Lagrangian}
  \hat{K}\ll m = \hat{P}\univ\ll {m+1}(\a) \equiv \text{coefficient of \(\JJM^{-m}\) in \(\log(\Gamma(\JJM + \alpha + 1))\)},
\end{equation}
for any value of $\a$ corresponding to a gauge theory
with marginal coupling.
These are essentially the Bernoulli polynomials of degree $m+1$: 
\begin{equation}
  \label{eq:universal-P}
  \hat{P}\univ_{m+1}(\alpha) = \frac{(-)^{m+1}}{m(m+1)}B_{m+1}(\alpha + 1) .
\end{equation}
For concreteness, we give the first several values
of $\hat{P}\ll{m+1}\upsns{\rm univ}(\a)$:
  \begin{align}
    \hat{P}\univ\ll 2(\a) &= \hh\a\sqd + \hh \a + {1\over{12}}\ , \\
    \hat{P}\univ\ll 3 (\a) &= - {1\over 6} \cc \a\uu 3
    - {1\over 4} \cc \a\sqd - {1\over {12}}\cc \a\ , \\
    \hat{P}\univ\ll 4(\a) &= +{1\over{12}}\cc\a\uu 4
    + {1\over 6} \cc \a\uu 3 + {1\over{12}}\cc\a\sqd - {1\over{360}} \\
    \hat{P}\univ\ll 5(\a) &= - {{1}\over{20}}\cc \a\uu 5
    - {1\over 8} \cc \a\uu 4 -  {1\over {12}} \cc \a\uu 3 
    + {1\over{120}}\cc \a\ , \\
    \hat{P}\univ\ll 6 (\a) &= +{1\over{30}}\cc\a\uu 6
    + {1\over{10}}\cc \a\uu 5 + {1\over {12}} \cc \a\uu 4
    - {1\over{60}}\cc \a\sqd + {1\over{1260}}\ , \\
    \hat{P}\univ\ll 7 (\a) &= - {1\over{42}}\cc\a\uu 7
    - {1\over{12}}\cc\a\uu 6 - {1\over{12}}\cc \a\uu 5
    + {1\over{36}} \cc\a\uu 3 - {1\over{252}}\cc\a\ ,
\end{align}
and so forth.
The coefficient of \(\alpha^{m+1}\) in \(\hat{P}_{m+1}\) is \((-1)^{m+1}/\pqty{m \pqty{m+1}}\). We show in Appendix~\ref{sec:saddle-point} that these values can be independently computed in our \ac{eft} without using to the recursion relations.
On the other hand, these polynomials satisfy an effective recursion relation
\begin{equation}
    \dv{\hat P\univ_{m+1}(\alpha)}{\alpha} = - \pqty{m-1} \hat P\univ_{m}(\alpha)
\end{equation}
which is independent of any notion of coupling.

Note that we do \rwa{not} use any information about
the sphere partition function in our derivation.  As emphasized in~\cite{Gerchkovitz2017}, the solution
to the recursion relations is not unique, and one needs
``boundary conditions'' of some kind to select the correct
solution.  For the case of rank-one superconformal
gauge theories, the sphere partition function uniquely
determines all the $q\ll n$.  Here we show
that, \rwa{without} using the sphere partition function
as an input, the large$-n$ asymptotics
corresponding to quantization of the Coulomb-branch
\ac{eft}, fix the correlators not completely uniquely, but uniquely
up to corrections smaller than any power of $n$.  
In the next Section, we shall present evidence that the
asymptotic expansion in inverse powers of $\JJM$
produces the physically correct answer, matching
the correlators extracted from the sphere partition
function, to exponentially fine accuracy as a function of $\JJM$. We
shall also return later to discuss the physical meaning
of these exponentially small corrections.

\subsection{Examples}
\label{sec:examples}

In this section we consider examples of gauge
theories with marginal couplings and the power-law
terms $K\ll m\cc
n\uu{-m}
= 2\uu{-m}\cc \hat{K}\ll m\cc n\uu{-m} = \hat{K}\ll m\cc \JJM\uu{-m}$ in their large-$\JJM$ expansions.

\paragraph{Abelian gauge theory without matter: $\a = 0$.}

The Abelian gauge theory with no matter is a gauge
theory with a marginal coupling $\t$.  This coupling
is a true parameter of the theory, affecting for instance
the spectrum of electric and magnetic flux states
on a spatial slice $\Sigma$ with a nonvanishing
second homology group; the Abelian gauge theory
is therefore a gauge theory with a marginal coupling,
which must obey the recursion relation in Eq.~\eqref{eq:recursion-relations} and
therefore have $\hat{K}\ll m = \hat{P}\ll {m+1}$.  

The Coulomb branch chiral ring is generated by $\phi$,
so $\co = \phi$ and $\D\ll\co = 1$.
The two-point functions $\ampname\ll n$ are
particularly easy to compute in this case
because the flux states are irrelevant to the computation of
correlation functions of local operators.  Therefore
$\t$ decouples completely from such correlators,
except through the normalization of the vector
multiplet scalar $\phi\lrm{holo}$, which drops out if
we consider correlators of $\phi \equiv \phi\lrm{unit}$.

The correlation function of $\phi\uu n$ with $\phb\uu n$
is thus given by
\bbb
\ampname\ll n = {\bf N}_\co \uu n \Gamma(n + 1) ={\bf N}_\co \uu n \Gamma(n + 1)\ ,
\eee
and so we have
\bbb
 K\ll m = 
 P\univ\ll{m+1} = \text{coefficient of \(\JJM^{-m}\) in \(\log(\Gamma(\JJM + 1))\),}
\eee
in agreement with formula in Eq.~\eqref{eq:coefficients-Lagrangian}.

\paragraph{${\cal N} = 4$ super-Yang--Mills with $g = su(2)$: $\alpha=1$.}

Now we consider the case of ${\cal N} = 4$ \ac{sym}
with $g = su(2)$. The generator of the chiral ring is ${\cal O} \propto  \Tr(\hat{A}\sqd)$, where $\hat{A}$ is the adjoint-valued
vector-multiplet scalar in the microscopic theory, giving $\co$ dimension $\D\ll\co = 2$, as in all rank-one gauge theories. The correlation functions in this case are
\bbb
\exp{q\ll n}  = Z\cc \ampname\ll n = {\bf N}\uu n\cc
\log[(2n+1)!] = \log(\Gamma(\JJM + 2))\ ,
\eee
and so the power-law corrections are
\begin{equation}
  \begin{aligned}
    \hat{K}\ll m &= \text{coefficient of \(\JJM^{-m}\) in \(\log(\Gamma(\JJM + 2))\)} = (-)^{m+1} \pqty{\frac{1}{m} + \frac{B_{m+1}}{m\pqty{m+1}}} \\
     &= \frac{(-)^{m+1}}{m \pqty{m+1}} B_{m+1}(2)\ ,
\end{aligned}
\end{equation}
where \(B_{m+1}(x)\) is the Bernoulli polynomial.
This agrees with formula in Eq.~\eqref{eq:coefficients-Lagrangian} for
the case $\a = \a\lrm{{\cal N} = 4,~G=SU(2)} = 1$.

\paragraph{${\cal N} = 2$ super-\acs{qcd}
with $g = su(2)$ and $N\lrm F = 4$: $\alpha=3/2$.}

Now we consider the more involved case of
conformal ${\cal N} = 2$ \ac{sqcd} with $g = su(2)$ and
four hypermultiplets in the ${\bf 2}$ representation of $SU(2)$.  Here, the coupling constant  dependence is complicated even for the sphere partition function without insertions; the correlation
functions $\ampname\ll n(\t,\tb)$ which are  obtained 
 from $Z(\t,\tb)$ by the recursion
relations, are more complicated still and the complication
grows quickly with $n$.

At large $n$, on the other hand, formula in Eq.~\eqref{eq:recursion-result-gamma} tells
us the dependence of
$\ampname\ll n$ on the coupling is just the trivial
$[\Im(\t)]\uu{-2n}$ geometric dependence, 
up to exponentially small corrections to $\log(\ampname\ll n)$.  In particular, the corrections
should obey the universal formula for power law
corrections for rank-one Coulomb-branch correlators, with $\D\ll\co = 2$ and $\a = \a\lrm{SQCD} = {3 / 2}$:
\begin{equation}
  \label{SQCDPowerLawCorrectionPrediction}
  \begin{aligned}
  K\ll m
  &= 2\uu{-m}\cc \hat{K}\ll m\cc\bigg |\ll{\a = {3\over 2}} = 2\uu{-m}\cc \hat{P}\ll{m+1}(\a) = \\
  &= \text{coefficient of \(\JJM^{-m}\) in \(\log(\Gamma(\JJM + 5/2))\)} = \frac{(-1)^m}{m \pqty{m+1}} B_{m+1}(5/2) .
\end{aligned}
\end{equation}
Lacking a closed-form expression, we instead compare
our prediction \rr{SQCDPowerLawCorrectionPrediction} 
with data from the numerical evaluation of
correlation functions, as in~\cite{Hellerman:2017sur}.  

We begin with the zero-instanton
approximation to the sphere partition function 
$Z = Z\ll 0 = \exp{q\ll 0}\to Z\upsns{\rm zero-instanton}$ and evolve up to $n = 40$ using the recursion relations
to get an approximate answer; we expect that the omission
of instanton effects should lead to errors no
larger than relative size $\exp{- 2\pi \cc \Im(\t)}$,
which is smaller than $2\times 10\uu{-3}$ for $\Im(\t) > 1$.

In the Appendix~\ref{sec:numerics} we present data for correlation
functions.  We take the second difference with respect to \(n\) in \(q_n\) to cancel the $n\uu 0$ and
$n\uu 1$ terms in $\ampname\ll n$,
\begin{equation}
  \label{eq:second-difference}
  \difference_n^2 q_n  = q_{n+2} - 2 q_{n+1} + q_{n},
\end{equation}
and we compare the $m\uu{-1}$ and smaller terms.

\begin{figure}
  \centering
  \begin{tikzpicture}
    \footnotesize
    \node at (0,0) {\includegraphics[width=.9\textwidth]{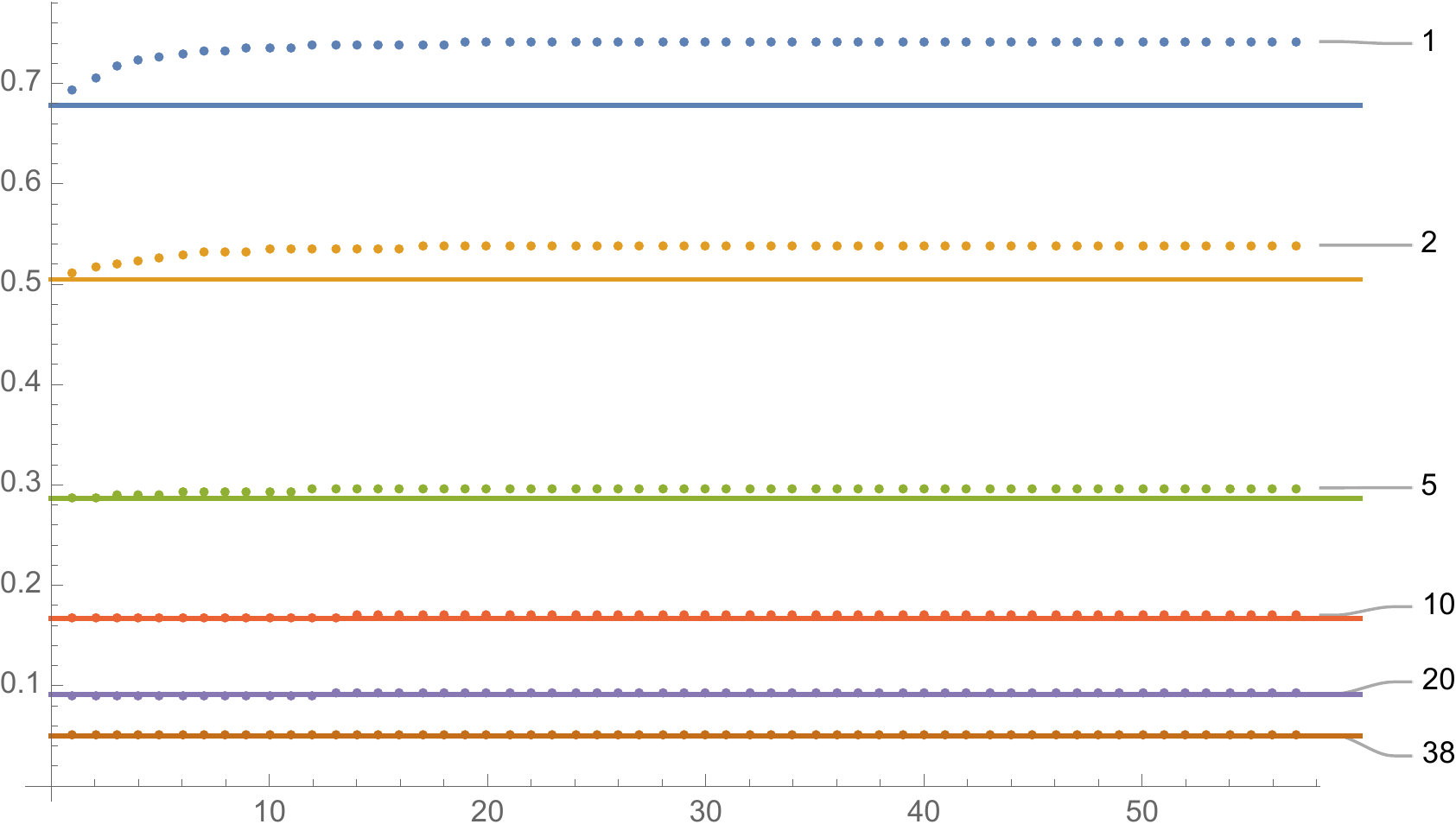}};
    \node at (-6,4) {\(\difference^2_n q_n\)};
    \node at (5,-3.5) {\(\Im \tau\)};
  \end{tikzpicture}
  \caption{Second difference in \(n\) for \(\difference^2_n q_n^{\text{(loc)}} \) (dots) and for \(\difference^2_n q_n^{EFT}\) (continuous lines) as function of \(\Im \tau\) at fixed values of \(n\). The numerical results quickly reach a \(\tau\)-independent value that is well approximated by the asymptotic formula when \(n\) is larger than \(n \gtrsim 5\).}
  \label{fig:delta-n-fixed-n}
\end{figure}

We find that the numerical values are in beautiful agreement with the prediction \rr{SQCDPowerLawCorrectionPrediction}. Quite rapidly, already for \(\tau \simeq 4i\), the \(\tau\)-dependence drops for all values of \(n\).
The asymptotic values is well approximated by our prediction \(\difference^2_n q_n^{\acs{eft}}\) for \(n\) larger that \(n \gtrsim 5\), where the discrepancy between the \ac{eft} result and the localization is of order \(1 - \eval{\difference^2_n q^{\acs{eft}}/ \difference^2_n q^{\text{(loc)}}}_{n = 5, \Im(\tau) \muchgreaterthan 1} \approx 1\%\).
Even for \(n = 1\), the discrepancy is only of order \(1 - \eval{\difference^2_n q^{\acs{eft}}/ \difference^2_n q^{\text{(loc)}}}_{n = 1, \Im(\tau) \muchgreaterthan 1} \approx 8\%\) (see Fig.~\ref{fig:delta-n-fixed-n}).
In Sec.~\ref{sec:SQCD}, we will estimate the behavior of the discrepancy as function of $\JJM$ and $\Im\tau$.

\section{Universal theory-independence of the $\JJM\uu{-m}$ corrections}
\label{Universality}

\subsection{Initial comments}

The derivation of the $1/n$ corrections in Section~\ref{sec:lagtheo}, has relied beyond first order on the recursion relations given in Eq.~\eqref{eq:recursion-relations}. 
These recursion relations,
as derived in~\cite{Papadodimas:2009eu}, apply only to theories with a marginal coupling.  Despite this, the actual formula for the
power-law corrections is completely independent of the marginal coupling, depending only on the $\a$-coefficient of the theory.  It is
tempting, therefore, to wonder whether the formula may also apply to rank-one theories with other values of $\a$ (for instance
those in the classification of Argyres {\it et al.},~\cite{Argyres:2015ffa, Argyres:2015gha, Argyres:2016xua, Argyres:2016xmc}),
most of which do not have a marginal coupling at all.  In this section we shall present arguments suggesting that the formula in Eq.~\eqref{eq:Reflando} for the power-law corrections should apply to any value of $\a$, for Coulomb-branch chiral-primary correlators
in any rank-one theory with any value of $\a$, whether or not it has a marginal coupling.

For infinitely many values of $\a$, the \ac{eft} has a nonunitary but superconformally invariant regulator obtained by
adding adjoint and fundamental hypermultiplets, of which either or both have the opposite statistics to that of a unitary matter field.
These nonunitary theories do have marginal couplings, and
we observe that the derivation of the recursion relations works equally well as in the unitary case
since this derivation does not
rely on unitarity at all.

\subsection{Ultraviolet regulators with marginal couplings}\label{sec:UVreg}

For three values of $\a$, namely $\a = 0,+1, +{3\over 2}$, 
the Coulomb-branch \ac{eft} has a unitary \ac{uv}
completion which in the first case is exactly
free and in the latter two cases has a marginal
coupling parameter.  These are the only
three unitary \acp{scft} with
marginal coupling and one-dimensional Coulomb branch.

We note, however, that unitarity\footnote{or more precisely,
reflection positivity in the Euclidean path integral}
appears to play no role in the derivation of the recursion
relations.  We can therefore consider gauge theories
with both nonunitary as well as unitary matter sectors.
If the matter is chosen so that the $\beta$-function vanishes
identically, these gauge theories will have marginal
gauge coupling $\t$ and values of the $\a$-coefficient
$\a = {5\over{12}}\cc {{{a\lrm{CFT} - a\lrm{EFT}}\over{a\lrm{free~U(1)~vector}}}}$.

The theories are constructed by adding ``ghost'' hypermultiplets -- with the same super-conformal transformations and R-symmetry quantum numbers as 
ordinary hypermultiplets, but with spin-statistics reversed -- that is, a multiplet whose lowest component is a scalar fermion, transforming in the ${\bf 2}$ of the $SU(2)$ R-symmetry group.

We give some details of these theories in  Appendix~\ref{GhostHyperAppendix}.
The relevant facts are the $\bfuc$-function cancellation
condition~\eqref{BetaFunctionCancellationWithGhostHypersFirstVersion}
  and the anomaly mismatch $\a$-coefficient of the Coulomb branch \ac{eft}.  If there
are $n\ll {\bf R}\uprm{hyper}$ in a representation
${\bf R}$ of the gauge group,
and some number of ghost hypers $n\ll {\bf R}\uprm{ghost~hyper}$ in a representation
of the gauge group, then the hypers contribute to the
$\b$-function and Weyl anomaly through the difference
between the two, 
\bbb
\tilde{n}\ll{\bf R}\uprm{hyper} \equiv 
n\ll{\bf R}\uprm{hyper} - n\ll{\bf R}\uprm{ghost~hyper}\ .
\eee

We can construct many interesting nonunitary
\acp{scft} this way, but the simplest choice is
to take only hypers and ghost hypers in the $\bf{2}$ and
$\bf{4}$ representations, in which case the $\b$-function
vanishes if
\begin{equation}
\tilde{n}\ll 2 \uprm{hyper}  = 8 - 10\cc \tilde{n}\ll{\bf 4}\uprm{hyper} \ ,
\end{equation}
and the theory is conformal.  There are no massless degrees of freedom on the Coulomb branch other than the massless $U(1)$ vector multiplet,
and the value of the $\a$-coefficient is
\begin{equation}
\a = {3\over 2} - {4\over 3}\tilde{n}\ll{\bf 4} .
\end{equation}
This shows that the Coulomb-branch \ac{eft} with
super-\ac{wz} $\a$-coefficient $\a = {3\over 2} - {4\over 3} \cc \tilde{n}\ll 4$ can be obtained with a manifestly
${\cal N} = 2$ superconformal regulator, analogous
to a Pauli--Villars regulator.  

Other values of $\a$
are obtainable using ${\cal N} = 2$ superconformal gauge theories with ordinary
and ghost hypermultiplets in
higher representations of $SU(2)$, but for the
present purposes it suffices to show how
an infinite number of values of $\a$ may be obtained
with such constructions.

Apart from the manifest superconformal invariance,
a second valuable feature of this regulator is
the fact that it possesses a marginal coupling and
thus obeys the recursion relations of~\cite{Papadodimas:2009eu} when considered as a full \ac{scft} rather than as a regulator for 
the vector multiplet \ac{eft}.  Though the recursion
relations were originally applied in the context of
unitary superconformal gauge theory, unitarity of the
\ac{cft} appears to play no essential role in the derivation of the
relations, and thus one expects ${\cal N} = 2$
superconformal gauge theories with ghost-hypers
to obey the same recursion relations as those with
ordinary matter.  We will now make use of this fact 
to write recursion relations for the power-law corrections
for more general values of the $\a$-coefficient.
\subsection{Universal polynomials}\label{sec:univPoly}

In the previous section, we have observed that there
appear to be an infinite number of rank-one $\mathcal{N} = 2$ superconformal gauge theories,
realized by $SU(2)$ gauge theory with combinations
of ordinary hypermultiplets with ghost hypermultiplets.
By taking such combinations, we find we can
regulate any one-dimensional Coulomb-branch \ac{eft} with $\a$-coefficient satisfying $\a \in {3\over 2} - {4\over 3} \cc \IZ$ by a superconformal $SU(2)$ gauge theory with
ordinary hypers and ghost hypers in the $2$- and
$4$-dimensional representations.  These
theories do have the marginal coupling constant $\t$ and must
therefore obey the recursion relations with respect
to $\t$, as in~\cite{Papadodimas:2009eu}.

As we have shown diagrammatically in Section~\ref{sec:alpha-dep}, the $1/n^m$ corrections to the logarithms of the two-point functions can
depend only on $\JJM$ and $\a$, and not on $\D\ll\co$
and $n$ individually, nor on marginal couplings, nor on any
other details of the microscopic completion, whether
it be a unitary quantum field theory or an artificial regulator,
so long as the regulator preserves exact ${\cal N} = 2$
superconformal symmetry and possesses an exact
conformal manifold parametrized by the gauge coupling 
$\t$. 
This means the terms
$K\ll m = P\ll m(\a)$ are universal polynomials,
common to any superconformal quantization of the \ac{eft},
whether Lagrangian or non-Lagrangian, unitary or
nonunitary.  At the same time, the recursion relations
establish that
\begin{equation}
\hat{P}\ll{m+1}(\a) = \hat{P}\univ\ll{m+1}(\a)
\end{equation}
for any superconformal \ac{uv}-completion obeying the recursion relations, with $\hat{P}\univ\ll{m+1}(\a)$ defined
in Eq.~\eqref{eq:universal-P}, whether the completion is unitary or not.
Since there appear to exist nonunitary superconformal
regulators for the Coulomb-branch \ac{eft} for an infinite number of distinct values of $\a$, it follows that 
$\hat{P}\ll{m+1}(\a) = \hat{P}\univ\ll{m+1}(\a)$
must agree for an infinite number of values, for 
any $m\geq 1$.  Any two polynomials of order $m+1$
that agree for $m+1$ or more values, must
agree identically, and we conclude
\begin{equation}
\hat{P}\ll{m+1}(\a) = \hat{P}\univ\ll{m+1}(\a) = \text{coefficient of \(\JJM^{-m}\) in \(\log(\Gamma(\JJM + \alpha + 1))\)}
\end{equation}
for all values of $\a$.

Assuming the accuracy of our inferences about the properties of the ghost-hyper regulators,
this establishes the formula for $\hat{P}\ll{m+1}(\a)$
for any value of $\a$, independent of any
reference to a \ac{uv}-completion.  In particular,
the polynomials $\hat{P}\ll{m+1}(\a)$ must
also give the power-law corrections $\hat{K}\ll m / \JJM\uu m$ for correlators of Coulomb-branch chiral primaries
in the non-Lagrangian theories in~\cite{Argyres:1995jj,Argyres:1995xn,Argyres:2015ffa, Argyres:2015gha, Argyres:2016xua, Argyres:2016xmc}. %
\section{Universal EFT behavior compared with $S\uu 4$ localization}
\label{sec:SQCD}

In Section~\ref{NonuniversalCorrectionDiscussion}, we showed that 
the \ac{eft} approximation to the $q\ll n$
must be universal to all orders in inverse powers
of $\JJM = n\D\ll\co$, but not universal nonperturbatively
in $n$.  In this section, we check this claim
in detail using the $q\ll n$ correlation functions in ${\cal N} = 2$ superconformal \ac{sqcd} with $G = SU(2)$ and
$N\ll F = 4$, as computed from the $S\uu 4$ partition
function as done in~\cite{Papadodimas:2009eu, Baggio:2014sna, Baggio:2014ioa, Baggio:2015vxa, Gerchkovitz2017}.

This method makes use of the \ac{bps} property of the
two-point function, using recursion relations to obtain
a closed-form expression for each $q\ll {n+2}(\t,\tb)$
in terms of the two previous amplitudes $q\ll {n+1}(\t,\tb)$
and $q\ll n(\t,\tb)$, for any rank-one theory with a
marginal coupling.  Since $q\ll 0(\t,\tb)$ can be computed
unambiguously (up to holomorphic scheme-dependence~\cite{Gerchkovitz:2014gta} which correspond to
Kahler transformations of the conformal-manifold Kahler
potential and which cancel out in the normalized correlation functions)
by supersymmetric localization, this gives in principle
a closed form expression for every $q\ll n$.  In practice,
even the sphere partition function $q\ll 0$ itself is
quite a complicated function of $\t$ and $\tb$,
and its evolution to higher $q\ll n$ by the recursion relations
grows rapidly in complexity with higher $n$,
making exact evaluation of large-order $q\ll n$ intractable
for $n$ moderately large.

We can evade this difficulty by 
taking only the perturbative piece of the sphere
partition function as an initial condition, and
considering only the second difference in Eq.~\eqref{eq:second-difference} \(\difference_n^2 q_n  = q_{n+2} - 2 q_{n+1} + q_{n},\) as in~\cite{Hellerman:2017sur}.
This has two advantages. First, the difference 
\(\difference_n^2 q_n \) removes
the theory-dependent coefficients ${\bf A}$ and ${\bf B}$
of the $\JJM\uu 1$ and $\JJM\uu 0$ terms from the
asymptotic expansion, and allows us to isolate the
power corrections and exponentially small terms.
Explicitly
\begin{equation}
  \difference_n^2 q^{\acs{eft}}_n = \log( \frac{\pqty{2n + \alpha + 3 } \pqty{ 2n + \alpha + 4}}{\pqty{2n + \alpha + 1} \pqty{ 2n + \alpha + 2}}) .
\end{equation}
Second,
we make use of the fact that our predicted values for
the power-corrections are universal, and therefore
coupling-independent. We can therefore take
a limit in which gauge instantons are unimportant.  
In particular, we can take a weak-coupling limit $\t\to
+ i \infty$.

We must be somewhat careful to take the limit in such
a way that the nonuniversal exponentially small
corrections are not enhanced by taking weak coupling:
The validity of the \ac{eft} depends on the ratio between the
infrared scale and the physical masses of the lowest massive excitations, such as the hypermultiplets and the
$W$-bosons.  If we set  this ratio too small, the validity
of the \ac{eft} would break down entirely, and in particular
we would expect the exponentially small corrections
$\exp{ - M / E\lrm{IR}}$ associated with macroscopic virtual
propagation of massive particles, to become large if we
were to take $\t\to +i\infty$ at fixed $\JJM$.
The clash of limits between $\t\to i\infty$
at fixed $\JJM$, and  $\JJM\to\infty$ at fixed $\t$  is inevitable for a simple reason: 
The Coulomb-branch \ac{eft} is obtained by integrating
out the massive excitations, the $W$-bosons and
hypermultiplets.  At fixed $|\phi\lrm{unit}| \propto \sqrt{\JJM} / r$ the masses of the lightest massive excitations
go as $g\lrm{YM} \cc |\phi\lrm{unit}|$.  If we fix $\JJM$,
at however large a value, taking the weak-coupling
limit always brings massive excitations to masses below
our Wilsonian cutoff $\L$, and the predictions of the
\ac{eft} are no longer valid.  
If on the other hand we fix $\Im \tau$ at however
large a value, and take $\JJM\to \infty$, the weak
coupling predictions are invalidated by the combinatorics
of the diagrams: For sufficiently many external legs, 
loop corrections to correlators
are enhanced by powers of $\JJM$ and gauge-theoretic
perturbation theory breaks down at arbitrarily weak
coupling $\JJM\uu{-1/2} \lesssim g\lrm{YM} \muchlessthan 1$.

The solution to this problem is to take the double-scaling limit of~\cite{Bourget:2018obm}, in which $\JJM$ is taken to
infinity with $\l = 2 \pi \, \JJM / \Im \tau$ held
fixed.  The ratio of the infrared scale to the 
mass of the heavy excitations is fixed in terms of $\l$,
so in this limit gauge instantons are suppressed while virtual
macroscopic massive propagation is suppressed exponentially as $\exp( - \text{const.} \times \lambda^{1/2}) $.

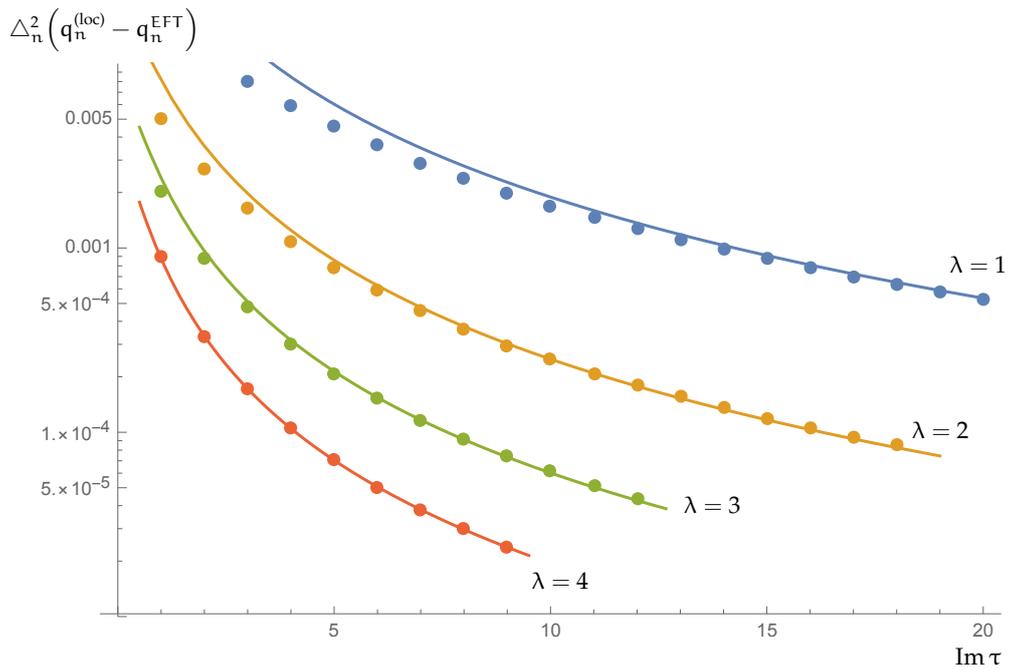
\begin{figure}
  \centering
  \begin{tikzpicture}
    \footnotesize
    \node at (0,0) {\includegraphics[width=.9\textwidth]{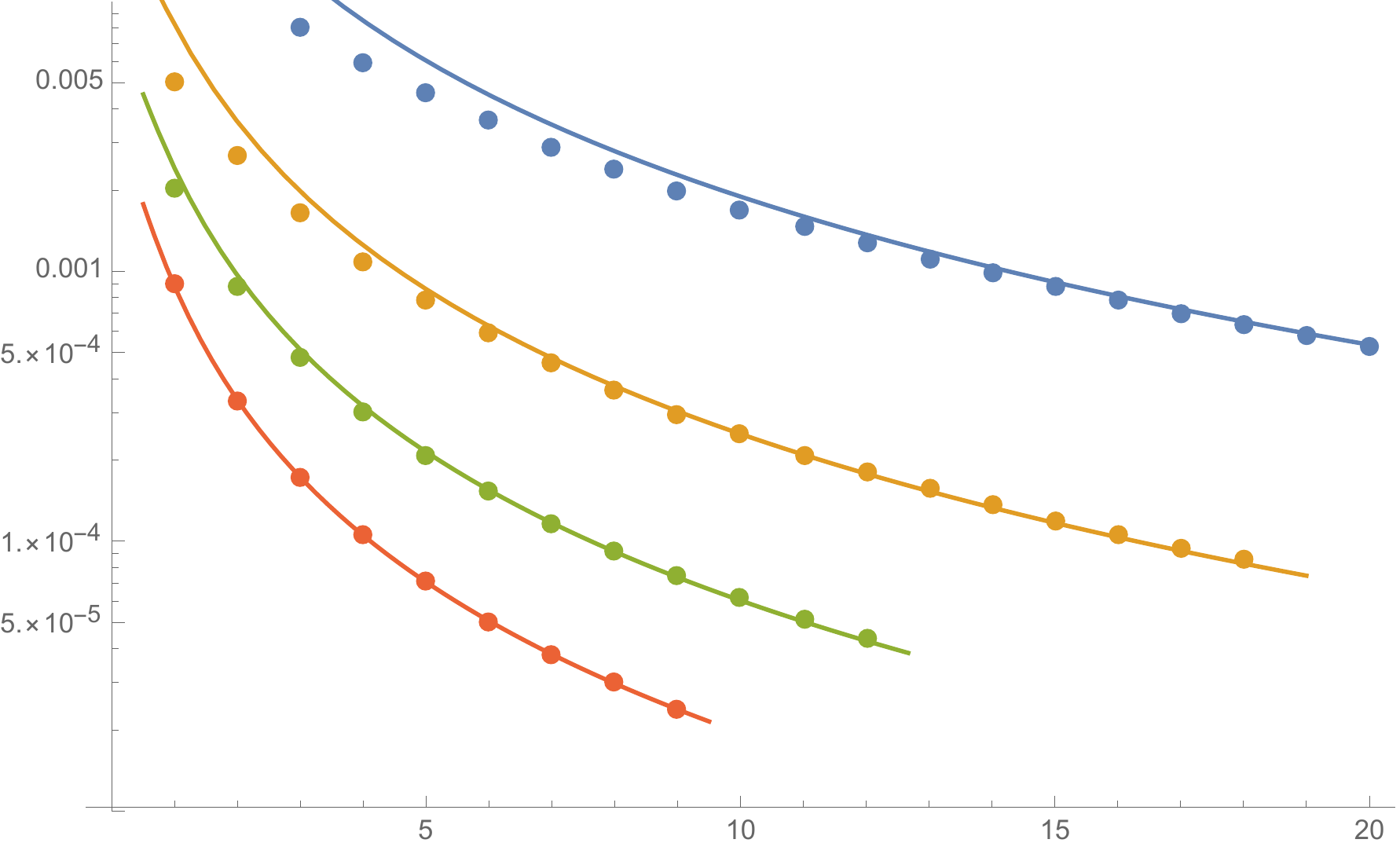}};
    \node at (-5.5,4.3) {\(\difference^2_n \pqty{q^{\text{(loc)}}_n - q^{EFT}_n}\)};
    \node at (6,-4) {\(\Im \tau\)};

    \node at (6,1.2) {\(\lambda = 1\)};
    \node at (5.5,-1) {\(\lambda = 2\)};
    \node at (2.5,-2) {\(\lambda = 3\)};
    \node at (.5,-3) {\(\lambda = 4\)};

  \end{tikzpicture}
  \caption{Second difference in \(n\) for the discrepancy between localization and EFT results \(\difference^2_n \pqty{q^{\text{(loc)}}_n - q^{EFT}_n}\) (dots) compared to \(\difference^2_n \pqty{ A_1  e^{- A_2 \sqrt{n/\Im \tau} }}\) (continuous lines) as functions of \(\Im \tau\) at fixed values of \(n/\Im \tau  \) at \(A_1 = 1.8 \), \(A_2 = 3.2 \). The agreement is quite good already for \(\lambda = 3\).}
  \label{fig:exponential-correction}
\end{figure}

\begin{figure}
  \centering
  \begin{tikzpicture}
    \footnotesize
    \node at (0,0) {\includegraphics[width=.9\textwidth]{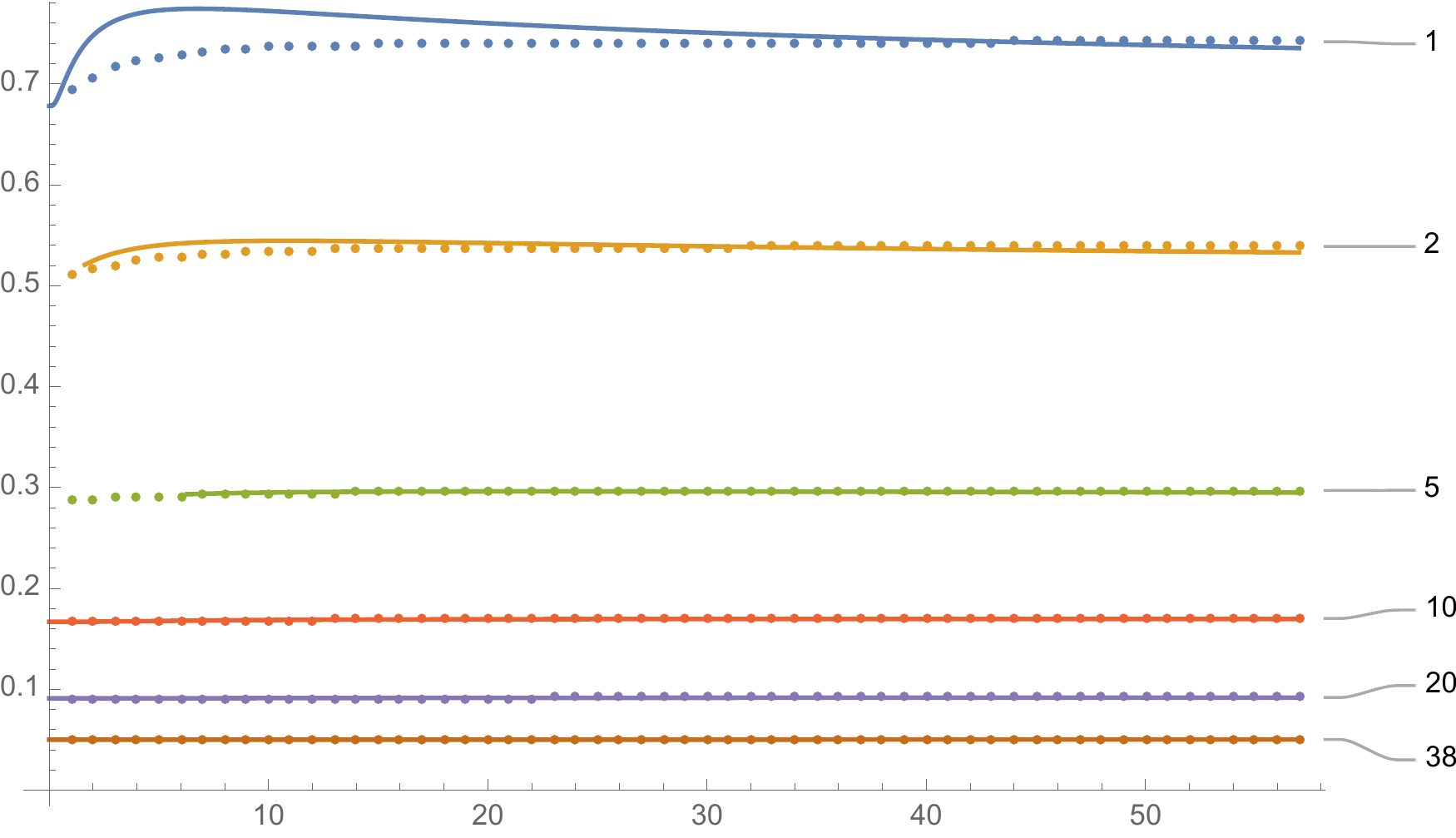}};
    \node at (-6,4) {\(\difference^2_n q_n\)};
    \node at (5,-3.5) {\(\Im \tau\)};
  \end{tikzpicture}
  \caption{Second difference in \(n\) for \(\difference^2_n q_n^{\text{(loc)}} \) (dots) and for \(\difference^2_n \pqty{q_n^{EFT} + A_1 e^{-A_2 \sqrt{n/\Im \tau}}}\) (continuous lines) as function of \(\Im \tau\) at fixed values of \(n\). The exponential term seems to account for most of the discrepancy at small values of \(n\) (compare with Figure~\ref{fig:delta-n-fixed-n}).}
  \label{fig:delta-n-fixed-n-instanton}
\end{figure}

In
Appendix~\ref{sec:numerics}, we compare the sum rule for the 
universal power law formula with the data of the
sphere partition function in the double scaling limit,
in the case of ${\cal N} = 2$ \ac{sqcd} with $G = SU(2)$ and
$N\ll f = 4$.
We find a remarkably accurate agreement and then can estimate the leading correction to be
\begin{equation}
  \eval{\log(\ampname\ll n)}^{\acs{sqcd}}_{\text{localization}}  - \eval{\log(\ampname\ll n)}^{\acs{sqcd}}_{\ac{eft}} \approx A_1 \, e^{- A_2 \sqrt{\JJM/(2\Im \tau)}} 
\end{equation}
where \(A_1 = 1.8(2) \) and \(A_2 = 3.2(1) \).
In Figure~\ref{fig:exponential-correction} we show how this simple form already reproduces the localization data for values of \(\lambda \approx 3\).
In Figure~\ref{fig:delta-n-fixed-n-instanton} we show how adding this contribution improves the agreement between our prediction and the localization data also at smaller values of \(n\) (this is to be compared with the purely perturbative results shown in Figure~\ref{fig:delta-n-fixed-n}).

\def\ERRORANALYSISDEFOUTB{ 
\heading{Approximation of the difference equation
by a differential equation}

For sufficiently large $n$, we can approximate the $n$-dependent linear
difference equation for $\d$ by an $n$-dependent linear differential equation,
\bbb
{{d\sqd}\over{dn\sqd}}\d\ll n = \rdots\ .
\een{WellKnownWhoeverEquation}
\shg{Note the error in approximating the difference equation
by a differential equation.}
This equation is the well-known $(\rdots)$ equation,
whose solution is a linear combination of the
two solutions:
\bbb
{\cal S}\ll A(n) \equiv \rdots
\een{WellKnownWhoeverEquationSolutionA}
and
\bbb
{\cal S}\ll B(n) \equiv \rdots
\een{WellKnownWhoeverEquationSolutionB}

\heading{Large $\JJM$ to larger $\JJM$}

Now we can try to understand how small perturbations
of initial data, propagate in the asymptotic large-$\JJM$
regime from large $\JJM$ to larger $\JJM$.  With
$n\ll i \muchgreaterthan 1$, we can match the perturbation
of the initial conditions precisely with the solutions
\rr{WellKnownWhoeverEquationSolutionA}, \rr{WellKnownWhoeverEquationSolutionB}
to the well-known $(\rdots)$ equation \rr{WellKnownWhoeverEquation}.

We have
\bbb
\d\ll n = {\bf C}\ll A\cc {\cal S}\ll A(n) +  {\bf C}\ll B\cc {\cal S}\ll B(n)
\eee

\shg{From here on, how I proceed depends on how
the solutions actually grow / don't grown.}

\heading{Imposition of $S$-duality symmetry}

Let us now consider the case where the theory possesses
and $S$-duality symmetry.

First, we note that this is equivalent to the statement
that $\exp{q\ll n}$ is a (nonholomorphic) modular form of degree $(\rdots)\D\ll\co + (\rdots)\a$ as a function
of $\t,\tb$.

\shg{Show this!}

\shg{IMPORTANT: Also note which $\t$ you are talking
about.  It is important we get straight the issue of
the difference between $\rdots$}

}
\section{Discussion}
\label{sec:conclusions}

In this article, we have studied $\mathcal{N}=2$ \aclp{scft} with a one-complex-dimensional Coulomb branch in a sector of fixed and large $R$-charge  $\JJM$.
Making use of the resulting \acl{eft} on the Coulomb branch at large $\JJM$, we have improved the results of~\cite{Hellerman:2017sur}, giving the \(1/\JJM\) expansion of the two-point functions $\ampname\ll n = \abs{x - y}^{ 2 n\D\ll\co} \ev{ (\mathcal{O}(x))^n (\bar{\mathcal{O}}(y))^n}$  to all orders. 
The absence of higher-order $F$-terms in our \ac{eft} on the Coulomb branch plays a critical role: it implies that the effective action on the Coulomb branch is given by the tree-level effective kinetic term, the supersymmetrized \ac{wz} term for the spontaneously broken Weyl invariance, and unknown D-terms which do not affect correlation functions of chiral primaries. 
	
\medskip
The first term in the large-$\JJM$ expansion is computed explicitly from the \ac{eft} in terms of Feynman diagrams. Then, we observe that for theories with a marginal coupling, we can compute all higher terms using recursion relations, arguing from the \ac{eft} that the higher terms do not depend on the marginal coupling.

Even though the recursion relations we solve apply only to \acp{scft} with a marginal coupling, we find effective recursion relations in the \ac{eft} for any theory with a one-dimensional Coulomb branch, whether it has a marginal coupling constant or not.  Based on this we argue that the correlators for \emph{any} four-dimensional $\mathcal{N}=2$ \ac{scft} with a one-dimensional Coulomb branch have a universal large-$\JJM$ behavior given by
\begin{equation}
    q\ll n \simeq \JJM\cc {\bf A} + {\bf B} + \log(\G(\JJM + \a + 1)) ,
\end{equation}
where ${\bf A}$ and ${\bf B}$ are theory-dependent constants and the \( \simeq\) indicates the presence of non-universal corrections that are exponentially small in \(\JJM\).

We discuss a number of concrete examples, such as \(su(2)\) $\mathcal{N}=2$ \ac{sqcd} in four dimensions with four flavors, which allows us a direct cross-check against numerical localization computations, verifying our results at high accuracy and allowing us even to see the leading exponential corrections to the asymptotic large-charge expansion.

\medskip

The results of this article highlight once more the universal applicability of the large-quantum number expansion. The constraints imposed on the models by supersymmetry and working at fixed large charge conspire and allow us to obtain analytic results of an unprecedented precision.

Also the importance of the underlying vacuum structure for the large-quantum number expansion is becoming increasingly obvious. The \acp{eft} at fixed charge display a \emph{universal behavior} depending on the nature of the ground state manifold, as evidenced by the classes of models studied so far with either a unique ground state, such as the $O(n)$ vector models and the $W=\Phi^3$ \ac{scft}~\cite{Hellerman:2015nra, Alvarez-Gaume:2016vff}, and the  $\mathcal{N}=2$ \acp{scft} with a one-dimensional Coulomb branch discussed in~\cite{Hellerman:2017sur} and this article.

\medskip

There are three obvious directions in which the present work can and
should be extended.  
\begin{itemize}
\item First, it would be very valuable to generalize our results
to correlation functions of operators in chiral rings of higher-dimensional
Coulomb branches.  These correlation functions likely lack the degree of
universality seen in the present work, because the Coulomb branch \acp{eft}
in rank greater than one, contain non-Goldstone excitations.  Nonetheless,
one would expect holomorphy and symmetries to sharply constrain
the possible higher-derivative F-terms, and thus lead to many relations
among correlators at large J.
\item Second, correlation functions at large $SU(2)$ R-charge could be computed
using \ac{eft} methods and related to computations by exact methods 
such as the ones used in~\cite{Beem:2013sza}.%
\item Third, the leading exponentially small correction to the power-law correlators
in the \(1/J\) expansion is quite interesting.  Numerical analysis and comparison with the universal power-law terms suggest strongly that the contribution has an interpretation in terms of propagation of a massive particle over macroscopic distance on the infrared scale.  It would be illuminating to identify the particle and its semiclassical trajectory, with a goal of matching the exponent and prefactor calculated numerically in the
case of ${\cal N} = 2$ \ac{sqcd} and generalizing the form of the exponentially small correction to the case of non-Lagrangian theories.
\end{itemize}

Another important future direction is to gain
a more intrinsic understanding of
the universality of the \(1/J^m\) power-law corrections.
In the present paper, we have proven our
universal formula directly for superconformal
gauge theories with marginal coupling,
and we have used an indirect argument to show
the formula must hold for all theories of rank one, 
including non-Lagrangian \acp{scft}.  We have also
independently computed an infinite series of coefficients
directly in the Coulomb-branch \ac{eft}, supporting the prediction
of the universal formula.  

We note that the recursion relations for Coulomb
branch correlators in ${\cal N} = 2$ gauge
theories are a particular kind of consistency condition among OPE coefficients, conceptually similar to the conformal bootstrap equations.  The simplification of the recursion relations for power-law corrections
from differential equations to algebraic equations, is intriguing
and suggests the possibility of a simpler derivation within the Coulomb-branch
EFT itself, avoiding the need to consider \ac{uv}-completions
of the EFT with marginal coupling.  A "algebraic version" of the recursion relations in EFT, not depending on the existence of
a marginal coupling, would have a closer formal similarity to the bootstrap
equations, with only a finite number of operators exchanged in either
channel.

 So far, the abstract "bootstrap" approach to CFT has been unable to prove the large-charge
behavior of CFT data, that is straightforwardly visible in the EFT picture.\footnote{Though see~\cite{Jafferis:2017zna} for progress in this direction.}. It would be
draw a more precise connection between the "EFT version" of the recursion
relations, and the conformal bootstrap equations, to derive
some aspects large-charge CFT data for higher-rank ${\cal N}=2$ theories, ${\cal N}=1$ theories, and for theories without vacuum manifolds, such as the $O(2)$ model. 

\newpage
\section*{Acknowledgements}

The authors would like to thank Ben Heidenreich for discussions and Zohar Komargodski, Gabriele Tartaglino--Mazzucchelli and Antoine Van Proeyen for correspondence.
DO~and SR~would like to thank the Kavli IPMU for hospitality during part of this work.
The work of SH~is supported by the World Premier International Research Center Initiative (\textsc{wpi} Initiative), \textsc{mext}, Japan; by the \textsc{jsps} Program
for Advancing Strategic International Networks to Accelerate the Circulation of Talented
Researchers; and also supported in part by \textsc{jsps kakenhi} Grant Numbers \textsc{jp22740153,
jp26400242}.  The authors particularly thank the \textsc{wpi} Initiative and Kavli IPMU for early funds supporting the meetings that generated this work.  SH thanks the University of Bern, the University of Torino, the Burke Institute at Caltech and the Galileo Galilei Institute for hospitality while this research was in progress.  SM~and~MW~acknowledge the support by \textsc{jsps} Research Fellowships for Young Scientists.
DO~acknowledges partial support  by the \textsc{nccr 51nf40-141869} ``The Mathematics of Physics'' (\textsc{swissmap}).
The work of SR~is supported by the Swiss National Science Foundation (\textsc{snf}) under grant number \textsc{pp00p2\_157571/1}.

\appendix

\section{Solving the recurrence equation}
\label{sec:solve-recurrence}

In~\cite{Papadodimas:2009eu, Baggio:2014sna, Baggio:2014ioa, Baggio:2015vxa, Gerchkovitz2017} it was observed that the correlations functions that we are interested in obey the Toda lattice equation
\begin{equation}
  \del \bar \del q_n(\tau, \bar \tau) = \exp[ q_{n+1}(\tau, \bar \tau) - q_n(\tau, \bar \tau)] - \exp[ q_{n}(\tau, \bar \tau) - q_{n-1}(\tau, \bar \tau)] .
\end{equation}
In this appendix we want to show how to solve this equation using the extra information coming from the \ac{eft} about the \(\tau\) dependence of the asymptotic expansion of \(q_n\) for large \(n\).

First, it is convenient to rewrite the second-order equation as a system of two first-order equations~\cite{Gibbon:1985dw}:
\begin{equation}
  \label{eq:first-order-system}
  \begin{cases}
    \del P_n(\tau, \bar \tau) = P_n(\tau, \bar \tau) \pqty{ Q_n(\tau, \bar \tau) - Q_{n-1} (\tau, \bar \tau)} \\
    \bar \del Q_n(\tau, \bar \tau) = P_{n+1}(\tau, \bar \tau) - P_n(\tau, \bar \tau) ,
  \end{cases}
\end{equation}
where
\begin{align}
  Q_n(\tau, \bar \tau) &= \del q_n(\tau, \bar \tau) ,&
                                                      P_n(\tau, \bar \tau) &= \exp[q_n(\tau, \bar \tau) - q_{n-1}(\tau, \bar \tau)] .
\end{align}
In Section~\ref{sec:diagr} we have seen that the dependence of  \(q_n(\tau, \tau \bar)\) on \(\tau\) is at most affine (\emph{i.e.} only the constant and linear in \(n\) terms depend on \(\tau\)).
We can separate this by writing
\begin{equation}
  q_n(\tau, \bar \tau) = n f(\tau, \bar \tau) + k_0(\tau, \bar \tau) + M_n .
\end{equation}
The variables \(Q_n\) and \(P_n\) then read
\begin{align}
  Q_n(\tau, \bar \tau) &= n \del f(\tau, \bar \tau) + \del k_0(\tau, \bar \tau) ,\\
  P_n(\tau, \bar \tau) &= e^{f(\tau, \bar \tau)} \exp[M_n - M_{n-1}] = e^{f(\tau, \bar \tau)} \Lambda_n  .
\end{align}
With this ansatz the first equation in Eq.~\eqref{eq:first-order-system} is identically satisfied and we only need to solve
\begin{equation}
  n \del \bar \del f(\tau, \bar \tau) + \del \bar \del k_0(\tau, \bar \tau) = e^{f(\tau, \bar \tau)} \pqty{ \Lambda_{n+1} - \Lambda_n }.
\end{equation}
If we isolate the terms that do not depend on  \(\tau, \bar \tau\) we can rewrite the equation as the system
\begin{align}
    \del {\bar \del} f(\tau, \bar \tau) = 2A  e^{f(\tau, \bar \tau)} ,\label{eq:Liouville} \\
    \del {\bar \del} k_0(\tau, \bar \tau) = B e^{f(\tau, \bar \tau)} , \label{eq:Poisson} \\
    \Lambda_{n+1} - \Lambda_n = 2A n + B,
\end{align}
where \(A \) and \(B \) are constants.
We see that  \(f(\tau, \bar \tau)\) obeys the Liouville equation~\eqref{eq:Liouville} on a hyperbolic plane of Gaussian curvature \(-4A\), and it sources the Poisson equation~\eqref{eq:Poisson} satisfied by \(k_0(\tau, \bar \tau)\).

The equation for \(\Lambda_n\) is easily solved and gives
\begin{equation}
  \Lambda_n = A n \pqty{n - 1} + B n + C' = A \pqty{n - n_+} \pqty{n - n_-},
\end{equation}
where \(C'\) is an integration constant and \(n_{\pm}\) are two numbers that satisfy
\begin{align}
  n_+ + n_- &= 1 - \frac{B}{A} , & n_+ n_- &= \frac{C'}{A} .
\end{align}
Using this expression we can solve for \(M_n\):
\begin{equation}
  e^{M_n - M_{n-1}} = \Lambda_n
\end{equation}
and find
\begin{equation}
  M(n) = D + n \log A + \log [\Gamma(n - n_- + 1) \Gamma( n - n_+ + 1)] ,
\end{equation}
where \(D\) is an integration constant.

Let us now consider the \(\tau\)-dependent equations.
The Liouville equation~\eqref{eq:Liouville} for \(f(\tau, \bar \tau)\) admits the general solution
\begin{equation}
  e^{f(\tau, \bar \tau)} =  \frac{1}{A} \frac{\abs{\del \phi(\tau)}^2}{\pqty{1 - \abs{\phi(\tau)}^2}^2} ,
\end{equation}
where \(\phi(\tau)\) is a meromorphic function.
Now that we have solved for \(f(\tau, \bar \tau)\), we can recast the equation for \(k_0(\tau, \bar \tau)\) as a Laplace equation:
\begin{equation}
  \del {\bar \del} \pqty{k_0(\tau, \bar \tau) + \frac{B}{2A} f(\tau, \bar \tau)} = 0
\end{equation}
so that \(k_0(\tau, \bar \tau)\) is given by
\begin{equation}
  k_0(\tau, \bar \tau) = -\frac{B}{2A} f(\tau, \bar \tau) + \psi(\tau) + \bar \psi (\bar \tau).
\end{equation}
We can now collect our results and write the final expression for \(q_n(\tau, \bar \tau)\):
\begin{equation}
  \label{eq:q-with-n-pm}
  \begin{aligned}
    q_n(\tau, \bar \tau) ={}& n f(\tau, \bar \tau) + k_0(\tau, \bar \tau) + M_n \\
    ={}& n\pqty{f(\tau, \bar \tau) + \log A} + k_0(\tau, \bar \tau) + D \\
    &+ \log \bqty{\Gamma(n - n_+ + 1) \Gamma(n - n_- + 1)}.
\end{aligned}
\end{equation}

Our solution depends on the constants, \(n_+\), \(n_-\).
They can be fixed in terms of the anomaly coefficient \(\alpha \) by comparing the large-\(n\) expansion of \(q_n(\tau, \bar \tau)\) with the results of the \ac{eft}.
Expanding the gamma function in the expression in Eq.~\eqref{eq:q-with-n-pm}:
\begin{multline}
  q_n(\tau, \bar \tau) = n\pqty{f(\tau, \bar \tau) + \log A - 2} + k_0(\tau, \bar \tau) + D + \log (2 \pi) -  \pqty{n_+ + n_- - 1} \log n + \\
  + \frac{1 + 3 n_+ \pqty{n_+ - 1 } + 3 n_- \pqty{n_- - 1 }}{6n} + \dots
\end{multline}
In~\cite{Hellerman:2017sur} it was shown that the coefficient of the term \(\log(n)\) is \(-\pqty{\alpha + 1/2}\), moreover in Eq.~\eqref{eq:K1} we have found the general form of the \(\order{1/n}\) term as function of \(\alpha\). We can use these two conditions to eliminate the constants \(n_\pm\):
\begin{equation}
  \begin{cases}
    - \pqty{ n_+ + n_- - 1} = \alpha + \frac{1}{2} \\
     \frac{1}{6} \pqty{1 + 3 n_+ \pqty{n_+ - 1 } + 3 n_- \pqty{n_- - 1 }} = \frac{1}{4} \pqty{\alpha^2 + \frac{\alpha}{2} + \frac{1}{6}},
  \end{cases}
\end{equation}
which gives us \(n_\pm\) as functions of \(\alpha\):
\begin{align}
  n_- &= -\frac{\alpha}{2}, & n_+ &= \frac{1}{2} - \frac{\alpha}{2} .
\end{align}
This allows us to use the duplication formula for the gamma function and to rewrite \(q_n(\tau, \bar \tau)\) in terms of the super-\ac{wz} coefficient \(\alpha\):
\begin{equation}
  \label{eq:recursion-result-gamma}
  q_n(\tau, \bar \tau) =  2n {\bf A}(\tau, \bar \tau) + {\bf B}(\tau, \bar \tau) + \log(\G(2n+ \a + 1)),
\end{equation}
where we have collected all the \(\tau \) dependence and constants in the two functions \(\mathbf{A}\) and \(\mathbf{B}\).
From this expression we can compute explicitly the large-\(\JJM\) expansion of \(q_n\) for example in the case of \(\mathcal{N} = 2\) \ac{sqcd} where \(\JJM = 2 n\).
We see that the coefficient of \(\JJM^{-m}\) is proportional to the Bernoulli polynomial \(B_{m+1}(\alpha + 1)\)~\cite{Luke:1969v1}:
\begin{multline}
  \label{eq:recursion-result-Bernoulli}
  q_n(\tau, \bar \tau) = \JJM\log(\JJM) + \pqty{\a + \hh} \cc\log(\JJM) + \pqty{{\bf A} - 1} \JJM + {\bf B} + \log \sqrt{2\pi} \\
  + \sum_{m=1}^N \frac{(-)^{m+1} B_{m+1}(\alpha + 1)}{m \pqty{m+1} \JJM^m} + \order{\frac{1}{\JJM^{N+1}}}.
\end{multline}
As is well known, the expansion in the last equation is asymptotic and there are corrections of order \(\exp[-2 \pi \, \JJM]\) that are subdominant with respect to the correction that we have discussed in Section~\ref{NonuniversalCorrectionDiscussion}.

\section{N=2 supersymmetrization of the Weyl anomaly action}
\label{sec:Weyl}

The \ac{wz} term for the Weyl anomaly is given in~\cite{Komargodski:2011vj}.  An ${\cal N} = 1$ supersymmetrization
of this term was given in~\cite{Bobev:2013vta}. This term it not the unique supersymmetrization preserving ${\cal N} = 1$ superconformal
symmetry: Alternate supersymmetrizations of the term can be obtained
by adding ${\cal N} = 1$ superconformally-invariant terms to the action, for instance involving a \(\mathcal{N}=1\) superconformal action for the gauge fields.  %

Using the extended ${\cal N} = 2$ superconformal invariance as an input simplifies the matter: As pointed out in~\cite{Dine:1997nq} there is a unique effective term in the Coulomb-branch dynamics of ${\cal N} = 2$ superconformal theories in four dimensions;
since the super-\ac{wz} term contains 
the ordinary \ac{wz} term in Eq.~\eqref{eq:KS-WZ} for Weyl
invariance, this fixes the coefficient unambiguously
as well.

In ${\cal N} = 2$ superspace, the term can be written formally as a full-superspace integral
\begin{equation}
{\cal L}\ll{{\cal N} = 2~{\rm super-}WZ} = ({\rm constant})
\times \int \cc \dd[4]{\th} \dd[4]{\thb} \log(\Phi / \m)
\cc \log(\Phb / \m)\ .
\end{equation}
We wish to write this in components, particularly
the terms involving the scalar $\phi,\phb$ and its
derivatives.

The full form of the ${\cal N} = 2$ super-\ac{wz} term is
easiest to write in terms of ${\cal N} = 1$ superfields,
as expressed in~\cite{deWit:2001brd, GonzalezRey:1997qh}. The ${\cal N} = 2$ vector multiplet $\Phi$ decomposes
into an ${\cal N} = 1$ superfield $\Phi\ls{{\cal N} = 1}$ and
an ${\cal N} = 1$ vector multiplet $V$ whose gauge-invariant super-field strength is ${\cal W}\ll \a$.  In ${\cal N}
= 1 $ superspace, the form of the term is

\begin{equation}
{\cal L}\ll{{\cal N} = 2~{\rm super-}WZ} =  \int \cc (\dd[4]{\th})\ll{{\cal N} = 1}  
\bigg [ C_1 \,  {\cal I}\ll 1\ups{{\cal N} = 1}
+ C_2  \, {\cal I}\ll 2\ups{{\cal N} = 1} \cc \bigg ] 
+ \text{(terms involving $W\ll\a$)}\ ,
\end{equation}
where $C_1$ and $C_2$ are constants and
\begin{align}
{\cal I}\ll 1 &\equiv {1\over{\phi\ll{{\cal N} = 1} \cc \phb
\ll{{\cal N} = 1}}} \cc (\pp\ll\m\phi\ll{{\cal N} = 1})(
\pp\uu\m\phb\ll{{\cal N} = 1})\ , \\
{\cal I}\ll 2 &\equiv {1\over{\phi\ll{{\cal N} = 1} \cc \phb
\ll{{\cal N} = 1}}} \cc (\e\uu{\a\b} D\ll\a D\ll\b \cc \phi\ll{{\cal N} = 1}) \cc (\e\uu{\ald\bed} \bar{D}\ll\ald \bar{D}\ll\bed \cc \phb\ll{{\cal N} = 1})\ .
\end{align}
The $D\ll\a$ and $\bar{D}\ll\ald$ are the spinorial
superspace covariant derivatives.  

In addition to the lowest component $\phi$, the ${\cal N} = 1$ superfield contains fermions and a complex auxiliary
field $F$, which is the lowest component of $D\sqd \Phi$.
In a generic ${\cal N} = 1$ action, we would have to keep
track of terms coupling $F$ to
$\phi$ and $\phb$: After eliminating the auxiliary fields,
these would become terms of order $\a\sqd$ which 
could contribute to the classical action.  However 
such terms cannot appear in a superconformal
effective action for a vector multiplet alone: The real
and imaginary parts of $F$ transform together
with the real auxiliary field $D$ of the ${\cal N} = 1$
vector multiplet, as a triplet under the $SU(2)$ $R$-symmetry in the superconformal algebra, and
any term coupling linearly to $F$ or $D$, would have to be
a triplet as well.  Since $\phi$ and $A\ll\m$ are
invariant under the $SU(2)$ R-symmetry, the only way
to build a scalar term coupling linearly in the auxiliary
fields, would be to include at least two fermions.
Therefore the values of the auxiliary fields after eliminating them by their \ac{eom}, can have no component
involving only the scalars and photon.  We are interested
in this section only in the classical action, and the fermions
only contribute quantum mechanically.  So we can
treat the auxiliary field $F$ as zero for purposes
of writing down the action for the scalars alone.
 
For the same reason, it will be unnecessary to keep track
of any component terms involving the ${\cal N} = 1$ vector superfield: In the classical solutions relevant to the
two-point function of $\phi$, the gauge field strength
vanishes, and so all terms involving the ${\cal N} = 1$
vector superfield vanish classically and contribute only
through their quantum effects.

So we need only consider the superspace integrals
of the two terms ${\cal I}\ll {1,2}$, and in particular
only the component terms containing no fermions
or auxiliary fields.

With attention restricted to such component terms,
the superspace integral of ${\cal I}\ll 2$ is easiest to compute.  In order to obtain a term involving
only scalars, we must take the $\thb\thb$ component
of $D\sqd \Phi\ll{{\cal N} = 1}$ and the 
$\th\th$ component of $\bar{D}\sqd\Phb\ll{{\cal N} = 1}$,
which are proportional to $\pp\sqd \phi$ and $\pp\sqd\phb$,
respectively.  So we have
\begin{equation}
\int \cc \dd[4]{\theta}\ll{{\cal N} = 1} \cc {\cal I}\ll 2
\simeq ({\rm constant}\pr)\ll 2 \cc |\phi|\uu{-2} \cc
(\pp\sqd\phi)(\pp\sqd\phb)\ ,
\end{equation}
where the $\simeq$ denotes the omission of terms
involving fermions and auxiliary fields.

The superspace integral of ${\cal I}\ll 1$ can be
evaluated easily using a trick: Treat
$\pp\ll\m\phi\ll{{\cal N} = 1}$ and its conjugate as
independent superfields $G\ll\m, G\dag\ll\m$, and
write the superspace integrand as a Kahler potential
for the five superfields $\chi\uu A\in \{ \phi\ll{{\cal N} = 1}, G\ll\m\}$ and their conjugates.  So
\bbb
{\cal I}\ll 1 = {\cal K}(\chi,\chi\dag) = |\phi\ll{{\cal N} = 1}|\uu{-2} \cc \eta\uu{\m\n} G\ll\m G\dag\ll\n\ .
\eee
Then the superspace integral is given by the usual
formula written in terms of the Kahler potential,
\begin{equation}
  \begin{aligned}
    \int   (\dd[4]{\theta})\ll{{\cal N} = 1} {\cal I}\ll 1
    &\simeq {\cal K}\ll{,\chi\uu A \chi\uu{B\dagger}}
    \pp\ll\n\chi\uu A \pp\uu\n \chi\uu {B\dagger}\\
    &= \begin{multlined}[t][.75\linewidth]
      {\cal K}\ll{,\phi\phb}
    \pp\ll\n\phi \pp\uu\n \phb
    + {\cal K}\ll{,\phi \cc G\dag\ll\m} \cc \pp\ll\n\phi\pp\uu\n G\dag\ll\m
    + {\cal K}\ll{, G\ll\m \phb} \cc \pp\uu\n G\ll\m \pp\ll\n\phb
    \\ + {\cal K}\ll{,G\ll\m G\dag\ll\r} \cc
    \pp\uu\n G\ll\m \pp\ll\n G\dag\ll\r
    \end{multlined}\\
    &=  \begin{multlined}[t][.75\linewidth]
      |\phi|\uu{-4} \cc G\uu\m G\dag\ll\m 
    \pp\ll\n\phi \pp\uu\n \phb  - \phi\uu{-2} \phb\uu{-1} G\uu\m
    \cc  \pp\ll\n\phi\pp\uu\n G\dag\ll\m \\
    -  \phb\uu{-2} \phi\uu{-1} G\dag\ll\m
    \cc  \pp\ll\n\phb\pp\uu\n G\uu\m
    + |\phi|\uu{-2} \cc \pp\uu\m G\uu\n \pp\ll\m G\dag\ll\n
    \end{multlined}\\
    &
    \begin{multlined}[t][.8\linewidth]= |\phi|\uu{-4} \cc (\pp\phi\pp\phb)\sqd
      - \phi\uu{-2} \phb\uu{-1} \cc (\pp\uu\m\pp\uu\n\phb)(\pp\ll\m\phi)(\pp\ll\n\phi) \\
      -  \phb\uu{-2} \phi\uu{-1} \cc (\pp\uu\m\pp\uu\n\phi)(\pp\ll\m\phb)(\pp\ll\n\phb) + |\phi|\uu{-2}\cc
      (\pp\uu\m\pp\uu\n\phi)(\pp\ll\m\pp\ll\n\phb)\ .
    \end{multlined}
  \end{aligned}
\end{equation}
Rewriting the two terms with the substitution
\bbb
\phi = \m\cc \exp{ - \dilaton - i \b}\ , \llsk\llsk
\phb = \m\cc\exp{- \dilaton + i \b} \ ,
\eee
we get
\begin{equation}
\int \cc  \dd[4]{\theta}\ll{{\cal N} = 1} \cc {\cal I}\ll 1 = ({\rm constant}) \times 
(\pp\uu\m\pp\uu\n\dilaton)(\pp\ll\m\pp\ll\n\dilaton)
+ \text{(fermions~and~auxiliary)}\ ,
\end{equation}
and
\begin{multline}
\int \cc  (\dd[4]{\theta})\ll{{\cal N} = 1} \cc {\cal I}\ll 2
=  (\text{constant}) \times \bigg [ 
(\pp\uu\m\pp\uu\n\dilaton)(\pp\ll\m\pp\ll\n\dilaton) - 2 \cc (\pp\uu\m\pp\uu\n\dilaton)(\pp\ll\m\dilaton)(\pp\ll\n\dilaton)
+ (\pp\dilaton)\uu 4
\cc \bigg ]\\
+ (\text{fermions and auxiliary})\ .
\end{multline}

Modulo total derivatives, and dropping the terms
involving fermions and auxiliary fields, this is
\begin{equation}
\int \cc  (\dd[4]{\theta})\ll{{\cal N} = 1} \cc {\cal I}\ll 1 = (\text{constant}) \times 
(\pp\sqd\dilaton)\sqd
+ \text{(fermions and auxiliary)}\ ,
\end{equation}
and
\begin{multline}
\int \cc  (\dd[4]{\theta})\ll{{\cal N} = 1} \cc {\cal I}\ll 2
=  (\text{constant}) \times \bigg [ 
(\pp\sqd\dilaton)\sqd - 2 \cc (\pp\dilaton)\sqd(\pp\sqd\dilaton)
+ (\pp\dilaton)\uu 4
\cc \bigg ] \\
+ (\text{fermions and auxiliary})\ .
\end{multline}

The coefficients are given in~\cite{deWit:2001brd, GonzalezRey:1997qh},  but it is simple to see what they must be.  The relative coefficient between the two terms must be $-1$: 
When $\b$ is set to a constant, we must recover the
usual \ac{wz} action for the Weyl symmetry
given in~\cite{Komargodski:2011vj},
which contains no term proportional to $(\pp\sqd\dilaton)\sqd$.  The absolute coefficient is also given by matching with~\cite{Komargodski:2011vj}, so that the
purely dilaton-dependent part of the super-\ac{wz} term
is equal to the \ac{ks} dilaton action:
\begin{equation}
 \text{const.} \times \int \cc  (\dd[4]{\theta})\ll{{\cal N} = 1} \cc 
\bqty{{\cal I}\ll 1 - {\cal I}\ll 2}  = \text{const.} \times \bqty{ 
 (\pp\dilaton)\uu 4
 - 2 \cc (\pp\dilaton)\sqd(\pp\sqd\dilaton)}
\end{equation}
where the constant is fixed by the anomaly.
This fixes the two coefficients;
so we conclude the dilaton and axion part of
the super-\ac{wz} term, and we get
\begin{multline}
{\cal L}\lrmns{super-WZ} \uprm{Euclidean}
= + 2 \cc (\Delta a)\ups{\acs{ks}}\cc \bigg [ \cc
(\pp\dilaton)\uu 4
- 2 (\pp\sqd\dilaton)(\pp\dilaton)\sqd 
+ 2 \cc (\pp\sqd\dilaton)(\pp\b)\sqd - 4 (\pp\dilaton\cdot\pp\b)
\cc (\pp\sqd\b) \\
 - 2 (\pp\dilaton)\sqd(\pp\b)\sqd + 4 (\pp\dilaton\cdot\pp\b)\sqd
+ (\pp\b)\uu 4  \cc\bigg ]\ ,
\end{multline}
where we have evaluated the term in flat space, and dropped
terms involving the gauge field, fermions, and auxiliary
fields.  Note that this agrees with the flat space expression given in~\cite{Bobev:2013vta}.

We would like to write this action in other conformal frames,
such as the sphere $S\uu 4$ or the cylinder $S\uu 3\times \IR$.  In order to do this, we need to include the appropriate curvature couplings that give the full action an appropriate
transformation law.
The transformation law for the curved-space super-\ac{wz} action must obey is itself nontrivial, because
the ${\cal N} = 2$ super-\ac{wz} action should not be conformally invariant: Indeed, it must have a nonvanishing additive transformation
under a Weyl transformation, in order to reproduce the 
anomalous quantum transformation of the logs
of the determinants
for the massive fields which have been integrated out.
However it can be decomposed into the \ac{ks} action~\eqref{eq:KS-WZ} itself, plus a remainder
term.   The former has the anomalous transformation law
dictated by the \ac{wz} consistency condition,
and so the remainder must be invariant under Weyl transformations. 

On flat space, the action breaks up as:
\begin{equation}
S[\dilaton,\b,g] =S\ls{\acs{ks}}[\dilaton,g] + S\ls{\rm remainder}[\dilaton,g]\ ,
\end{equation}
where
\begin{equation}
  \label{RemainderTermOnFlatSpace}
  \begin{aligned}
    S\ls{\text{KS}}[\dilaton,g] &\equiv + 2  (\Delta a)\ups{\acs{ks}} \bigg [ \cc
    (\pp\dilaton)\uu 4
    - 2 (\pp\sqd\dilaton)(\pp\dilaton)\sqd \cc\bigg ] \\
    S\ls{\text{remainder}}[\dilaton,g] &\equiv
    \begin{multlined}[t][.65\linewidth]
      + 2 (\Delta a)\ups{\acs{ks}} \bigg [ 
      + 2 (\pp\sqd\dilaton)(\pp\b)\sqd - 4 (\pp\dilaton\cdot\pp\b)
      \cc (\pp\sqd\b) \\
      - 2 (\pp\dilaton)\sqd(\pp\b)\sqd + 4 (\pp\dilaton\cdot\pp\b)\sqd
      + (\pp\b)\uu 4  \bigg ]\ .
    \end{multlined}
  \end{aligned}
\end{equation}

The remainder term is covariant. For any conformally flat space, the covariantization is unique,
and given by turning the flat metric into the dressed hatted metric:
\begin{equation}
  \hat{g}\ll{\m\n} \equiv \exp[-2\dilaton]  g\ll{\m\n} .
\end{equation}
On flat space, we have already worked out the remainder, which 
fixes its covariantization.  The covariant action for the axiodilaton alone in
a general conformally flat metric, has been given in~\cite{Bobev:2013vta}. Therefore
the axiodilaton part of the remainder term is given by the difference between this action and the \ac{ks} dilation action.
In the ${\cal N} = 2$ theory, there are also terms involving gauge field strengths and fermions, which we omit, because we will not need them: the gauge fields and fermions make no contribution up to and including order $1/\JJM$, and we have derived the higher power-law corrections on general grounds without the need to use the other terms in the action explicitly.

The covariantization of the remainder term  is given in~\cite{Bobev:2013vta} as
\begin{equation}
  \label{BEORemainderTermCovariant}
{\cal L}\lrmns{{{\rm Remainder}\atop{BEO}}}\uprm{Lorentzian}
 \equiv - 4 (\Delta a\ups{\acs{ks}})\cc
{{\sqrt{- \hat{g}}}\over{\sqrt{-g}}}
\cc \bigg [ \cc
 \Big ( \hat{R}^{\mu\nu} - \frac{1}{6}\hat{R}\,\hat{g}^{\mu\nu} \Big )  \cc\gg\ll\m\b\gg\ll\n\b  + \frac{1}{2}\, \Big(\hat{g}^{\mu\nu}\,\gg\ll\m\b\gg\ll\n\b\Big)^2
\cc\bigg ] . %
\end{equation}
In four dimensions
the Riemann curvature is given by a sum of
the Ricci tensor and Weyl tensor, and so any term
vanishing in conformally flat space, must be proportional
to at least one power of the Weyl tensor and its derivatives.
Since we are only ever considering conformally flat geometries in this paper, we will henceforth drop all terms
involving the Weyl tensor. Therefore on a general space we have
\bbb
{\cal L}\lrmns{Remainder}\uprm{Lorentzian}
= {\cal L}\lrmns{{{\rm Remainder}\atop{BEO}}}\uprm{Lorentzian}
+ \bigg (\text{covariant terms involving the Weyl tensor} \bigg )\ .
\een{BEORemainderTermCovariantOnConformallyNonflatSpaces}
and of course the pure Weyl-anomaly term is given by the KS-action given in Eq.~\eqref{eq:KS-WZ}.

So, all in all we have
\begin{equation}
\begin{aligned}
S\lrmns{super-WZ,~curved} \uprm{lorentzian} 
 ={}&
 \begin{multlined}[t][.7\textwidth]
-\Da\ups{\acs{ks}} \int \dd[4]{x}  \sqrt{-{g}}
 \bigg[ \dilaton\,E_4\ups{\acs{ks}} +  \bigg(4\,\Big(R^{\mu\nu}-\frac{1}{2}R\,g^{\mu\nu}\Big)\nabla_\mu\dilaton\,\nabla_\nu\dilaton  \\ - 2\,(\nabla\dilaton)^2\Big(2\,\Box\dilaton - (\nabla\dilaton)^2\Big) \bigg) \bigg]
\end{multlined}
\\
 &
 + 4\sqrt{-\hat{g}}\,\bigg[ \Big ( \hat{R}^{\m\n} - \frac{1}{6}\hat{R}\,\hat{g}^{\m\n} \Big )  \cc\gg\ll\m\b\gg\ll\n\b  + \frac{1}{2}\, \Big(\hat{g}^{\mu\nu}\,\gg\ll\m\b\gg\ll\n\b\Big)^2 \bigg] 
\\
&+ \bigg (\text{covariant terms invoving the Weyl tensor} \bigg )
\label{RestrictedBEORECAPRECAPRECAP}
\end{aligned}
\end{equation}
It is convenient to separate the dependence on the powers of dilaton and axion:
\begin{equation}
  \label{RestrictedBEOActionInPieces}
{\cal L}\lrmns{super-WZ}\uprm{Lorentzian} =
 {\cal L}\ll{\dilaton\uu 1}
+ {\cal L}\ll{\dilaton\uu 2} + {\cal L}\ll{\dilaton\uu 3}
+ {\cal L}\ll{\dilaton\uu 4}
+ {\cal L}_{\beta\sqd} + {\cal L}\ll{\dilaton\uu 1\b\sqd}
+ {\cal L}\ll{\dilaton\sqd\b\sqd} + {\cal L}_{\beta\uu 4}\ ,
\end{equation}
where
\begin{align}
  {\cal L}\ll{\dilaton\uu 1} &= -(\Delta a)\cc \dilaton \cc E_4\ , \label{SuperWZB} \\
  {\cal L}\ll{\dilaton\uu 2} &= - 4 \cc (\Delta a) \cc
 \bigg [ \cc R^{\mu\nu}-\hh\cc {\tt Ric}\ll 4\cc g\uu{\m\n} \cc\bigg ]
                               \gg\ll\mu\dilaton\,\gg\ll\nu\dilaton \ , \label{SuperWZC} \\
 {\cal L}\ll{\dilaton\uu 3} &= + 4 (\Delta a)\cc(\gg\dilaton)\sqd
                              \cc (\gg\sqd\dilaton) \label{SuperWZD}\\
 {\cal L}\ll{\dilaton\uu 4} &= - 2 (\Delta a) \cc (\gg\dilaton)\uu 4  \label{SuperWZE} \\
  {\cal L}_{\b\sqd} &= - 4 (\Delta a) \cc
  \bigg [ \cc R\uu{\m\n} - {1\over 6} \cc 
{\tt Ric}\ll 4 \cc g\uu{\m\n} \cc \bigg ]
\cc (\gg\ll\m\b)(\gg\ll\n\b) \label{SuperWZF} \\
  {\cal L} \ll{\dilaton\uu 1\b\sqd} &=  - 8 \cc (\Delta a)\cc
  (\gg\uu\m\gg\uu\n\dilaton)\cc \gg\ll\m\b\cc\gg\ll\n\b\ , \label{SuperWZG} \\
  {\cal L}\ll{\dilaton\sqd\b\sqd}  &= - 4 \cc (\Delta a)\cc
  \bigg [ \cc 2\cc (\gg\dilaton\cdot\gg\b)\sqd - (\gg\dilaton)\sqd
  (\gg\b)\sqd \cc \bigg ]\ , \label{SuperWZH}\\
  {\cal L}_{\b\uu 4}  &= - 2\cc (\Delta a) \cc (\gg\b)\uu 4\ .  
\end{align}

\section{Nonexistence of higher-derivative $F$-terms on conformally flat space}\label{sec:noF}

In general ${\cal N} = 2$ supersymmetric
gauge theories, the
effective action on the Coulomb branch has
higher derivative $F$-terms, of which those with
few derivatives have been partially classified by~\cite{Argyres:2003tg, Argyres:2004kg, Argyres:2004yp, Argyres:2008rna}.  In the case of superconformal gauge theories with rank one, the
remarkable simplifications of the dynamics of the Coulomb
branch have to do with the absence of such terms. \ok

More precisely, the only half-superspace integrands consistent with superconformal symmetry on a general curved background, are the tree-level kinetic term proportional
to $\Phi\sqd$, and terms involving the background
Weyl multiplet, which contains the $U(1)\lrm R$-photon
background and the self-dual part of the Weyl
tensor. \ok

\heading{Vanishing of higher-derivative terms on
(superconformally) flat space}

Consider the effective action of a single Abelian vector
multiplet in a superconformally invariant theory.
The symmetries of a ${\cal N} = 2$ superconformal theory include dilatation invariance and $U(1)$ R-symmetry,
which act on a vector multiplet $\phi$ by rescalings and complex phase rotations respectively, both in the underlying
microscopic \ac{cft} and in the \ac{eft} of the Coulomb branch.  The Weyl symmetry acts as
\bbb
\phi\to \exp{\r} \phi\ ,
\eee
and the $U(1)\ll R$ acts as 
\bbb
\phi\to \exp{i\g}\phi\ .
\eee
One can combine the Weyl and $U(1)\ll R$ parameters into a single complex parameter $\s\equiv - \r - i \g$, which
acts as
\bbb
\phi\to \exp{- \s}\phi\ , \llsk\llsk \phb\to \exp{- \bar{\s}}\phb\ .
\eee
In a superconformal theory it is natural to promote $\s$ to a local function of superspace rather than just 
the $x$ coordinates.  In order to preserve the chirality constraint $\bar{Q}\ll \ald \uu i \cdot \phi = 0$ 
we can require $\s$ to obey the same chirality constraint $\bar{Q}\ll\ald\uu i \cdot \s = 0$.   Invariance of a superconformal theory under
super-Weyl transformations parametrized by a chiral superfield has been studied previously (see~\cite{Wess:1992cp} and references within for ${\cal N} = 1$ theories and~\cite{Freedman:2012zz} and references within for ${\cal N} = 2$ theories).  

The chiral superfield Weyl parameter $\s$ consists of a complex scalar, 
fermions, and a vector parameter $\hat{\l}\ll\m$, and other components which act only
on the auxiliary fields.  The scalar and fermionic members of the parameter superfield 
implement Weyl, $U(1)\ll R$, and local supersymmetry transformations,
respectively; the $\l\ll\m$ transformations shift the gauge field as
\bbb
\d A\ll\m = \phi\cc \hat{\l}\ll\m\ .
\eee

Local transformations
are not themselves symmetries of the dynamical fields
alone, but can be understood as ``spurionic'' symmetries,
that preserve the action for dynamical variables
together with a set of background fields, when the
background fields are transformed appropriately.  In
the case of local dilatation and local $U(1)\ll R$ transformations, the corresponding background fields are the metric and the $U(1)\ll R$ gauge field, which transform
by Weyl transformations and local $U(1)\ll R$ gauge
transformations, respectively.  The $\l\ll\m$-transformations can be thought of as shifting a background
antisymmetric tensor field $B\ll{\m\n}$ by a gauge transformation
\bbb
B\ll{\m\n} \to (d\hat{\l})\ll{\m\n} = \pp\ll\m \hat{\l}\ll\n - (\m\leftrightarrow \n)\ .
\een{DefinitionOfLambdaHatTransformations}
There is no unique or canonical formulation of supergravity off-shell, even 
${\cal N} = 1$ SUGRA, and the variety of off-shell formulations of ${\cal N} = 2$
SUGRA is even larger. The action of minimal super-Weyl invariance on the vector multiplet is necessarily the same in any off-shell formalism, since it can be expressed
directly in terms of physical currents and their operator products with physical
vector multiplet degrees of freedom.  So, the transformation of the superfield
$\phi$ under a super-Weyl transformation paramerized by the chiral
superfield $\s$, is independent of the ${\cal N} = 2$ SUGRA formalism and
set of additional compensators and auxiliary fields needed to give a complete
off-shell formulation.

\heading{Constraints on the EFT from super-Weyl invariance}

We observe that the Wilsonian effective action is super-Weyl
invariant if the underlying \ac{cft} is super-Weyl invariant: The
\ac{eft} on moduli space inherits this property
directly from the \ac{cft}.  Super-Weyl invariance of
the \ac{cft} is automatic if the theory is supersymmetric and conformal.  
Weyl invariance can be seen explicitly at the Lagrangian
level for superconformal gauge theories with hypermultiplets
and vanishing $\b$-function; super-Weyl invariance also acts on the vector-multiplet
action in a transparent way.

The \ac{eft} inherits the super-Weyl invariance of the underlying \ac{cft}, so we can now consider what possible terms one might write in a supersymmetric ${\cal N} = 2$ \ac{eft} consistent with
super-Weyl invariance. 

For a single $U(1)\ll{\rm gauge}$
vector multiplet, the Weyl and local $U(1)\ll R$
transformations give enough freedom to set the
complex scalar $\phi $ equal to some fixed nonzero
value, say $\m$, everywhere
that it is nonvanishing:  By choosing $\s = + {\tt log}(\phi / \m)$ we can
fix the "gauge" $\phi = \m$.  

The fermions $\psi\ll\a\uu i$ in the Abelian vector
multiplet are superpartners of $\phi$, and supersymmetry
implies that if $\phi - \m$ can be made to vanish with
a local transformation, then $\psi\ll\a\uu i$
can be made to vanish as well.  And, indeed,
superconformal transformations can 
be promoted to local transformations as well: by
integrating the supercurrents against general functions $\eta\ll\a\uu i(x)$ of space, we have enough freedom 
to set to zero the fermions $\psi\ll \a$ at the cost of turning on a nonzero but flat background for the (spurionic) gravitini.

The freedom to make $\hat{\l}\ll\m$-transformations \rr{DefinitionOfLambdaHatTransformations} allows us to set the gauge field to zero as well, and so the
entire vector multiplet in the \ac{eft} can be gauged away.  It follows that there can be no super-Weyl-invariant terms
containing only the metric and no background curvatures.

Certainly there may be many terms involving background curvatures, but we are considering only the maximally
supersymmetric background $\IR\uu 4$ and backgrounds equivalent to it such as the sphere $S\uu 4$
and the cylinder $S\uu 3 \times \IR$.  We therefore need only consider couplings involving the Ricci curvature
and its derivatives, since the Weyl curvature and R-symmetry gauge flux vanish in the backgrounds we consider. 

For D-terms there are many such terms one can construct: The dressed metric $\hat{g}\ll{\m\n} \equiv |\phi|\sqd\cc g\ll{\m\n}$
is Weyl-invariant and its superspace extension is super-Weyl-invariant by construction.  So any term 
constructed from these has Weyl weight zero and is suitable for addition to the action as a $D$-term (\it i.e., \rm full ${\cal N} = 2$
superspace $d\uu 4 \th \cc d\uu 4 \bar{\th}$ integrand) consistent with super-Weyl invariance.  

The correlators we consider in the present paper are computable by localization and insensitive to $D$-terms; we need
therefore consider only super-Weyl-invariant $F$-term contributions to the effective action.\footnote{In addition to
the familiar $D$-terms and $F$-terms, ${\cal N} = 2$ supersymmetric effective theories with hypermultiplets may have
terms that can be represented as ${3\over 4}$-superspace integrals but not true $F$-terms.  Some such terms
have been worked out in \cite{Argyres:2003tg, Argyres:2004kg, Argyres:2004yp, Argyres:2008rna}.  However
we can restrict our attention to theories with only a pure Coulomb branch and no massless neutral hypers,
rather than an enhanced Coulomb branch.  For theories with no hypermultiplets we may consider only
the usual $F$-terms and $D$-terms.}

Such terms must be of the form 
\bbb
\Delta {\cal L} = \int \cc d\uu 4 \th \cc \phi\sqd \cc {\cal I}\ll 0\ ,
\eee
where ${\cal I}\ll 0$ is a super-Weyl-invariant term that is also a chiral primary field, \it i.e., \rm annihilated by
all the $\bar{D}\ll{\dot{\a}}\uu i$ superderivatives.  As we have pointed out above, such terms
must be constructed from Ricci curvatures of the hatted metric.  However the hatted metric
is not a chiral field, nor is the Ricci curvature or any of its derivatives.  One can see this easily from
its definition: Acting with $\bar{D}\ll {\dot{\a}}\uu i$ on $\hat{g}\ll{\m\n}$ gives 
\bbb
\bar{D}\ll {\dot{\a}}\uu i (\hat{g}\ll{\m\n}) = \o\cc \hat{g}\ll{\m\n}\ , \llsk\llsk \o\equiv \bar{\psi}\uu i\ll{\dot{\a}} / \bar{\phi}\ .
\eee
In other words, even though $\hat{g}\ll{\m\n}$ is Weyl-invariant, acting with the antichiral supersapce derivative on
$\hat{g}\ll{\m\n}$ is equivalent to infinitesimally Weyl-transfroming the \rwa{hatted} metric by a Weyl parameter proportional
to $\bar{\psi}\uu i\ll{\dot{\a}} / \bar{\phi}$, which does not vanish identically, obviously.  The only quantities 
that can be constructed from $\hat{g}\ll{\m\n}$ are exactly the same as the Weyl-invariant quantities that
could be constructed from the \rwa{unhatted} background metric $g\ll{\m\n}$, with $g\ll{\m\n}$ replaced by
$\hat{g}\ll{\m\n}$.  But since these quantities are Weyl-invariant, the replacement has no effect and
 they are exactly the same as the ones constructed from $g\ll{\m\n}$, that is, the Weyl curvature and various powers
 of it and its Weyl-covariantized derivatives.  
 
 As mentioned earlier, many such terms can be constructed, and would contribute to $F$-terms on a non-conformally-flat
 background; however for a background with vanishing Weyl curvature, all such terms vanish.   We therefore conclude that all higher-derivative $F$-terms vanish identically on a conformally flat background,
 in the effective theory of a single Abelian vector multiplet.    Adding a flat
 background connection for the $U(1)\ll R$-symmetry, allows more terms to be written but does not
 change the conclusion: There are no superconformally invariant higher-derivative $F$-terms that can be written for
 a single vector multiplet, even with an R-symmetry connection included, so long as the flux vanishes and
 the metric is conformally flat.

\heading{More comments on the currents}

The action of the super-Weyl transformations on the physical fields is generated by currents with
protected integer operator dimensions living in a single current multiplet; for the
case of dilatations the generating operator
is the trace of the stress tensor
with dimension $4$ and for $U(1)$ transformations
the generating current is the $U(1)\ll R$-current with dimension $3$.
The field generating the
$\l$-transformations is an antisymmetric tensor of weight $3$ (see for instance~\cite{Dolan:2002zh,Cordova:2016emh}) which is not a conserved current
but whose curl is the weight-$4$ topological current that integrates to
the central charge ${\cal Z}$.

Since these currents are local, they can be integrated against arbitrary functions to
generate well-defined
local transformations of the fields. This is the physical basis of the super-Weyl transformation:
An infinitesimal change of the \textsc{sugra} background is equivalent to an infinitesimal transformation
of the physical degrees of freedom, which in turn is equivalent to inserting integrated currents into
the path integral.  For instance an infinitesimal change in the background metric is equivalent to
\bbb
S\to S + \int \cc \sqrt{|g|} \cc (\d g\ll{\m\n}) \cc T\uu{\m\n}\cc ;
\eee 
an infinitesimal change in the R-symmetry gauge connection is equivalent to
\bbb
S\to S + \int \cc \sqrt{|g|} \cc (\d {\cal A}\uu{U(1)\ll R}\ll\m) \cc J\uu\m\ll{U(1)\ll R}\cc ;
\eee
and an infinitesimal change in the antisymmetric tensor background is equivalent to
\bbb
S \to S + \int \cc \sqrt{|g|} \cc (\d {\cal B}\ll{\m\n}) \cc {\cal Z}\uu{\m\n} + ({\rm c.c.})\ .
\eee

Diffeomorphism and Weyl invariance are equivalent to the statements that $T\ll{\m\n}$ is divergenceless and
traceless, respectively; $U(1)\ll {\rm R}$ invariance is equivalent to the statement that
$J\uu\m\ll{U(1)\ll R}$ is divergenceless. There is no simple analogous statement about
the ${\cal Z}$-current, which 
sits in the (short) stress tensor multiplet as an anti-self-dual tensor with conformal dimension 3.  At the free-field
level, the ${\cal Z}$-current is proportional to $\phi F^{(-)}_{\mu\nu}$, where $F^{(-)}$ is the anti-self-dual part
of the gauge field strength.  Its complex conjugate generates $\hat{\lambda}$-transformations on the vector multiplet when
integrated against $\hat{\lambda}$.

Unlike the R-current and stress tensor, its divergence
does not vanish.  Correspondingly, the coupling of ${\cal Z}\uu{\m\n}$ to
the background ${\cal B}\ll{\m\n}$-field is somewhat subtle; the $\hat{\l}$ one-form transformations
act on other background fields in addition to the ${\cal B}\ll{\m\n}$-field.  The coupling of the ${\cal Z}$-current to the \textsc{sugra} 
background is formalism-dependent, as the ${\cal B}$-field is not part of the minimal ${\cal N} = 2$ \textsc{sugra} multiplet
and the details have not been worked out in the \textsc{sugra} literature.  One can infer the physically relevant
properties of the coupling by considering the current directly, whose properties are formalism-independent.

The ${\cal Z}\ll{\m\n}$ current, which generates the $\l\ll\m$-transformations which
shift the gauge field in the vector multiplet, is less well-studied than the other members of its multiplet, the
stress tensor and $R$-current. Since the super-Weyl transformation generated in part by ${\cal Z}\ll{\m\n}$ 
plays a role in forbidding higher-derivative $F$-terms for one-dimensional Coulomb-branch \acp{eft},
we comment briefly on properties of this current for the sake of context~\cite{Dolan:2002zh,Cordova:2016emh,Dumitrescu:2011iu}.

The ${\cal Z}\ll{\m\n}$ current is similar to the line-charge symmetry that shifts the photon in a weakly-coupled
Maxwell gauge theory~\cite{Drukker:2009tz, Aharony:2013hda},
but it is a different sort of current.  The line-charge current in four-dimensional
Abelian gauge theory has dimension
approximately two at weak coupling rather than three, and cannot be exactly conserved
unless the dimension is exactly two and Maxwell field is exactly free, in analogy with the
parallel Sugawara theorem for spin-one currents in two dimensions~\cite{Hofman:2018lfz}. 

By contrast the ${\cal Z}$-current has dimension three and is not divergenceless.  Indeed, the divergence of the
${\cal Z}$-current contributes to the central charge in the ${\cal N} = 2$ supersymmetry algebra.  That is,
\bbb
\{ Q\ll \a\uu i, Q\ll\b\uu j\} = 2 \cc \e\uu{ij} \cc \e\ll{\a\b} \cc Z\ ,
\xxn
Z\ni \int \cc d\uu 3 \cc {\cal N}\uu\m \cc \gg\uu\n \cc {\cal Z}\ll{\m\n}\ ,
\xxn
{\cal Z}\ll{\m\n} \propto \e\ll{ij} \cc ([\s\ll\m,\s\ll\n])\uu{\a\b}\cc \e\ll{ij} \cc Q\ll\a\uu i\cdot 
Q\ll\b \uu j \cdot {\bf J}\lrm{scalar}\ ,
\eee
where $ {\bf J}\lrm{scalar}$ is the lowest component of the stress tensor multiplet, a scalar primary of dimension
$\D = 2$ transforming trivially under the $R$ symmetry and equal to $\phi\phb$ in the Coulomb-branch \ac{eft}~\cite{Dolan:2002zh,Cordova:2016emh}.

In a superconformal ${\cal N} = 2$ theory without marginal operators, this current is the \rwa{only} contribution to the 
central charge; there are no other currents of dimension three and the correct
quantum numbers to appear in the \ac{susy} algebra.    The normalization of
the central charge $Z$ is therefore determined by the three-point function of the current multiplet in such
theories, which means its value is fixed entirely by the anomaly coefficients $a$ and $c$.
This has interesting implications for the \ac{bps} dyon spectrum on the Coulomb branch of 
non-Lagrangian ${\cal N} = 2$ \ac{scft}.

In a superconformal ${\cal N} = 2$ theory with marginal operators, there is a second independent
component of the central charge, also a total derivative, of a current which we
shall call ${\cal Y}\ll{\m\n}$:
\bbb
Z\ni \int \cc d\uu 3 \cc {\cal N}\uu\m \cc \gg\uu\n \cc {\cal Y}\ll{\m\n}\ ,
\xxn
{\cal Y}\ll{\m\n} = \sum\ll A y\ll A \cc \e\ll{ij} \cc ([\s\ll\m,\s\ll\n])\uu{\a\b}\cc \e\ll{ij} \cc Q\ll\a\uu i\cdot 
Q\ll\b \uu j \cdot {\cal O}\uu A\ ,
\eee
where $A$ runs over all marginal operators ${\cal O}\uu A$ and $y\ll A$ can vary over the conformal manifold.
All the dependence of the central charge on the marginal directions is through the ${\cal Y}$-current contribution.

\section{N=2 superconformal gauge dynamics with ghost hypermultiplets}
\label{GhostHyperAppendix}

\heading{Weyl anomalies and $\b$-functions for
${\cal N} = 2$ gauge theory with $G=SU(2)$}

Consider for instance the case of an ${\cal N} = 2$ gauge theory with $G = SU(2)$ and
ordinary hypermultiplets.  A hypermultiplet in a representation ${\bf R}$ of $SU(2)$ contributes to the
$\b$-function as %
\begin{equation}
\bfuc\lrm{{\rm ordinary~hypermultiplet~in~{\bf R}}} = + {{g\lrm{YM}\uu 3}\over{16\pi\sqd}} {\Tr}\ll{\bf R}(t\uu A \cc t\uu A)\ ,
\end{equation}
where $A = 1,2,3$ and the representation matrices $t\uu A$ are taken to be Hermitean and normalized
so that the level spacing of $t\uu {A=3}$ is differences of $1$.  So if ${\bf R}$ is the $k$-dimensional
representation then $t\uu 3$ has eigenvalues $\{- {{k-1}\over 2}, -{{k-3}\over 2},\cdots, + {{k-3}\over 2}, +{{k-1}\over 2} \}$,
so
\bbb
{\tt Tr}\ll{\bf R}(t\uu A \cc t\uu A)= 3\times\cc
{\tt tr}\ll{\bf R}((t\uu {A=3})\sqd) = {{k(k\sqd - 1)}\over 4}\ .
\eee
In terms of the largest eigenvalue $\ell \equiv k - \hh$
of $t\uu{A = 3}$, this is just the dimension $k = 2\ell + 1$
of the representation, times the quadratic Casimir
$\ell(\ell + 1) = {1\over 4}(k\sqd - 1)$.

So the $\bfuc$-function of an ordinary hypermultiplet is
\begin{equation}
\bfuc\lrm{{\rm ordinary~hypermultiplet~in~{\bf R}\ll k}} = +{{g\lrm{YM}\uu 3}\over{16\pi\sqd}} {{k(k\sqd - 1)}\over 4}\ .
\end{equation}
The $\b$-function in ${\cal N} = 2$ theories comes entirely from one loop.  

\heading{Ghost hypermultiplets}

If we were to couple hypermultiplets
in representation ${\bf R}$ with spin-statistics opposite to the usual ones, then the $\b$ function 
would be of the same magnitude and opposite sign as for ordinary matter.  Such opposite-statistics ``ghost matter'' in supersymmetric gauge theory as been considered elsewhere in a similar spirit~\cite{Evans:2006eq, Okuda:2006fb, Dijkgraaf:2016lym, Buican:2017rya}.
So
\begin{equation}
\bfuc\lrm{{\rm ghost~hypermultiplet~in~{\bf R}\ll k}} = -{{g\lrm{YM}\uu 3}\over{16\pi\sqd}} {{k(k\sqd - 1)}\over 4}\ .
\end{equation}
The $SU(2)$ vector multiplet contribution to the $\b$-function is
\begin{equation}
\bfuc\ll{\text{$SU(2)$ vector multiplet}} = -6 \ ,
\end{equation}
so the condition for the cancellation of the $\b$-function is
\begin{equation}
\label{BetaFunctionCancellationWithGhostHypersFirstVersion}
\sum\ll k {{k(k\sqd - 1)}\over 4} \times (n\ll k\uprm{hyper} - n\ll k\uprm{ghost~hyper}) = +6\ ,
\end{equation}
where $n\ll k\uprm{hyper} $ and $n\ll k\uprm{ghost~hyper} $ are the numbers of ordinary hypermultiplets and
ghost hypermultiplets, respectively, in the $k$-dimensional representation of $SU(2)$.

The $\bfuc$-function depends only on the differences
$\tilde{n}\ll k\uprm{hyper} \equiv n\ll k\uprm{hyper}
- n\ll k\uprm{ghost~hyper}$, and so we can write
formula \rr{BetaFunctionCancellationWithGhostHypersFirstVersion}
as
\begin{equation}
\label{BetaFunctionCancellationWithGhostHypersSecondVersion}
\sum\ll k {{k(k\sqd - 1)}\over 4} \times \tilde{n}\ll k\uprm{hyper}  = +6\ .
\end{equation}
This is just the generalization of the usual $\bfuc$-function
formula to negative numbers of hypermultiplets;
the path integral with ghost hypers gives this generalized
formula a concrete physical
interpretation, at least in terms of a superconformal
statistical system in four euclidean dimensions,
if not a quantum theory in $3+1$ spacetime dimensions.

Our only intended use for this system is to serve as
a nonunitary regulator for the effective vector multiplet
action with various values of the $\a$-coefficient of
the super-\ac{wz} term.

Since we only wish to define the effective theory up
to the scale $\L \muchlessthan |\phi|$, the
nonunitary nature of the ghost hypers is irrelevant
since all hypermultiplet degrees of freedom
are massive at the scale set by $|\phi|$: So long
as the ghost hypers satisfy this condition, then they
just serve as a nice regulator for the \ac{wz} action that
has the useful property of preserving
the spontaneously broken ${\cal N} = 2$ superconformal
symmetry.  Similar regulators for ${\cal N} = 4$
theories involving
ghost matter have been considered elsewhere~\cite{Evans:2006eq, Okuda:2006fb, Dijkgraaf:2016lym, Buican:2017rya}.
The present ghost regulators are similar to
those of~\cite{Buican:2017rya}, which are
simpler than those of~\cite{Evans:2006eq, Okuda:2006fb, Dijkgraaf:2016lym}, in that the latter theories considered there involved
nonunitary degrees of freedom in the gauge
sector as well as in the matter sector, necessarily so
in order to preserve the full ${\cal N} = 4$ supersymmetry.
Our regulating theories, like those of~\cite{Buican:2017rya},
have nonunitarity only in the matter sector.

We therefore need to engineer a vacuum manifold consisting solely of an Abelian vector multiplet, with no additional massless
degrees of freedom from the hypers when the vector multiplet scalar has a nonzero vev.  That is, we wish to exclude
the case of an ``enhanced'' Coulomb branch or its ghost generalization.  To achieve this, it is necessary and sufficient
to choose all the representations to be even-dimensional.  Then the mass matrix for the hypers, $t\uu A\ll{\bf R} \phi\uu A$,
has no vanishing eigenvalues for nonzero $\phi\uu A$, and the vacuum manifold is a pure Coulomb branch.  So we will
restrict our representation content to $k$ even.  With
this criterion, all ghost degrees of freedom have
masses of order $|\phi|$ and are
above the cutoff $\L$.

Now let us write an expression for the $a$-anomaly
of the underlying \ac{cft}.  So long as the $\b$-function
vanishes, the gauge coupling $\t$ is
marginal and the anomaly is $\t$-independent,
and we can compute the Weyl anomaly accurately
in free field theory.  Just as for the gauge anomaly,
the ghost hypermultiplets contribute to the Weyl anomaly
oppositely to the ordinary hypermultiplets in the
same representation.  Thus we have the total $a$-
coefficient
\bbb
a\ups{\acs{aefj}}\lrm{CFT} = a\upsns{\ac{aefj}}\lrm{SU(2)~vector~multiplet} + 
a\upsns{\ac{aefj}}\lrm{matter} =
{5\over 8} + {1\over{24}}\times \sum\ll k\cc k\cc \tilde{n}\ll k\uprm{hyper}\ .
\eee
If we have chosen all the $k$ to be even, then
there are no massless degrees of freedom on
the Coulomb branch other than the vector multiplet,
and so the Coulomb branch \ac{eft} has
\bbb
a\ups{\acs{aefj}}\lrm{EFT} = a\upsns{\ac{aefj}}\lrm{U(1)~vector~multiplet}  = + {5\over{24}}\ .
\eee
Then the anomaly mismatch in \ac{aefj} units is
\begin{equation}
(\Delta a)\ups{\acs{aefj}} \equiv a\ups{\acs{aefj}}\lrm{EFT} - a\ups{\acs{aefj}}\lrm{EFT} = {5\over {12}} + {1\over{24}}\times \sum\ll k\cc k\cc \tilde{n}\ll k\uprm{hyper}
\end{equation}
and the $\a$-coefficient then comes out to
\begin{equation}
  \label{AlphaFormulaFromHyperRepMultiplicities}
\a \equiv 2 \times (\Delta a)\ups{\acs{aefj}} = {5\over 6} + {1\over{12}} \sum\ll k\cc k\cc \tilde{n}\ll k\uprm{hyper}
\end{equation}

We include only even $k$ in the sum, but other
than that there is no restriction on the $\tilde{n}\ll k$
other than the requirement \rr{BetaFunctionCancellationWithGhostHypersSecondVersion} that the
$\bfuc$-function vanishes.

\heading{Conformal combinations of matter and ghost matter}
Since the $\tilde{n}\ll k$ can be positive or negative,
there are many ways to satisfy equation \rr{BetaFunctionCancellationWithGhostHypersSecondVersion} while giving different values for $\a$
as determined by equation \rr{AlphaFormulaFromHyperRepMultiplicities}.
For instance, for $\tilde{n}\ll 4\uprm{hyper}$ any integer, we can take
\begin{equation}
  \begin{cases}
    \tilde{n}\ll 2\uprm{hyper} = 4 - 10 \cc \tilde{n}\ll 4\uprm{hyper}\ ,\\
\tilde{n}\ll k = 0 & \forall k \neq 2,4\ .
  \end{cases}
\end{equation}
Then the $\bfuc$-function cancellation equation
\rr{BetaFunctionCancellationWithGhostHypersSecondVersion} is satisfied,
and the value of $\a$ is
\bbb
\a = {3\over 2} - {4\over 3} \cc \tilde{n}\ll 4\uprm{hyper}\ .
\eee

\heading{Super-Weyl invariance of the ghost-hyper theories}

The vanishing of the $\bfuc$-function means
that these theories are scale-invariant.  However
we can see that they are not only
scale-invariant, they are Weyl-invariant on
curved space and therefore super-Weyl-invariant
on curved superspace~\cite{Festuccia:2011ws,Dumitrescu:2012ha}.

The action for ghost hypers is Weyl-invariant at the
Lagrangian level exactly as it is for ordinary hypers:
For both types of multiplet, the action
is exactly quadratic in hypermultiplet degrees of freedom,
and the ghost hypers are taken to have exactly the
same super-Weyl transformation laws as the ordinary
hypers.  So even though nonunitary scale-invariant
theories are not Weyl-invariant in general,
the ghost-hyper \acp{scft} are special cases which are
in fact super-Weyl invariant.  This is important to
emphasize, because we will use super-Weyl invariance,
not just scale invariance, as a symmetry to eliminate
higher-derivative $F$-terms in the Coulomb branch \ac{eft}
of the ghost-hyper theories.

For vector multiplet actions, ${\cal N} = 2$ super-Weyl invariance follows automatically from Weyl-invariance
and ${\cal N} = 2$ \ac{susy} because the ${\cal N} = 2$
supergravity background has a superspace 
formalism which couples naturally to half-superspace
$F$-terms for vector multiplets as well as full-superspace
terms.  For hypermultiplet $F$-terms, maintaining
manifest supersymmetry off-shell is more subtle, 
requiring more sophisticated superspace formalisms
such as harmonic superspace %
or projective superspace, %
to which we know
of no currently developed formalism for coupling to a
curved superbackground.

However it is possible to see directly 
that the action for ghost
hypermultiplets must be super-Weyl-invariant, if
the action for ordinary hypermultiplets is super-Weyl-invariant.  There are two more or less equivalent ways to see
this, one ``on-shell'' and one ``off-shell''.  Both
forms of the proof use the fact that the action
for hypermultiplets, both ghost type and ordinary
type, is exactly quadratic in the hypermultiplet
fields.

The on-shell, operator argument is as follows.
Since the action is exactly quadratic in the
hypermultiplet degrees of freedom, so must
be the stress tensor, supersymmetry generators,
and other currents.  In particular, the virial current
would have to be quadratic in hypermultiplet degrees of
freedom, and there is no candidate virial current
that is quadratic in hypermultiplet degrees of freedom.

This operator proof translates into an off-shell argument in component fields, as follows:

Given an ${\cal N} = 2$ supergravity background
and a fixed (not necessarily supersymmetric or
on-shell) background for the vector multiplet,
we can perform a super-Weyl transformation on the
metric and vector-multiplet degrees of freedom.  

The full action for vector and hypermultiplets
is super-Weyl invariant, and thus for an arbitrary
super-Weyl transformation of the background metric and
dynamical $SU(2)$ vector multiplet, there must exist
a corresponding transformation on the components
of the hypermultiplet that leaves the Lagrangian
invariant, not just up to a total derivative or local \ac{susy} transformation, but invariant exactly, since the virial
current must vanish.  The action is exactly
quadratic, and the transformation of the off-shell
hypermultiplet
component fields under the super-Weyl transformation
is linear.

The exact same super-Weyl
transformation can be applied as a linear
transformation to the off-shell ghost
hypermultiplet component fields, and the action will
necessarily still be invariant: For a quadratic action for a complex field, a linear transformation on a bosonic
field leaves the action invariant if and only if the corresponding action for a fermionic field also does so:
For a quadratic action for a complex field,
the statistics of the field
are irrelevant to the invariance of the action so long
as the transformation is linear.

We therefore conclude that the fixed
points with ghost-hypermultiplets are invariant under the same super-Weyl transformations as the \acp{scft} with  ordinary unitary hypermultiplets.

\section{Saddle point value of the classical action}
\label{sec:saddle-point}

In Section~\ref{Universality} we have given an indirect proof that the power law corrections
$\hat{K}\ll m / \JJM$ to the $q\ll n$ for any one-dimensional Coulomb
branch, must be given by the universal polynomials
$\hat{P}\univ\ll{m+1}(\a)$ given in Eq.~\eqref{eq:universal-P}, even
in a non-Lagrangian theory with no marginal coupling and no evident
reason to obey the gauge-theoretic recursion relations from which the
$\hat{P}\ll {m+1}(\a)$ were derived. Our proof was somewhat abstract,
and relies on the well-definiteness of unfamiliar path integrals
involving nonunitary matter in the hypermultiplet sector.

Due to the absence of higher-derivative $F$-terms,
the $\hat{K}\ll m$ are certainly well-defined universal, and
computable within the Coulomb-branch \ac{eft} itself,
independent of any data other than the $\a$-coefficient.
It is therefore possible in principle to check our result
directly by computing correlators in the \ac{eft}.

In this Appendix we perform the simplest possible
check of deriving the leading coefficients
of the $\hat{P}\ll{m+1}$ polynomials, which
give the terms of order $\a\uu{m+1} / \JJM\uu m$
in the expansion of $q\ll n$.

The terms of order $\a\uu{m+1} / \JJM\uu m$,
\it i.e., \rm the leading terms in the polynomials
$\hat{P}\ll {m+1}(\a)$, are most conveniently
computed by using the \ac{bps} helical property of the
classical solution:   The only change in the classical helical solution, at any
 order in $\a$, is the equilibrium value of $|\phi|$ at 
 fixed $\JJM$ and fixed frequency $\o = {1/r}$.

Modulo $\JJM$-independent normalization constants
for the $S\uu 4$ partition function and the operator
$\co$ itself, the $q\ll n$ are simply given by
the partition function with sources:
\bbb
q\ll n = \log \big [ \cc
Z\ll 0\cc\ampname\ll n \cc \big ] = \log( Z\ll n) = \exp{- W\ll n}\ ,
\eee
where $Z\ll n$ is the path integral over the action
\begin{equation}
S\ll n \equiv S_{\text{free + super-\acs{wz}}} + S\lrm{sources} .
\end{equation}

The sum of tree diagrams contributing to $W\ll n$ is
simply the classical action at the saddle point,
including the free action, sources,
and Wess-Zumino term.  Therefore the full $\a$-dependent
expression for the classical action, will give us
the leading terms of all the polynomials in $\hat{P}\ll{m+1}$:
\bbb
\sum\ll m \cc q\ll n\cc \bigg | \ll{\a\uu {m+1} / n\uu m~{\rm term}} = - S\ll n\cc \bigg |\lrm{saddle~point~value} .
\eee

In the $\dilaton, \b$ variables, the (Lorentz mostly-plus signature) classical action on the cylinder
$S\uu 3\times {\tt time}$ is
\begin{equation}
  S\uprm{Lorentzian} = \int \cc d\uu 4 x\cc \sqrt{|-g|} \cc {\cal L}\uprm{Lorentzian}\ ,
\end{equation}
which we can decompose into the kinetic and \ac{wz} term:
\begin{equation}
{\cal L}\uprm{Lorentzian} = {\cal L}\lrmns{kin}\uprm{Lorentzian} + {\cal L}\lrmns{super-WZ}\uprm{Lorentzian}\ ,  
\end{equation}
The kinetic term is simply
\begin{equation}
{\cal L}\lrmns{kin}\uprm{Lorentzian} \equiv -\m\cc \exp{-2\dilaton} \cc \bigg [ \cc (\pp\dilaton)\sqd + (\pp\b)\sqd - \frac{1}{6} \cc {\tt Ric}\ll 4 \cc \bigg ]
\label{FreeKineticTerm}
\end{equation}
where the Ricci scalar is given by \({\tt Ric}\ll 4 = {\tt Ric}\ll 3 =  6/r^2\).

We have written the \ac{wz} term in Eq.~\eqref{RestrictedBEOActionInPieces}.
The sources are at infinity and do not affect
the helical solution at all, except insofar as they
set the value $\JJM$ of the $R$-charge
for the solution.
The frequency of a \ac{bps} helical
solution is in general fixed by the \ac{bps} property (for a realization of this fact in the contest of string theory see~\cite{Hellerman:2011mv}) 
and indeed the \ac{eom} $\dot{\b}\sqd = 1/{r\sqd}$ for $\dilaton$ tells us immediately
that \(\beta = \pm 1/r\) is the only allowed frequency for a helical solution,
since the undifferentiated $\dilaton$ appears
only in the kinetic term ${\cal L}\lrm{kin}$ and its
variational equation fixes $\b\sqd = 1 /{r\sqd}$,
with the one sign for $\dot{\b}$ corresponding to the \ac{bps} helical solution and the other sign corresponding to an anti-\ac{bps} helical
solution.

For any given $\dilaton$, the $R$-charge density is simply the derivative of the Lagrangian density with respect
to $\dot{\b}$:
\begin{equation}
  \begin{aligned}
    \r &= {{\d{\cal L}}\over{\d\dot{\b}}} =  2 \m\sqd\cc \exp{- 2\dilaton} \cc \dot{\b} - {{8\cc (\Delta a)\ups{\acs{ks}}}\over{r\sqd}}\cc \dot{\b} - 8\cc (\Delta a)\ups{\acs{ks}} \cc \dot{\b}\uu 3 \\
    &= {\tt sgn}(\dot{\b}) \cc \bigg [ \cc 
    2 \m\sqd\cc \exp{- 2\dilaton} \cc r\uu{-1} - {{8\cc (\Delta a)\ups{\acs{ks}}}\over{r\uu 3}}  - 8\cc (\Delta a)\ups{\acs{ks}} \cc r\uu{-3}
    \cc\bigg ] \\
    &= {{2\cc {\tt sgn}(\dot{\b})}\over r} \cc \bigg [ \cc 
 \m\sqd\cc \exp{- 2\dilaton} - {{8\cc (\Delta a)\ups{\acs{ks}}}\over{r\uu 2}}
 \cc\bigg ] \\
    &= {{2\cc {\tt sgn}(\dot{\b})}\over r} \cc \bigg [ \cc 
  |\phi|\sqd  - {{8\cc (\Delta a)\ups{\acs{ks}}}\over{r\uu 2}}
 \cc\bigg ] \\
&= {1\over{{\cal A}\ll{S\uu 3}}} \cc \JJM = {1\over{2\pi\sqd \cc r\uu 3}}\times \JJM,
  \end{aligned}
\end{equation}
where in the last equality we have used
\bbb
\JJM = \mathcal{A}\ll{S\uu 3} \times \r = 2\pi\sqd \cc r\uu 3 \times \r\ .
\eee 
We see that ${\tt sgn}(\dot{\b}) = {\tt sgn}(\JJM)$ and $\JJM/{\dot{\b}} = \abs{\JJM}$, so
\bbb
 \m\sqd\cc \exp{- 2\dilaton}   - {{8\cc (\Delta a)\ups{\acs{ks}}}\over{r\uu 2}} =  |\phi|\sqd  - {{8\cc (\Delta a)\ups{\acs{ks}}}\over{r\uu 2}} = {1\over{4\pi\sqd \cc r\uu 2}}\times |\JJM|.
\eee
Solving for $|\phi|\sqd$ gives
\begin{equation}
\m\sqd\cc \exp{- 2\dilaton}  = |\phi|\sqd = {1\over{4\pi\sqd \cc r\uu 2}}\times |\JJM| +  {{8\cc (\Delta a)\ups{\acs{ks}}}\over{r\uu 2}} .
\end{equation}
Using the translation between the \ac{aefj} normalization and the \ac{aefj} normalization in~\cite{Anselmi:1997am} for the $a$-anomaly, given in Appendix~\ref{sec:solve-recurrence} of~\cite{Hellerman:2017sur} as
\begin{equation}
a\ups{\acs{ks}} = {1\over{16\pi\sqd}}\cc a\ups{\acs{aefj}}\ , \llsk\llsk a\ups{\acs{aefj}} = 16\pi\sqd \cc a\ups{\acs{ks}}\ ,
\end{equation}
we have
\begin{equation}
\m\sqd\cc \exp{- 2\dilaton}  = |\phi|\sqd = {1\over{4\pi\sqd \cc r\uu 2}} |\JJM| +  {{(\Delta a)\ups{\acs{aefj}}}\over{2\pi\sqd\cc r\uu 2}}\ , 
\end{equation}
and using the definition of the $\a$-coefficient,
\bbb
\a =  2\cc (\Delta a)\ups{\acs{aefj}}\ ,
\eee
we have
\begin{equation}
\m\sqd\cc \exp[- 2\dilaton]  = |\phi|\sqd = {1\over{4\pi\sqd \cc r\uu 2}} |\JJM| +  {{\a}\over{4\pi\sqd\cc r\uu 2}} 
= {1\over{4\pi\sqd}} \cc \pqty{|\JJM| + \a}\ ,
\end{equation}
as an exact statement in the classical solution.
Comparing the value at $\a = 0$, we have
\begin{equation}
  \frac{\abs{\phi}^2}{\eval{\abs{\phi}^2}_{\alpha = 0}} = f^2(\alpha)
\end{equation}
with
\begin{equation}
 f(\a) = \sqrt{1+ {{\a}\over{|\JJM|}}} = \sqrt{1 + \hat{\a}}\ .
\end{equation}
Substituting $\dot{\b} \to {1\over r}$, and solving for $\dilaton$ in terms of $\JJM$ yields
\begin{equation}
 |\phi\ll\a|\sqd = {1\over{4\pi\sqd \cc r\uu 2}}\times |\JJM| +  {{\a}\over{4\pi\sqd\cc r\uu 2}} 
= {1\over{4\pi\sqd}} \cc (|\JJM| + \a) = f\sqd(\a) 
|\phi\ll 0|\sqd\ ,
\end{equation}
where $\phi\ll\a$ is the solution for nonzero $\a$
and $\phi\ll 0$ is the solution at $\a = 0$.

The ratio $ |\phi| / |\phi|\ll 0$ is Weyl-invariant
and is also constant over space (equal to $f(\a)$)
in cylinder frame; therefore it is
constant in all conformal frames.

Modulo $\phi$-independent terms proportional to $n\uu 1$, the saddle-point
value of the source term in the action is
\bbb
S\lrm{source}[\a] = - \JJM\cc \log[\phi\ll\a(x)] 
- \JJM\cc \log[\phb\ll\a(y)] 
\xxn
= S\lrm{source}[\a = 0] 
- \JJM\cc \log[f\sqd(\a)]
\xxn
=  S\lrm{source}[\hat{\a} = 0] 
- \JJM\cc \log[1 + \hat{\a}].
\eee

The kinetic term is $R$-symmetry invariant
and exactly quadratic in $|\phi|$.
In~\cite{Hellerman:2017sur} it was found that 
\begin{equation}
  \abs{\del \phi(x)}^2 = \frac{\JJM}{2} \delta^{(4)} (x - x_1) + \frac{\JJM}{2} \delta^{(4)} (x - x_2)
\end{equation}
so the saddle-point value of the free kinetic term on flat
space, including $\d$-function contributions 
at the insertion points, is
\begin{equation}
  \begin{aligned}
    S\lrm{kinetic}[\a] &= f(\a)\sqd \cc S\lrm{kinetic}[\a = 0]
    =  (1 + \hat{\a}) \cc S\lrm{kinetic}[\a = 0] \\
    &= (1 + \hat{\a}) \cc S\lrm{kinetic}[\hat{\a} = 0]
    = \JJM \cc (1 + \hat{\a})\ .
  \end{aligned}
\end{equation}

Finally, the super-\ac{wz} term contains only gradients of $\b$ and differentiated
logarithms of $|\phb|$, and therefore
$f(\a)$ drops out of the super-\ac{wz} term altogether,
except for the Euler-density piece:
\begin{equation}
  \label{ClassicalWZActionAtSaddlePointValue}
  \begin{aligned}
    S\lrmns{super-WZ}\uprm{Euclidean}[\a] ={}& S\lrmns{super-WZ}\uprm{Euclidean}\cc \bigg |\ll{\phi = \phi\ll\a} = S\lrmns{super-WZ}\uprm{Euclidean}\cc \bigg |\ll{\phi = \phi\ll 0} 
    - \alpha \log(f(\a)) \int \dd[4]x \sqrt{-g} E\ll 4^{\IZ} \\
    ={}&
\bigg ( {\rm order~}\a\uu 1\cc \log(\JJM)~{\rm term}
\bigg ) +  \bigg ( {\rm order~}\a\uu 1\cc \JJM\uu 0~{\rm term}
\bigg ) \\
& - \alpha \log(f(\a)) \int \dd[4]x \sqrt{-g} E\ll 4^{\IZ} .
  \end{aligned}
\end{equation}
Of these, the first was already computed in~\cite{Hellerman:2017sur}
and is equal to $-\a\cc \log(\JJM)$ and thus contributes to $q\ll n$ as $+ \a\cc\log(\JJM)$.
For purposes of the computation in this section we are really only
interested in terms that are order $\hat{\a}\sqd$ and higher, and so we ignore both the first two terms
in the \ac{wz} action.
These two contain only terms linear in $\a$; only
the third term contains terms of order $\a\sqd$ and
larger.

As in~\cite{Hellerman:2017sur} we convert the \ac{ks} normalization of the Euler density into 
the integer-normalization of the Euler density, 
\emph{i.e.} the one normalized so that
\begin{equation}
\int \cc d\uu 4 x \cc \sqrt{|g|} \cc E\ll 4\uu \IZ = \chi({\bf X}_4) \in \IZ
\end{equation}
with the convention \(\chi(S^4) = 2\).
The proportionality constant is~\cite{Hellerman:2017sur}
\bbb
E\ll 4\ups{\acs{ks}} = 32\pi\sqd\cc E\ll 4\uu \IZ\cc \ , 
\llsk\llsk
E\ll 4\uu \IZ = {1\over{32\pi\sqd}}\cc  E\ll 4\ups{\acs{ks}}
\eee
so the third term of \rr{ClassicalWZActionAtSaddlePointValue} is
\begin{equation}
  \begin{aligned}
  {\cal L}\lrmns{super-WZ}\uprm{Euclidean}[\a] \cc \bigg |\ll{E\ll 4 \cc \log[f(\a)]} &= (\Delta a)\ups{\acs{ks}}\cc \dilaton \cc E_4\ups{\acs{ks}}
= {1\over{{16\pi\sqd}}}\cc (\Delta a)\ups{\acs{aefj}} 
\times (32\pi\sqd) \cc \dilaton \cc  E_4\ups{\IZ} \\
&= 2\times (\Delta a)\ups{\acs{aefj}} \times \dilaton \cc  E_4\ups{\IZ}
=  \a\times \dilaton \cc  E_4\ups{\IZ}\ ,
\end{aligned}
\end{equation}
so for the four-sphere
\begin{equation}
  \begin{aligned}
    S\lrmns{super-WZ}\uprm{Euclidean}[\a] \cc \bigg |\ll{E\ll 4 \cc \log[f(\a)]} ={}& 2\a\dilaton
    = - 2\a\cc \log(|\phi|/\m)
= - \a\times \log(|\phi|\sqd/\m\sqd) \\
={}& - \JJM\hat{\a} \times \log(|\phi|\sqd/\m\sqd)\\
={}& - \JJM\hat{\a} \times \log(|\phi\ll 0|\sqd/\m\sqd)
- \JJM\hat{\a} \times \log(|\phi\ll a|\sqd/|\phi\ll 0|\sqd) \\
={}&   - \JJM\hat{\a} \times \log(|\phi\ll 0|\sqd/\m\sqd) - \JJM\hat{\a} \times \log[f\sqd(\a)] \\
 &- \JJM\hat{\a} \times \log(|\phi\ll 0|\sqd/\m\sqd) - \JJM\hat{\a} \times \log[1 + \hat{\a}].
  \end{aligned}
\end{equation}
We can write this as
\begin{equation}
  \begin{aligned}
S\lrmns{super-WZ}\uprm{Euclidean}[\a] \cc \bigg |\ll{E\ll 4 \cc \log[f(\a)]} &= - \JJM\hat{\a} \times \log(|\phi\ll 0|\sqd/\m\sqd) - \JJM\hat{\a} \times \log[1 + \hat{\a}] \\
&= \bigg ( \cc {\rm order~}\JJM\uu 1\cc\hat{\a}\uu 1 \cc\bigg )
- \JJM\hat{\a} \times \log[1 + \hat{\a}]\ .
\end{aligned}
\end{equation}
So putting it all together, we have
\begin{align}
S\lrmns{kinetic}\uprm{Euclidean}[\a]
&= \bigg ( \cc {\rm affine~in~}\a \cc \bigg )\ ,\\
S\lrmns{source}\uprm{Euclidean}[\a] &= 
\bigg ( \cc {\rm order~}\JJM\uu 1 \hat{\a}\uu 0 \cc \bigg )
- \JJM\cc \log(1 + \hat{\a})\ ,\\
S\lrmns{super-WZ}\uprm{Euclidean}[\a] &= 
\bigg ( \cc {\rm affine~in~}\a \cc \bigg ) - \JJM\cc \hat{\a} \cc 
 \log(1 + \hat{\a})\ ,
\end{align}
and the complete saddle point action is
\begin{equation}
  \begin{aligned}
S\lrmns{saddle~point,~total}\uprm{Euclidean}[\a] &=
\bigg ( \cc {\rm affine~in~}\a \cc \bigg ) - 
\JJM\cc (1 + \hat{\a} )\cc 
 \log(1 + \hat{\a})\\
 &= \bigg ( \cc {\rm affine~in~}\a \cc \bigg ) -
\JJM\times \bigg [ \cc \a + \sum\ll{m \geq 1} \cc (-1)\uu {m+1}\cc {{\hat{\a}\uu{m+1}}\over{m(m+1)}} \cc \bigg ]\\
&= \bigg ( \cc {\rm affine~in~}\a \cc \bigg ) -
\JJM\times \bigg [ \cc \sum\ll{m \geq 1} \cc (-1)\uu {m+1}\cc {{\hat{\a}\uu{m+1}}\over{m(m+1)}} \cc \bigg ]
\end{aligned}
\end{equation}
so for all $m\geq 1$ we have
\begin{equation}
  \label{eq:universal-saddle-value}
  \begin{aligned}
    K\ll n \cc n\uu{-m}\cc \bigg |\ll{{\rm order~}\a\uu{m+1}} &= \hat{K}\ll m \JJM\uu{-m}\cc \bigg |\ll{{\rm order~}\hat{\a}\uu{m+1}} 
 = \JJM\uu{+1}\cc \hat{P}\ll{m+1}(\hat{\a}) \cc \bigg |\ll{{\rm order~}\hat{\a}\uu{m+1}} \\
 &=  q\ll n \cc \bigg |\ll{{\rm order~}\hat{\a}\uu {m+1}\cc \JJM\uu {+1}}
= - S\ll{n,~{\rm saddle~point,~total}}\uprm{Euclidean}[\a]
 \cc \bigg |\ll{{\rm order~}\hat{\a}\uu {m+1}\cc \JJM\uu {+1}}\\
 &=
+ \JJM\times (-1)\uu {m+1}\cc {{\hat{\a}\uu{m+1}}\over{m(m+1)}}\ ,
\llsk \forall m \geq 1\ .
\end{aligned}
\end{equation}

This result is to be compared with the formula for the polynomials $\hat{K}\ll m = \hat{P}\ll{m+1}(\a)$ in Eq.~\eqref{eq:universal-P} and gives an infinite number of direct computations consistent with our 
universal formula .  It is a
direct calculation in the \ac{eft} and it is independent of the ghost-hyper argument and independent of any
\ac{uv}-completion of the \ac{eft}.
This infinite number of agreeing coefficients supports our argument that our
formula for the power-law corrections $\hat{K}\univ\ll m$ is universal among all theories with a given value of $\a$,
including non-Lagrangian theories.

\section{Numerics}
\label{sec:numerics}

In the case of \(\mathcal{N}=2\) \ac{sqcd} with \(4\) flavors the correlators that we discuss can be computed via localization~\cite{Gerchkovitz2017}.
The function \(G_{2n}\) is the ratio of two determinants:
\begin{equation}
  G_{2n} = 4^{2n} \frac{\det(M_n)}{\det( M_{n-1})},
\end{equation}
where \(M_n\) is the upper-left \(\pqty{n-1} \times \pqty{n-1}\) submatrix of the (normalized) matrix of derivatives \(M\) of the partition function \(Z_0\):
\begin{equation}
  \eval{M}_{m,n} = \frac{1}{Z_0} \del^n \bar \del^m Z_0 .
\end{equation}
The partition function for \(\mathcal{N}=2\) \ac{sqcd} is written in terms of the Barnes \(G\)-function~\cite{barnes}:
\begin{equation}
  Z_0 = Z_{S^4}^{\acs{sqcd}}(\tau, \bar \tau) = \int_{-\infty}^\infty \dd{a} a^2 e^{-4 a \Im \tau} \frac{\abs{G(1 + 2 i a )}^4}{\abs{G(1 + i a)}^{16}} \abs{Z_{\text{inst}}(i a, \tau)}^2,
\end{equation}
where \(Z_{\text{inst}}\) is the instanton partition function~\cite{Nekrasov:2002qd,Alday:2009aq}:
\begin{equation}
  Z_{\text{inst}}(a, \tau) = 1 + \frac{1}{2} \pqty{a^2 - 3} e^{2 \pi i \tau } + \order{ e^{4 \pi i \tau}}.
\end{equation}
For simplicity we will consider the regime \(\Im \tau > 1\) and ignore the instanton corrections.
Note that in this approximation the partition function \(Z_{\text{inst}}(a, \tau)\) is independent of \(\Re(\tau)\).

Since we want to evolve the recursion relations numerically
starting from an approximate initial condition for the
$S\uu 4$ partition function,  we need to estimate the sensitivity of large-$\JJM$ correlation functions 
to imprecise initial conditions.

We may wish to start at some initial value $n\ll i$ greater
than $0$.  The recursion relation is second order, so in order to define initial conditions, we need to define both
$q\ll {n\ll i}$ and $q\ll{n\ll i + 1}$.  These initial
conditions are of course functions of $\t$ and $\tb$,
but we will suppress in this section the dependence on the
arguments $\t,\tb$ in our notation.

It is useful to write
the rank-one recursion relations
in their ``deterministic'' form.  Given any initial
conditions at $n\ll i, n\ll i + 1$, 
there is always a unique solution to the recursion relations for $n \geq n\ll i + 2$.  One can consider two nearby
solutions, separated by a small amount $\d\ll n$, 
and analyze how the linearized deviation propagates
to larger values of $n$.  The deviation propagation equation is:
\def\ErrorAnalysisDefoutA{ %
\bbb
E\ll{n + 2} = E\ll{n+1} \cc \pp\ll\t\pp\ll\tb\cc\bigg [
\cc \log(E\ll{n+1}) \cc \bigg ] + {{E\ll{n+1}\sqd}\over
{E\ll n}}\ ,
\xxn
E\ll n \equiv \exp{q\ll n}\ .
\een{RecursionRelationsDeterministicFormForExponential}
In terms of $q\ll n$ itself this is
\bbb
q\ll{n+2} = \log(E\ll {n+2}) = \log\cc\bigg [ \cc
\exp{2 q\ll {n+1} - q\ll n} + \exp{q\ll{n+1}} \cc \pp\ll\t
\pp\ll\tb q\ll{n+1}
\cc\bigg ]
\xxn
= 2 q\ll {n+1} - q\ll n + 
 \log\cc\bigg [ \cc
1 + \exp{q\ll n - q\ll{n+1}} \cc \pp\ll\t
\pp\ll\tb q\ll{n+1}
\cc\bigg ]
\een{RecursionRelationsDeterministicFormForqItself}
which we can write as
\bbb
q\ll{n+2} - 2q\ll {n+1} + q\ll n =
\log\cc\bigg [ \cc
1 + \exp{q\ll n - q\ll{n+1}} \cc \pp\ll\t
\pp\ll\tb q\ll{n+1} \cc\bigg ] .
\een{RecursionRelationsSecondOrderForm}

\heading{General evolution equation for small perturbations}

Now consider some
small perturbations $\d\ll {n\ll i}, \d\ll{n\ll i + 1}$ of the initial
conditions that are small relative 
to the initial conditions $q\ll{n\ll i}, q\ll{n\ll i + 1}$  themselves:
\bbb
r\ll {n\ll i}, r\ll{n\ll i + 1}  \muchlessthan 1 \ ,
\llsk\llsk
r\ll n \equiv {{\d\ll n}\over{q\ll n}}\ .
\een{RelativeErrorSmall}
Within this assumption, the propagation equation
for $r\ll n$ can always be treated linearly.

Suppose we have an unperturbed solution to the recursion relations \rr{RecursionRelationsDeterministicFormForqItself}, for $n \geq n\ll i + 2$,
and we perturb it with some small change to the
initial conditions $\d\ll {n\ll i}, \d\ll {n\ll i} $, so
\bbb
q\ll {n\ll i}\uprm{new} = q\ll {n\ll i} \uprm{unp}  + \d\ll{n\ll i}\ ,
\xxn
q\ll {n\ll i + 1}\uprm{new} = q\ll {n\ll i + 1} \uprm{unp}  + \d\ll{n\ll i+1}\ ,
\een{InitialConditionRelativeErrorSmall}
with $\d\ll {n\ll i}, \d\ll{n\ll i + 1}$ satisfying \rr{RelativeErrorSmall}.

The recursion relation is of course continuous as a function
of its initial conditions, so the smallness of the perturbation of the initial conditions \rr{InitialConditionRelativeErrorSmall}
implies the smallness of the perturbation \rr{RelativeErrorSmall} for $n - n\ll i$ not too large.

If both $q\uprmns{new}\ll n$ and $q\uprmns{unp}\ll n$
obey the recursion relations exactly for $n\geq n\ll i + 2$,
then the recursion relation for $\d\ll n$ is
\bbb
\d\ll{n+2} - 2\d\ll {n+1} + \d\ll n 
\xxn
=
\bigg [ \cc
1 + {{E\ll n\uprm{unp}}\over{E\ll{n+1}\uprm{unp}}} \cc \pp\ll\t
\pp\ll\tb q\uprmns{unp}\ll{n+1} \cc\bigg ]\uu{-1}
\times \bigg ( \cc 
{{E\ll n\uprm{unp}}\over{E\ll{n+1}\uprm{unp}}} \cc \pp\ll\t
\pp\ll\tb \d\ll{n+1}
+ {{E\ll n\uprm{unp}}\over{E\ll{n+1}\uprm{unp}}} 
\times (\d\ll n - \d\ll{n+1})\cc \pp\ll\t
\pp\ll\tb q\uprmns{unp}\ll{n+1} 
\cc\bigg )
+
 O(\d\sqd)
 \xxn
 = 
 \bigg [ \cc
1 + {{E\ll n\uprm{unp}}\over{E\ll{n+1}\uprm{unp}}} \cc \pp\ll\t
\pp\ll\tb q\uprmns{unp}\ll{n+1} \cc\bigg ]\uu{-1}
\times {{E\ll n\uprm{unp}}\over{E\ll{n+1}\uprm{unp}}} \times
 \bigg ( \cc  \pp\ll\t
\pp\ll\tb \d\ll{n+1}
+  (\d\ll n - \d\ll{n+1})\cc \pp\ll\t
\pp\ll\tb q\uprmns{unp}\ll{n+1} 
\cc\bigg )
+
 O(\d\sqd)
 \xxn
  = 
 \bigg [ \cc
 {{E\ll {n+1}\uprm{unp}}\over{E\ll n\uprm{unp}}} + \pp\ll\t
\pp\ll\tb q\uprmns{unp}\ll{n+1} \cc\bigg ]\uu{-1}
\times \bigg ( \cc  \pp\ll\t
\pp\ll\tb \d\ll{n+1}
+  (\d\ll n - \d\ll{n+1})\cc \pp\ll\t
\pp\ll\tb q\uprmns{unp}\ll{n+1} 
\cc\bigg )
+
 O(\d\sqd)
 \ , \llsk\forall n\geq n\ll i\ .
\een{PerturbationEvolutionEquationGeneralv1}
Since we will be always assuming the small-deviation condition $\rdots$, we will discard the $O(\d\sqd)$
term, in which case we can ignore the distinction
between $q\uprm{unp}$ and $q\uprm{new}$ in
the perturbation evolution equation \rr{PerturbationEvolutionEquationGeneralv1}.  So 
treating $\d$ as a purely infinitesimal quantity we can write:
\bbb
\hskip-.75in
\d\ll{n+2} - 2\d\ll {n+1} + \d\ll n = 
\bigg [ \cc
 {{E\ll {n+1}}\over{E\ll n}} + \pp\ll\t
\pp\ll\tb q\ll{n+1} \cc\bigg ]\uu{-1}
\times \bigg ( \cc  \pp\ll\t
\pp\ll\tb \d\ll{n+1}
+  (\d\ll n - \d\ll{n+1})\cc \pp\ll\t
\pp\ll\tb q\ll{n+1} 
\cc\bigg ) \ , \llsk\forall n\geq n\ll i\ . ~~~
\een{PerturbationEvolutionEquationGeneralv2}

We shall explore the question of the rate of growth
of $\d\ll n$ for different
regimes of $n,\t,$ and so forth, always taking
the perturbation to obey 
\rr{RelativeErrorSmall}.

\heading{Evolution of perturbations of solutions with EFT asymptotics at large $n$}

Now let us examine a solution to the recursion
equations in which the $q\ll{n\ll i}\uprm{unp}$ 
asymptotes to a solution with the large-$n$ asymptotics considered here, \it i.e., \rm
\bbb
q\ll{n\ll i}\uprm{unp} = q\ll{n\ll i}\uprm{EFT} + {\cal W}\ll n(\t,\tb)\ ,
\eee
where we have defined
\bbb
q\ll n\uprm{EFT} \equiv
{\bf A}(\t,\tb)\cc n + {\bf B}(\t,\tb) + \log\bigg [\G(2n + \a + 1) \cc \bigg ]\ ,
\xxn
 {\cal W}\ll n(\t,\tb) = \bigg ({\rm exponentially~small~in~} n\bigg )\ ,
\een{UniversalEFTSolutionForm}
with
\bbb
{\bf A}(\t,\tb) = -2 \cc \log[\Im(\t)]- 4 \log(2)\ ,
\xxn
{\bf B}(\t,\tb) = - (\a + \hh) \cc  \log[\Im(\t)] + k\ll 0\ ,
\een{UniversalEFTNormalizationConstantForm}
with $k\ll 0$ independent of $\t$ and $\tb$.

For these asymptotics, the error propagation equation
becomes
\bbb
\d\ll{n + 2} + \d\ll n - 2 \d\ll{n+1} = \rdots +  \bigg ({\rm exponentially~small~in~} n\bigg )\ .
\eee

Now we observe, as \opt{noted}{not noted} earlier
in the paper, that the form \rr{UniversalEFTSolutionForm},
\rr{UniversalEFTNormalizationConstantForm}
with ${\cal W}\ll n = 0$ identically, is an exact
solution to the recursion relations for $n \geq (\rdots)$.
This is equivalent to setting
\bbb
q\ll n\uprm{unp} \to q\ll n\uprm{EFT}\ , \llsk\llsk
{\cal W}\ll n(\t,\tb) \to 0
\eee
identically as the definition of the unperturbed solution,
and absorbing ${\cal W}\ll n(\t,\tb)$ into the definition of
the deviation $\d\ll n$,
\bbb
\d\ll n \to \d\ll n + {\cal W}\ll n\ ,
\eee
 which is consistent long as we are at $n$ which is large enough that ${\cal W}\ll n(\t,\tb)$ 
can be treated as small.  Then we have
\bbb
E\ll n = \big [ \cc \exp{{\bf A}\ll n} \cc \big ]\uu n\times
\big [ \cc \exp{{\bf B}} \cc \big ] \times \G(2n+ \a + 1)
\xxn
= \big [ \Im(\t) \big ]\uu{-2n - \a - \hh} \times 
2\uu{-4n} \times \exp{k\ll 0} \times \G(2n+ \a + 1)
\xxn
{{E\ll {n+1}}\over{E\ll n}} = 
{1\over{16\cc [\Im(\t)]\sqd}}\cc {{\G(2n+\a + 3)}\over{\G(2n+\a + 1)}} = {1\over{16\cc [\Im(\t)]\sqd}}\cc
{{(2n+\a + 2)!}\over{(2n+a)!}}
\xxn
= {{(2n+\a + 2)(2n+\a + 1)}\over{16\cc  [\Im(\t)]\sqd}}\ ,
\xxn
{{E\ll n}\over{E\ll{n-1}}} = {{(2n+\a )(2n+\a - 1)}\over{16\cc  [\Im(\t)]\sqd}}\ ,
\xxn
\pp\ll\t\pp\ll\tb q\ll n = -(2n + \a + \hh) \cc \pp\ll\t\pp\ll\tb
\cc\bigg (  \log[{\rm Im}(\t)] \bigg ) = -(2n + \a + \hh) 
\times (\t - \tb)\uu{-2} 
\xxn
= 
-(2n + \a + \hh) 
\times \bigg ( \cc 2 i\cc [\Im(\t)] \cc \bigg )\uu{-2}
= + {1\over 4} \cc (2n + \a + \hh) \cc [\Im(\t)] \uu{-2}
\xxn
\pp\ll\t\pp\ll\tb q\ll {n+1} = + {1\over 4} \cc (2n + \a + {5\over 2}) \cc [\Im(\t)] \uu{-2}\ ,
\eee
and 
the propagation
equation for the deviation becomes exactly
}%
\begin{multline}
  \d\ll{n + 2} + \d\ll n - 2 \d\ll{n+1} = {{16  \pqty{\Im(\t)}\sqd}\over{(2n+3+\a)(2n+4+\a)}} \cc \pp\ll\t\pp\ll\tb\cc \d\ll{n+1}\\
- {{2\cc (4n + 2\a + 5)}\over{(2n + \a + 3)(2n+\a + 4)}} \pqty{\d\ll {n+1} - \d\ll n} .
\end{multline}
Even at the linearized level, this equation is nontrivial,
and depends on the decomposition of the
error into eigenvalues of the Laplacian on the upper half
plane or its quotient under the modular group.
We do not
analyze the propagation of errors for general perturbations.
Instead, we use the fact that the perturbative 
piece of $Z\ll 0(\t,\tb)$ is a good approximation
at weak coupling.  As pointed out in~\cite{Bourget:2018obm},
the clash between weak coupling and large $\JJM$
can be avoided if one considers the limit $\JJM\to\infty$
while taking $\l\equiv  2\pi \,\JJM / \Im(\t)$
fixed.  Since our formula for the power-law corrections
is $\t$-independent for rank-one theories,
these two limits coincide for the power-law piece
$\log(\G(\JJM + \a +1))$, differing
only in the behavior of the nonuniversal
exponential correction.
We can therefore isolate this correction easily
in the fixed-$\l$ limit, in which the instanton
contributions to $Z\ll 0(\t,\tb)$ go to zero exponentially
in $n$.  

One might expect the exponentially small corrections
to be associated with the breakdown of the \ac{eft} altogether,
capturing the leading effects of massive states
propagating over distances on the infrared scale, as
discussed in Sec.~\ref{NonuniversalCorrectionDiscussion}.
One would therefore anticipate exponentially small
corrections proportional to
$ \propto \exp[ - \kk \lambda^{1/2}]$, with $\kk$ some fixed number depending on the geometry
of the virtual propagation, but not on $n$ or $\t,\tb$.
Numerically, we find a remarkably accurate match
to such an exponential, with $\kk = {{\sqrt{\pi}}/ 2}$,
as shown below.

\bigskip

In Section~\ref{sec:alpha-dep} we have seen that only the coefficients of \(n^0\) and \(n^1\) in the asymptotic expansion of \(q_n(\tau, \bar \tau)\) are expected to depend on \(\tau\).
This means that the second variation in \(n\) of \(q_n(\tau, \bar \tau)\) is \(\tau\)-independent.
Let \(\difference\) be the difference operator \(\Delta_n q_n = q_{n+1} - q_n\).
We want to compute the second difference
\begin{equation}
  \difference_n^2 q^{\text{(loc)}}_n (\tau, \bar \tau) = q^{\text{(loc)}}_{n+2}(\tau, \bar \tau) - 2 q^{\text{(loc)}}_{n+1}(\tau, \bar \tau) + q^{\text{(loc)}}_{n}(\tau, \bar \tau) 
\end{equation}
and compare it with the result in Eq.~\eqref{eq:recursion-result-gamma}:
\begin{equation}
  \difference_n^2 q^{\acs{eft}}_n = \log( \frac{\pqty{2n + \alpha + 3 } \pqty{ 2n + \alpha + 4}}{\pqty{2n + \alpha + 1} \pqty{ 2n + \alpha + 2}}) .
\end{equation}

Figure~\ref{fig:delta-n-fixed-n} shows the results of a numerical computation for imaginary values of \(\tau\) between \(1\) and \(60\) and for \(n\) between \(1\) and \(40\), representing the values of \(\difference^2_n q^{\text{(loc)}}\) as function of \(\tau\) at fixed values of \(n\).
We see that quite rapidly, already for \(\tau \simeq 4i\), the \(\tau\)-dependence drops for all values of \(n\).
The asymptotic value is well approximated by \(\difference^2_n q_n^{\acs{eft}}\) for \(n\) larger that \(n \gtrsim 5\), where the discrepancy is of order \(1 - \eval{\difference^2_n q^{\acs{eft}}/ \difference^2_n q^{\text{(loc)}}}_{n = 5, \tau \muchgreaterthan 1} \approx 1\%\).
At \(n = 1\), the discrepancy is of order \(1 - \eval{\difference^2_n q^{\acs{eft}}/ \difference^2_n q^{\text{(loc)}}}_{n = 1, \tau \muchgreaterthan 1} \approx 8\%\).

The numerical data can help us estimate the \(\tau\) and \(n\) dependence of the difference \(\difference^2_n( q^{\text{loc}}_n - q^{\text{us}}_n )\).
As discussed in Section~\ref{NonuniversalCorrectionDiscussion}, we expect the leading contribution to the difference to have the form
\begin{equation}
  q^{\text{(loc)}}_n - q^{\acs{eft}}_n \sim f_n(\tau, \bar \tau) e^{-\kappa \sqrt{n/\Im \tau}} = f_n(\tau, \bar \tau) e^{-\kappa \sqrt{\lambda/(4 \pi)} } ,
\end{equation}
where \(\lambda = n g^2 = 4 \pi n / \Im \tau\).
To verify this conjecture and estimate the proportionality factor \(f_n(\tau , \bar \tau)\) and the coefficient \(\kappa\) we have computed the difference as a function of \(\tau\), keeping the ratio \(n / \Im \tau = \lambda / (4 \pi)\) fixed (see Figure~\ref{fig:exponential-correction}).
The numerical data is consistent with \(f_n(\tau, \bar \tau)\) being a constant approximately equal to \(f_n(\tau, \bar \tau) \approx 1.6\) and \(\kappa \approx \pi\).
Already for \(\tau \approx 3\) our conjecture seems to reproduce the localization data to high accuracy.
Interestingly, this single exponential term to our solution \(q^{(us)}_n (\tau, \bar \tau)\) seems to account for the discrepancy \(\difference^2_n \pqty{q^{\text{(loc)}}_n - q^{\acs{eft}}_n}\) both in the small-\(\tau\), large-\(n\) (\emph{i.e.} large-\(\lambda\)) regime and in the large-\(\tau\) regime (see Figure~\ref{fig:delta-n-fixed-n-instanton}).

\begin{small}
  \dosserif
  \bibliography{references-susy-two-point}{}
  \bibliographystyle{JHEP}
\end{small}

\end{document}